\documentclass[aps,prd,showkeys,nofootinbib,preprintnumbers,superscriptaddress,floatfix]{revtex4}

\usepackage{epsfig}
\usepackage{bm}
\usepackage{amssymb}
\usepackage{amsmath}
\usepackage{color}
\usepackage{subfigure}
\usepackage{dcolumn}

\usepackage[utf8]{inputenc}

\usepackage{amsmath}
\usepackage{amssymb}
\usepackage{amsthm}
\usepackage{amscd, amsfonts, mathrsfs}
\usepackage{cases}
\usepackage{verbatim} 
\usepackage{epsf}
\usepackage[bookmarksnumbered,pdfpagelabels=true,plainpages=false,colorlinks=true,
            linkcolor=black,citecolor=black,urlcolor=black]{hyperref}

\theoremstyle{plain}
\newtheorem{theorem}{Theorem}

\theoremstyle{definition}

\allowdisplaybreaks

\newcommand{\cA}{\mathcal A}

\newcommand{\cD}{\mathcal D}

\newcommand{\cI}{\mathcal I}

\newcommand{\cQ}{\mathcal Q}

\newcommand{\cT}{\mathcal T}

\newcommand{\al}{\alpha}
\newcommand{\be}{\beta}

\newcommand{\ep}{\epsilon}
\newcommand{\la}{\lambda}

\newcommand{\Si}{\Sigma}

\newcommand{\RR}{\mathbb R}

\newcommand{\cqd}{\hfill $\qed$\\ \medskip}
\newcommand{\rar}{\rightarrow}

\newcommand{\beq}{\begin{equation}}
\newcommand{\eeq}{\end{equation}}
\newcommand{\bea}{\begin{eqnarray}}
\newcommand{\eea}{\end{eqnarray}}

\newcommand{\nn}{N}
\newcommand{\mass}{\mathcal{M}}
\newcommand{\eqent}{S}

\begin{document}


\title{Causality and existence of solutions of relativistic viscous fluid dynamics with gravity}
\date{\today}

\author{F\'abio S.\ Bemfica}
\affiliation{Escola de Ci\^encias e Tecnologia, Universidade Federal do Rio Grande do Norte, 59072-970, Natal, RN, Brazil}
\email{fabio.bemfica@ect.ufrn.br}

\author{Marcelo M.\ Disconzi}
\affiliation{Department of Mathematics,
Vanderbilt University, Nashville, TN, USA}
\email{marcelo.disconzi@vanderbilt.edu}

\author{Jorge Noronha}
\affiliation{Instituto de F\'isica, Universidade de S\~ao Paulo, Rua do Mat\~ao,
1371, Butant\~a, CEP 05508-090, S\~ao Paulo, SP, Brazil}
\email{noronha@if.usp.br}


\begin{abstract}
A new approach is described to help improve the foundations of relativistic viscous fluid dynamics and its coupling to general relativity. Focusing on neutral conformal fluids constructed solely in terms of hydrodynamic variables, we derive the most general viscous energy-momentum tensor yielding equations of motion of second order in the derivatives, which is shown to provide a novel type of generalization of the relativistic Navier-Stokes equations for which causality holds. We show how this energy-momentum tensor may be derived from conformal kinetic theory. We rigorously prove existence, uniqueness, and causality of solutions of this theory (in the full nonlinear regime) both in a Minkowski background and also when the fluid is dynamically coupled to Einstein's equations. Linearized disturbances around equilibrium in Minkowski spacetime are stable in this causal theory. A numerical study reveals the presence of an out-of-equilibrium hydrodynamic attractor for a rapidly expanding fluid. Further properties are also studied and a brief discussion of how this approach can be generalized to non-conformal fluids is presented.
\end{abstract}

\keywords{Relativistic viscous fluid dynamics, causality, stability, existence of solutions, 
conformal fluids, quark-gluon plasma, hydrodynamic attractor.}


\maketitle

\section{Introduction\label{section_introduction}} 

Relativistic fluid dynamics is an essential tool in high-energy nuclear physics \cite{Heinz:2013th}, cosmology \cite{WeinbergCosmology}, and astrophysics \cite{RezzollaZanottiBookRelHydro}. For instance, it has been instrumental in the discovery of the nearly perfect fluidity of the quark-gluon plasma formed in ultrarelativistic heavy ion collisions \cite{Gyulassy:2004zy} and also in the modeling of complex phenomena involved in binary neutron star mergers \cite{Baiotti:2016qnr}. Its power stems directly from the conservation laws and the presence of a hierarchy among energy scales, which allows one to investigate the regular macroscopic motions of the conserved quantities without specifying the fate of the system's microscopic degrees of freedom. Thus, it is widely accepted that relativistic hydrodynamics may be formulated as an effective theory \cite{Baier:2007ix}.

Absent other conserved currents, ideal relativistic fluid dynamics is described by the equations of motion for the flow velocity $u_\mu$ (with $u_\mu u^\mu=-1$) and the energy density $\epsilon$ obtained via the conservation law $\nabla_\mu T_{ideal}^{\mu\nu}=0$, where 
$T^{ideal}_{\mu\nu} = \epsilon\, u_\mu u_\nu +P(\epsilon)\Delta_{\mu\nu}$
is the energy-momentum tensor. The pressure $P(\epsilon)$ is given by an equation of the state, which is determined from the microscopic dynamics or phenomenologically, 
$g_{\mu\nu}$ is the spacetime metric, and $\Delta_{\mu\nu}=g_{\mu\nu}+u_\mu u_\nu$ is the projector orthogonal to $u^\mu$. The fluid equations of motion in this case are of 1st order in spacetime derivatives and the initial value problem is well-posed, i.e., given suitable initial data for $\epsilon$ and $u_\mu$, one can prove that the system admits a unique solution (see below
for a precise definition and discussion of well-posedness); this is true both in the case of a Minkowski background \cite{AnileBook} as well as when the fluid equations are dynamically coupled to Einstein's equations \cite{Choquet-BruhatFluidsExistence, DisconziRemarksEinsteinEuler}. In both cases, the solutions are causal, i.e., the field values at a point $x$ in spacetime are completely determined by the region in spacetime that is in the past of and causally connected to $x$ (see a precise
definition below). The physical meaning of causality is that information cannot propagate at superluminal speeds. 

\begin{figure}[th]
\includegraphics[width=0.6\textwidth]{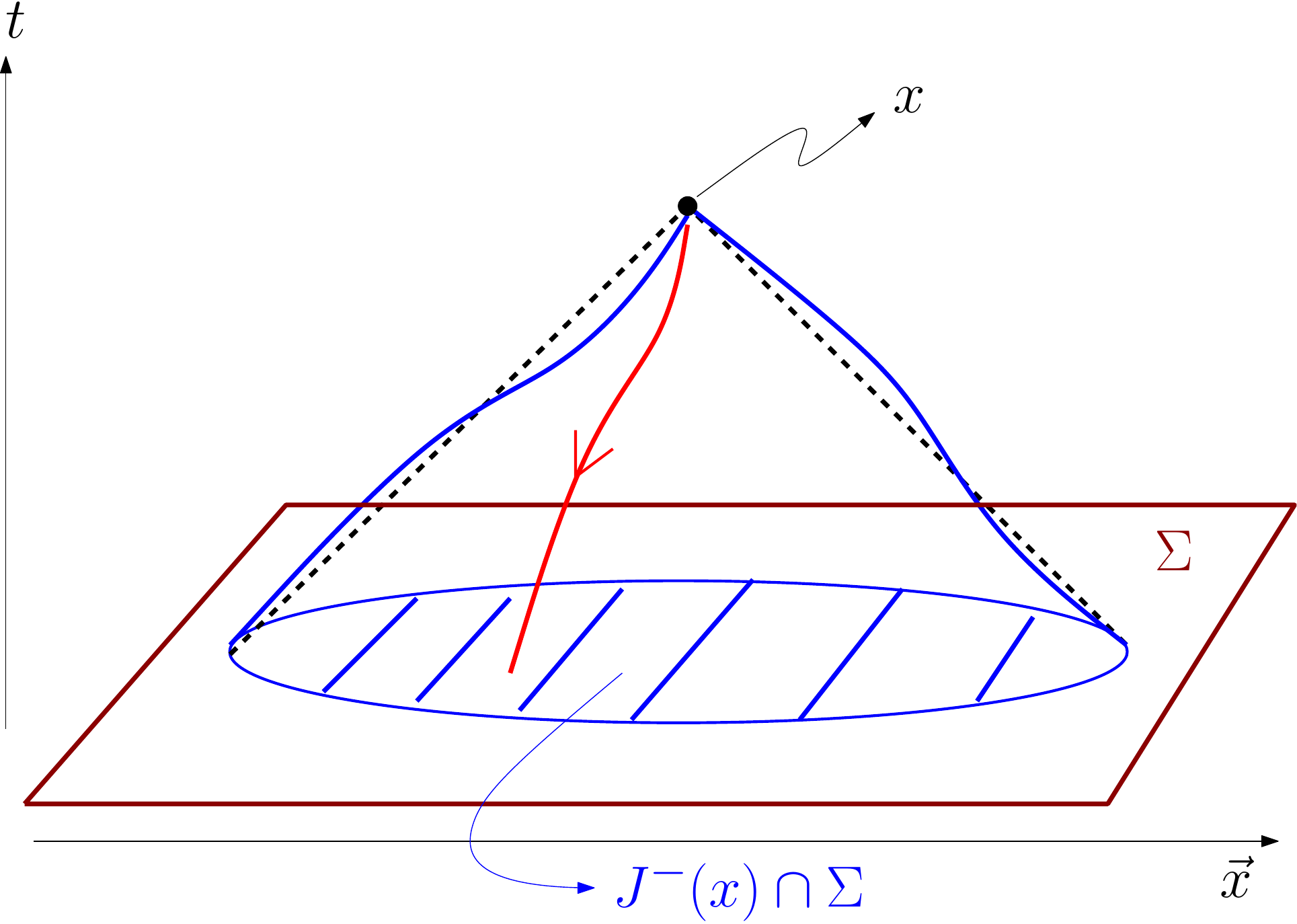}
\caption{(color online) Illustration of causality. In curved spacetime $J^-(x)$ looks like a distorted light-cone opening to the past (blue region); in flat spacetime the cone would be straight (dotted line). Points inside $J^-(x)$ can be joined to a point $x$ in spacetime by a causal past directed curve (e.g. the red line). The Cauchy surface $\Si$ supports the initial data and the value of the field $\varphi(x)$ depends only on the initial data on $J^-(x) \cap \Si$.}
\label{fig1}
\end{figure}

As the concept of causality is central in this paper, here we recall its precise definition. Let $(M,g)$ be a globally hyperbolic Lorentzian manifold\footnote{Global hyperbolicity forbids several pathologies that would otherwise complicate our analysis.} (both Minkowski spacetime and spacetimes that arise as solutions to the initial-value problem for Einstein's equations are globally hyperbolic, so this assumption covers most cases of interest). Consider on $M$ a system of (linear or non-linear) partial differential equations, which we write as  $P^I_K \varphi^K  = 0$, $I,K=1,\dots, N$, where $\{\varphi^K\}_{K=1}^N$ are the unknowns. Let $\Si \subset M$ be a Cauchy surface where initial data is prescribed. The system is causal if for any point $x$ in the future of $\Si$, $\varphi^K(x)$ depends only on the initial data on $J^-(x) \cap \Si$, where $J^-(x)$ is the causal past of $x$ \cite[page 620]{ChoquetBruhatGRBook}  \cite[Theorem 10.1.3]{WaldBookGR1984} (see Fig.\ \ref{fig1}). 
In particular, causality implies that $\varphi^K(x)$ remains unchanged if the initial data is 
altered\footnote{Causality can be equivalently stated in the following manner. 
If $\{ \varphi^K_0 \}_{K=1}^N$ and 
$\{ \widetilde{\varphi}^K_0 \}_{K=1}^N$ are two sets of initial data for the system such that 
$\varphi_0^K = \widetilde{\varphi}_0^K$ on a subset $S \subset \Si$, and $\varphi^K$ and 
$\widetilde{\varphi}^K$ are the 
corresponding solutions to the equations, then $\varphi^K = \widetilde{\varphi}^K$ on $D_g^+(S)$, where 
$D^+_g(S)$ is the future domain of dependence of $S$ \cite[Theorem 10.1.3]{WaldBookGR1984}.} only outside $J^-(x) \cap \Si$. 

Causality lies at the foundation of relativity theory, so
the matter sector in Einstein's equations (i.e., the fluid) must be compatible with this general principle.   
Hence, in the regime where an effective hydrodynamic theory is expected to provide an accurate description of the system's dynamics, such a theory must be causal, even in Minkowski background, if it is to describe fully \emph{relativistic} phenomena.

To be more specific, in practice a given effective theory description may be allowed to violate causality if these violations lead only to \emph{unobservable} 
phenomena \cite{Geroch-RelativisticDissipative, Herrera:2001if, Kostadt:2001rr, Kostadt:2000ty, LindblomRelaxationEffect}. Such a scenario, however, poses undesirable features. Firstly, one needs to have a precise and quantitative understanding of causality violations in order to trace its consequences and prove that they are always unobservable. It is not clear how such a task may be performed in hydrodynamics at the full nonlinear level. Secondly, the property that causality violations are unobservable might depend on particular modeling choices, preventing one from drawing general conclusions. Thus, the possibility of constructing a simple and robust formalism that can be applied to the study of relativistic viscous fluids is seriously hindered when causality is lost. A safer and more straightforward path is to devise effective theories that remain fully causal in their regime of applicability. 

Despite its importance in relativity theory, causality has proven to be a difficult feature
to accomplish in standard theories of relativistic viscous hydrodynamics. The original relativistic
Navier-Stokes (NS) equations put forward by Eckart \cite{EckartViscous}
and Landau \cite{LandauLifshitzFluids} decades ago have been shown to be acausal
both at the linearized \cite{Hiscock_Lindblom_instability_1985} and
non-linear level \cite{PichonViscous}. Currently, the most widely 
used theoretical framework for the study of relativistic viscous fluids is 
due to Israel and Stewart (IS) \cite{MIS-2,MIS-6}, together with the so-called resummed BRSSS theory \cite{Baier:2007ix}, and formulations derived from the relativistic Boltzmann equation, such as the DNMR equations \cite{Denicol:2012cn}. While these theories have been instrumental in the construction of models that provide us with great insight into the physics of viscous relativistic fluids, causality has been established in these theories only in the case of linear disturbances around equilibrium
and for certain values of the dynamic variables
\cite{Baier:2007ix,Hiscock_Lindblom_stability_1983}. 
These observations, added to the fact that heat flow in IS theory is known to display acausal behavior far from equilibrium 
\cite{Hiscock_Lindblom_pathologies_1988}, show that causality in
 IS-like theories is a delicate matter. This leaves open
the possibility of causality violations (even near equilibrium) when the the full non-linear dynamics is studied
or a wide range of field values is considered. 

Furthermore, causality is not the only unsettled question in IS, BRSSS, and DNMR theories. Questions regarding the existence (and uniqueness) of solutions, including the case when the fluid is dynamically coupled to gravity\footnote{With the exception of highly symmetric situations
such as FRW cosmologies \cite{Maartens:1995wt}.}, 
remain open for these theories.  We stress that this is not a mere academic
question. With exception of simple toy-model explicit solutions, the study
of relativistic hydrodynamics relies widely on numerically solving the equations of motion\footnote{
See \cite{Jeon:2015dfa} for a discussion of numerical approaches to relativistic hydrodynamics and Refs.\ \cite{Romatschke:2007mq,Song:2007ux,Luzum:2008cw,Schenke:2010rr,Niemi:2011ix,Noronha-Hostler:2013gga,DelZanna:2013eua,Noronha-Hostler:2014dqa,Habich:2014jna,Shen:2014vra,Romatschke:2015gxa,Ryu:2015vwa,Niemi:2015qia,Bazow:2016yra,Okamoto:2017ukz} for examples of numerical simulations where Israel-Stewart-like equations of motion were solved in the context of heavy-ion collisions.}.
Without knowing that the equations of motion admit solutions, the reliability
of numerical results may be called into question\footnote{Naturally, 
existence of solutions is not the only criteria to judge the reliability of a numerical simulations.
Issues of discretization, numerical stability, etc., are also important. But these become
relevant only when the underlying equations have (or are assumed to have) solutions.}.
In contrast, other areas of physics that also rely heavily on numerical simulations,
involving, for instance, ideal relativistic hydrodynamics \cite{RezzollaZanottiBookRelHydro},
typically deal with equations (e.g., Einstein or the relativistic Euler equations)
for which the problem of existence of solutions is already well understood.
(We return to this, giving a more precise definition of existence and uniqueness and discussing
its relevance for this work, in section \ref{section_about_existence}.)

Moreover, not only is the ability of coupling a theory to gravity essential from a foundational
point of view (effective theories must allow interaction with gravity in their regime of 
validity), but it is of course crucial for the study of many astrophysical phenomena.
This is equally true in the case of viscous hydrodynamics, especially given increasing 
awareness of the importance of dissipative phenomena in the study of heavily dense
atrophysical objects such as neutron stars \cite{DuezetallEinsteinNavierStokes, Herr_axially,HerreraFullCausal, Alford:2017rxf,RezzollaZanottiBookRelHydro}.
 
Further properties usually required in a fluid theory are linear stability
around equilibrium
(in the sense of \cite{Hiscock_Lindblom_instability_1985}) and 
non-negative entropy production. For instance, IS, DNMR, and the resummed BRSSS theory satisfy 
both of these conditions. In contrast, the relativistic
NS equations have non-negative entropy production but are linearly unstable
\cite{Hiscock_Lindblom_instability_1985}.

Finally, it is also important to connect a given fluid model with a microscopic approach.
More precisely, we would like to show that a given fluid model arises from a microscopic
description following some coarse graining procedure. The reason
to work with a fluid model is that one cannot in practice solve the full microscopic dynamics.
But the physics is ultimately determined by the latter, and the physical significance 
of a fluid model becomes fuzzy absent a connection with
this fundamental physical description. It is important to point out that the aforementioned
difficulties with theories of relativistic viscous fluids (possible lack of causality
or existence of solutions) are not present in standard microscopic descriptions
such as kinetic theory based on the Boltzmann equation\footnote{In fact, the local Cauchy problem is well-posed for the Einstein-Boltzmann system, as proven by Bancel and Choquet-Bruhat \cite{Bancel,BancelBruhat}.}.
We see, therefore, that these
pathologies are an artifact of a particular coarse graining method.
In fact, both the IS and Landau's theories are derivable from relativistic kinetic theory using different methods 
\cite{degroot}, but their causality and stability properties are
drastically distinct.

To the best of our knowledge, this manuscript gives the first example
of a theory of viscous fluid dynamics satisfying all of the above properties, i.e., the theory we shall present is
causal, existence and uniqueness of its solutions has been established
(in the full non-linear regime)
both in Minkowski background and when coupled to Einstein's equations,
the theory is linearly stable, derivable from microscopic theory, 
while also satisfying the second law of thermodynamics, and at the same time producing meaningful physical results in widely used test models such as the Gubser and Bjorken flows.

In Minkowski background, our theory is determined by four evolution equations
of second order, which can be rewritten as eight first order evolution equations.
For comparison, conformal IS theory has nine equations of motion. However,
four of our eight equations are simple field redefinitions used
to recast the second order system as a first order one and, in this sense, are trivial.
From a computational point of view, the complexity of our theory is reduced to four first order
equations, hence simpler than IS\footnote{In this paper we work with the equations written in second order form. The mention of reducing to a first order system was for comparison with IS theory only.}.

Here, we focus on conformal fluids because of their simplicity and  immediate relevance
for applications in the description of the quark-gluon plasma (see below). However, our constructions can be generalized to non-conformal relativistic fluids. We shall return to this point at the end of this paper.

The discussion above focused on IS-like approaches because of their wide use 
in the high energy nuclear physics and cosmology communities. Before we present our new approach, here we briefly mention other theories of relativistic viscous fluids that attempt to overcome the acausality and instability issues present in relativistic NS theory. One causal theory of relativistic viscous hydrodynamics, defined solely in terms of the hydrodynamic fields and
applicable, in particular to pure radiation fluids and ideal
gases, has been recently proposed in \cite{TempleViscous,TempleViscous2,TempleViscous3}.
This theory is well-posed
and linear stability has been verified in the fluid's rest frame.
However, as far as we know,
investigations addressing the stability of the equations of motion
in a Lorentz boosted frame and coupling to Einstein's equations
have not appeared in the literature for the energy-momentum tensor
introduced in \cite{TempleViscous,TempleViscous2,TempleViscous3}.
It is not known whether \cite{TempleViscous,TempleViscous2,TempleViscous3}
can be derived from kinetic theory.
Another proposal was put forward by Lichnerowicz back in the 50's
\cite{LichnerowiczBookGR}, but only recently it has been shown
to yield a theory that is causal and well-posed,
including when dynamically coupled to gravity, at least in the cases of irrotational fluids
\cite{DisconziViscousFluidsNonlinearity,DisconziKephartScherrerNew} or with restrictions
on the initial data \cite{DisconziCzubakNonzero}. 
Applications of Lichnerowcz's theory to cosmology appeared in 
\cite{Disconzi_Kephart_Scherrer_2015,DisconziKephartScherrerNew,MontaniLichnerowiczViscosityBianchi}.
It is not known whether 
Lichnerowicz's theory is linearly stable around equilibrium, nor whether it can
be derived from kinetic theory.

A large class of fluid theories can be constructed from the formalism
of divergence-type (DT) theories \cite{GerochLindblomDivergenceType, GerochLindblomCausal,LiuMullerRuggeri-RelThermoGases,MuellerRuggeriBook}.
While this formalism per se does not guarantee any of the aforementioned
properties (causality, linear stability, well-posedness, or coupling to gravity)
\cite{RezzollaZanottiBookRelHydro}, it has
been successfully applied to the construction of theories that 
are causal near equilibrium \cite{RamosCalzettaDT1, RamosCalzettaDT2,Lehner:2017yes} (see also \cite{Nagy_et_all-Hyperbolic_parabolic_limit,Kreiss_et_al,Reula_et_al-CausalStatistical}).
In fact, DT theories provide a very general formalism for the study
of fluid dynamics that can be showed to be compatible with kinetic theory,
 but there is no prescription of how to determine a particular
set of fields and equations of motion for the study of concrete problems.
Hence, despite their flexibility (or perhaps because of it), applications of
DT theories in the study of nuclear physics, astrophysics, and cosmology 
have so far been limited.

In Ref.\ \cite{VanFirstOrder} the authors construct a linearly stable theory involving
only the hydrodynamic variables. Their theory is derived from kinetic
theory, although well-posedness and causality remain open. A similar statement holds
for the theory introduced in \cite{Tsumura:2011cj}. Last but not least, motivated by the rapid expansion and the highly anisotropic initial state of the quark-gluon plasma formed in heavy ion collisions, a new set of fluid dynamic equations has been studied defining the so-called anisotropic hydrodynamics formalism \cite{Florkowski:2010cf,Martinez:2010sc}. This subject is still under development \cite{Strickland:2014pga} and statements regarding stability, causality, and existence of solutions are not yet available.

\subsection{About existence and uniqueness of solutions\label{section_about_existence}}

Before continuing our discussion of relativistic viscous fluids, 
we recall the definition of local well-posedness for a system of partial 
differential equations and discuss its significance. We chose to highlight 
the importance of this concept, which captures the idea of existence and uniqueness of solutions, for the following reason. For most traditional physical theories, existence
and uniqueness of solutions has been long established\footnote{See, e.g., \cite{ChoquetBruhatGRBook}, for a discussion of several physical models that are
locally well-posed.}. Therefore,
we can extract physical consequences of the equations of motion without worrying
whether such consequences are based on a vacuous assumption (e.g., on equations
without solutions). For relativistic viscous fluids, however, this is not the case. 
As discussed above, very little is known about local well-posedness
for relativistic models with viscosity.
Therefore, the question of working with equations that admit solutions to begin with
becomes of primary importance.

Consider in $\RR \times \RR^n$ a (linear or non-linear)
$k^{th}$ order partial differential equation for a function $\varphi$, 
which we write as $P \varphi  = 0$ (for instance, $P$ could be the wave operator). 
We think of $\RR \times \RR^n$ as a parametrization
of spacetime in terms of a time variable $t \in \RR$ and spatial variables 
$x \in \RR^n$. Let $\mathsf{X}$ be a function space (typically, but not necessarily, a Banach or Hilbert space) of functions defined on $\RR^n$. 
For example, one could have
$\mathsf{X} = C^\infty(\RR^n)$, the space
of infinitely differentiable functions on $\RR^n$. We say that the partial differential
equation is locally well-posed in $\mathsf{X}$ if the following holds. Given $k$
functions $\varphi_\ell$, $\ell=0, \dots k-1$, 
there exist a $\mathsf{T}>0$ and a function $\varphi$ defined on $[0,\mathsf{T})\times \RR^n$,
such that $\varphi$ satisfies the differential equation on $[0,\mathsf{T})\times \RR^n$, 
$\varphi(0,x) =\varphi_0(x),\dots, \partial_t^{k-1}\varphi(0,x) = \varphi_{k-1}(x)$
for all $x \in \RR^n$, where $\partial_t$ is differentiation with respect to the 
first coordinate in $\RR \times \RR^n$, and, for each $t \in [0,\mathsf{T})$, $\varphi(t,\cdot) \in \mathsf{X}$;
moreover, $\varphi$ is the only function defined on $[0,\mathsf{T})\times \RR^n$ satisfying these properties. We considered a scalar partial differential equation
in $\RR \times \RR^n$ for concreteness, but the definition of local well-posedness, as well
as the discussion below, naturally generalizes to systems and equations defined on 
manifolds\footnote{For geometric equations such as Einstein's equations, uniqueness
is understood in a geometric sense, i.e., up to changes by diffeomorphisms. See, e.g.,
\cite[Theorem 10.2.2]{WaldBookGR1984}.}; see, e.g., 
\cite[Definition 1.2.2]{KlainermanNicoloBook}.

Naturally, the functions $\{\varphi_\ell \}_{\ell=1}^k$ correspond to the
 initial conditions for the partial differential equation. 
Thus, roughly speaking,
local well-posedness says that given initial conditions, there exists a unique
solution to the equation taking the given initial data at time zero\footnote{Strictly
speaking, we are defining here local well-posedness of the initial value problem, 
which is the relevant notion of existence and uniqueness for evolution problems. We
can also define local well-posedness for boundary value problems, etc.}. 
 Since
$\varphi(t,\cdot) \in \mathsf{X}$ and $\varphi(0,x) = \varphi_0(x)$, we must
have $\varphi_0 \in \mathsf{X}$. The condition $\varphi(t,\cdot) \in \mathsf{X}$ can be
interpreted as saying that the solution
does not ``lose information" with respect to the given initial conditions (e.g., if the initial
conditions are square integrable, so is the solution).

One can also consider variations of the above. For instance, considering
that each $\varphi_k$ belongs a priori to some function space $\mathsf{X}_k$
($\mathsf{X}_0 \equiv \mathsf{X}$ in the previous notation), we could demand
that $\partial_t^k \varphi(t,\cdot) \in \mathsf{X}_k$, and in this case we talk
about local well-posedness in $\mathsf{X}_0 \times \cdots \times \mathsf{X}_{k-1}$.
Many authors include in the definition of local well-posedness the requirement that 
solutions vary continuously with the initial data, meaning that 
the map $\varphi_0 \mapsto \varphi$ is continuous\footnote{In the mathematical literature, continuity
with respect to the initial data is sometimes also referred to as
stability, but we stress that this is entirely different from the notion of stability 
which is discussed in this paper (which
follows the notion of stability introduced in \cite{Hiscock_Lindblom_instability_1985},
see section \ref{section_linear_stability}). For example,
the ordinary differential equation $\dot{x} = x$, $x(0) = x_0$ has solution
$x(t) = x_0 e^t$, which varies continuously with $x_0$. However, the trivial solution
$x_{trivial}(t) \equiv 0$ corresponding to $x_0 = 0$ is unstable in the terminology
of this paper in that
for any $x_0 \neq 0$, $x(t)$ will diverge exponentially from $x_{trivial}$.}
 with respect to the topology
of $\mathsf{X}$. One often talks about local well-posedness in the sense of Hadamard
(see, e.g., \cite{MisiolekYonedaIllPosednessEuler, MisiolekYonedaIllPosednessQuasiGeostrophic} and references therein)
when one wants to stress that continuity with respect to the initial data is also
taken into account in the definition of local well-posedness. Here, for simplicity,
we will not include such continuity requirement in the definition of local 
well-posedness, since establishing it usually requires substantial further technical
work that would distract us from the main point of local well-posedness, which is 
to guarantee that solutions exist, as we now discuss.

Local well-posedness furnishes the basic starting point for the validation of a theory
from a theoretical point of view, as it would be hard to imagine what it means
to develop a formalism based on equations that have no solution. It would be equally
puzzling to work with equations that admit more than one solution (for the same initial data).
In this regard, it is extremely important to stress that local well-posedness
is always local well-posedness \emph{in} $\mathsf{X}$, i.e., it depends on the class
of functions we choose to work with. For instance, it can happen that an equation
admits no solution for arbitrary initial data in certain function space $\mathsf{X}$,
but that a solution exists and is unique if we restrict the initial data to lie
in some subspace $\mathsf{X}^\prime \subset \mathsf{X}$ 
(see, for example, \cite[Chapter 4]{MizohataCauchyProblem}). It can also happen that
solutions exist but are not unique for in initial data in some function space but they 
exist and are unique for initial data in some other function space
(see, for instance, \cite{CC10,CzubakChanRemarks}).
The choice of $\mathsf{X}$ is typically tied to some physical requirement, 
e.g., functions that are square integrable in quantum mechanics or
vector fields that are divergence-free in the classical incompressible
Navier-Stokes equations. But as this discussion and the previous references indicate,
the choice of $\mathsf{X}$ may also be based on available mathematical techniques
or in the plain fact that some equations are not locally well-posed in some function
spaces\footnote{For instance, the (non-relativistic) incompressible Euler equations
are locally well-posed in the Sobolev spaces $H^s$ for $s>\frac{n}{2}+1$ \cite{Taylor3},
but are not locally well-posed in $H^s$ for $s=\frac{n}{2}+1$ \cite{BourgainLiIllPosedEuler}.}.

Moreover, local well-posedness is also very important for the validation of numerical
codes. Typically, when constructing a numerical algorithm one would like 
to show that it converges. Broadly speaking, this means that the sequence of numerical
solutions obtained by discretization converges to the actual solution of the differential
equation when the ``size" of the discretization approaches zero. 
Obviously, this is predicated on the idea that the differential equation is locally 
well-posed.

Furthermore, local well-posedness guarantees a solution to exist and to be unique for a finite
time interval $[0,\mathsf{T})$ and some function space $\mathsf{X}$. 
It is natural to ask how large $\mathsf{T}$ can be, and in particular
whether one can have $\mathsf{T}=\infty$, i.e., if solutions exist and are unique for all time.
When the latter happens, we say that the partial differential equation is 
\emph{globally} well-posed. Questions of global well-posedness naturally 
arise for non-linear
equations since they tend to develop
singularities. For example, one has the famous singularity theorems for Einstein's
equations\footnote{Although, for initial data near Minkowski, Einstein's
equations are globally well-posed \cite{ChristodouKlainermanStability}. 
See \cite{LindbladRodnianskiStabilityMinkowski,MR3120746,SpeckNonlinearStability,MR3101792}
for related results.} \cite{HawkingEllisBook}, 
or the formation of shock waves for fluid dynamic equations (see \cite{ChristodoulouShocks,ChristodoulouMiaoShocks}
and references therein), or yet
blow-up phenomena for non-linear wave equations (see \cite{SpeckBook} and references therein). From a physical perspective, when
the equations are not globally well-posed it becomes important to understand the nature
of the singularity. For example, the presence of a singularity might simply indicate a
limitation of the effective description. On the other hand, absent a better effective description,
one may attempt to enlarge the function space $\mathsf{X}$ to allow for functions
with singularities, e.g., distributions. Questions of this type are typically very challenging
and are beyond the scope of this work\footnote{As a matter of fact, the incompressible non-relativistic Navier-Stokes equations are locally well-posed but the question whether global well-posedness also holds in this case is one of the Millennium Prize problems in mathematics.}. Henceforth, we will refer to well-posedness to mean
local well-posedness throughout, although we will make some brief observations about global well-posedness
in section \ref{section_energy_conditions}.

\subsection{Organization of the paper}
The remaining of the paper is organized as follows. 
In section \ref{section_conformal} we introduce conformal viscous hydrodynamics (starting from a discussion of the non-conformal case).
In section \ref{section_new_tensor} we introduce our new tensor and discuss some of its properties.
Causality and well-posedness are proved in section \ref{section_well_posedness_causality}
whereas linear stability is established in section \ref{section_linear_stability}.
In section \ref{section_kinetic_theory} we show how our tensor can be derived from relativistic kinetic
theory. Section \ref{section_applications} provides applications of this new theory and a brief discussion on the choice of initial
conditions.
Section \ref{section_discussion} discusses our results, including a critical analysis of the theory's
limitations and open questions, and possible generalizations. Conclusions are presented in section
\ref{section_conclusions}.

\subsection{Conventions}
For the rest of the paper, we work in units such that $c = \hbar = k_B = 1$.
Our convention for the spacetime metric is $(-+++)$. All indices are lowered and raised
with the spacetime metric. Einstein's summation convention is adopted, Greek indices run from $0$ to $3$, and Latin indices from $1$ to $3$.

\section{Conformal viscous hydrodynamics\label{section_conformal}} 

For completeness, we begin this section with a discussion about the more general case of a non-conformal relativistic fluid in the absence of conserved vector currents (e.g.\ the baryon number current). The corresponding conformal limit, which is the focus of this paper, will be discussed in detail below. 

In general, one may always decompose \cite{degroot} the energy-momentum tensor of a fluid 
as follows\footnote{Provided the weak energy condition is satisfied, see section \ref{section_energy_conditions} and Ref.\ \cite{Arnold:2014jva}.}
\begin{equation}
T^{\mu\nu}  =\mathcal{E}u^\mu u^\nu +\mathcal{P}\Delta^{\mu\nu}
+ \pi^{\mu\nu} ,
\label{mostgeneralnc}
\end{equation}
where $\pi^{\mu\nu}$ is the symmetric traceless viscous tensor contribution 
orthogonal to the flow, $\mathcal{E}=u_\mu u_\nu T^{\mu\nu}$ is the energy density measured by a comoving observer, and $\mathcal{P}=\Delta_{\mu\nu}T^{\mu\nu}/3$ is the fluid's total pressure. In the standard approach by Landau and Eckart, assumed upon writing \eqref{mostgeneralnc} (see below), the quantity $\mathcal{E}$ in an out-of-equilibrium state is matched to the equilibrium energy density of an auxiliary (fictitious) system, with which one may define the local temperature $T$ and the local equilibrium pressure of the system via the thermodynamical equation of state $P=P\left(\mathcal{E}\right)$. In fact, for a non-conformal fluid the total pressure of the out-of-equilibrium system may be defined as $\mathcal{P} = P+\Pi$, where $\Pi$ is the bulk scalar, which encodes all the out-of-equilibrium corrections to the pressure. The fluid description only holds if, besides the condition $\mathcal{E}\geq 0$, the out-of-equilibrium correction $\Pi$ is such that the overall $\mathcal{P}$ is non-negative. 

The original ten independent degrees of freedom in $T^{\mu\nu}$ are thus parametrized in \eqref{mostgeneralnc} by the quantities $\{\mathcal{E},\Pi,u^\mu, \pi^{\mu\nu}\}$. In this case, the flow velocity of the system was defined by the condition $u_\mu T^{\mu\nu} = -\mathcal{E}u^\nu$, which was first introduced by Landau \cite{LandauLifshitzFluids}. However, differently than the case of an ideal fluid, for a system that is out of equilibrium quantities such as local temperature and flow velocity are not uniquely defined \cite{MIS-6,Kovtun:2012rj}. As a matter of fact, different choices lead to local temperature and velocity fields that differ from each other by gradients of the hydrodynamic variables, each particular choice being called a 
\emph{frame}\footnote{This meaning of the word frame has nothing to do with ``rest" and ``boosted frames." Unfortunately, these terminologies
are too widespread to be changed here. Hence, we use the word frame to refer to both a choice of local 
temperature and velocity, e.g., the Landau frame, and in the usual sense of relativity, e.g.,
the rest frame. The difference between both uses will be clear from the context. We also note
that frame, in the sense of a choice of local variables, has been used unevenly in the literature.
In \cite{Kovtun:2012rj}, for instance, frame is used in the same sense as employed here. 
In \cite{Denicol:2012cn}, the authors employ
frame, or, more specifically, hydrodynamic frame, to refer solely to the choice that determines the local flow velocity, while the choices
that determine the local temperature and chemical potential are called matching conditions.}.
Several frame choices have been pursued over the years, starting with Eckart \cite{EckartViscous}, Landau \cite{LandauLifshitzFluids}, Stewart \cite{Stewart:1972hg}, and others (for a discussion, see \cite{Tsumura:2006hn}). In section \ref{section_kinetic_theory}
we discuss the role played by such frame choices in the derivation of the hydrodynamic equations from kinetic theory.

An alternative decomposition for the energy-momentum tensor can be written using a different definition of the flow velocity, namely
\begin{equation}
T^{\mu\nu}  =\left(\epsilon+\mathcal{A}\right)u^\mu u^\nu +\left(P(\epsilon)+\Pi\right)\Delta^{\mu\nu}
+ \pi^{\mu\nu} +\mathcal{Q}^\mu u^\nu + \mathcal{Q}^\nu u^\mu,
\label{mostgeneralnc1}
\end{equation}
where now $\epsilon$ is matched to the corresponding expression for the energy density in equilibrium, $\mathcal{A}$ is the non-equilibrium correction to the energy density, $P(\epsilon)$ is the equilibrium pressure defined by the equilibrium equation of state, $\Pi$ is again the out-of-equilibrium correction to the pressure, and $\mathcal{Q}^\mu = -\Delta_\nu^\mu T^{\nu \alpha}u_\alpha$ is the flow of energy (heat flow)\footnote{We note that even though we used the same variable for the flow velocity and $\Pi$ in Eqs.\ \eqref{mostgeneralnc} and \eqref{mostgeneralnc1}, these quantities are not the same.}. These dissipative contributions $\cA$, $\cQ^\mu$, $\Pi$, and $\pi^{\mu\nu}$ to the energy-momentum tensor are such that they vanish in equilibrium. And, in fact, in this description deviations from local equilibrium to the energy density and pressure are treated in equal footing, with $\mathcal{A}$ playing a role in the total energy density of the system analogous to what $\Pi$ represents to the total pressure. This can lead to further insight on how relativistic fluids behave out-of-equilibrium and we note that such a decomposition was recently employed in Ref.\ \cite{Monnai:2018rgs} in applications to heavy-ion collisions. It is clear, however, that the decomposition of the energy-momentum tensor in terms of new set of variables $\{\epsilon,\mathcal{A},\Pi,u^\mu,\mathcal{Q}^\mu,\pi^{\mu\nu}\}$ is underdetermined, i.e., four extra conditions must be imposed to take into account the fact that there are only ten independent variables in $T^{\mu\nu}$. Such conditions may be derived using either the guidance of a microscopic description, such as kinetic theory, or via assumptions regarding the definition of the entropy current out-of-equilibrium in the sense of Israel and Stewart \cite{Monnai:2018rgs}.    

In this paper we focus on the case of \emph{conformal} hydrodynamics \cite{Baier:2007ix,Bhattacharyya:2008jc}, which provides the simplest set of assumptions regarding the properties of the underlying microscopic theory that can be used to study relativistic hydrodynamic phenomena. In this case $T^\mu_\mu=0$ and, thus $\epsilon = 3P$ (i.e., $\epsilon \propto T^4$ with $T$ being the temperature), and the equations of motion $\nabla_\mu T^{\mu\nu}=0$ change covariantly under a Weyl transformation of the metric, i.e.,
under $g_{\mu\nu} \to e^{-2\Omega}g_{\mu\nu}$, with $\Omega$ being an arbitrary Lorentz scalar. Since the quark-gluon plasma is approximately conformal at sufficiently large temperatures \cite{Borsanyi:2016ksw}, conformal fluids with their enhanced set of symmetries provide a testbed for numerical investigations in relativistic hydrodynamics, as shown in \cite{Luzum:2008cw}. However, we note that conformal invariance fixes the equation of state but it does not fully determine the dissipative corrections to the energy-momentum tensor, which must be specified by further assumptions. Nevertheless, conformal invariance allows us to write the most general energy-momentum tensor as 
\begin{equation}
T^{\mu\nu}  =\left(\epsilon+\mathcal{A}\right)\left(u^\mu u^\nu +\frac{\Delta^{\mu\nu}}{3}\right)
+ \pi^{\mu\nu} +\mathcal{Q}^\mu u^\nu + \mathcal{Q}^\nu u^\mu.
\label{mostgeneral}
\end{equation}
The conformal $T^{\mu\nu}$ has 9 independent components and the decomposition above in terms of $\{\epsilon,u^\mu,\mathcal{A},\mathcal{Q}^\mu,\pi^{\mu\nu}\}$ has 13 independent degrees of freedom at this level. Therefore, again a choice must be made to eliminate 4 extra degrees of freedom and fully specify the system's dynamics. In order to guarantee a smooth transition to the ideal fluid limit, it is natural to assume that such a choice involves the variables $\{\mathcal{A},\mathcal{Q}^\mu,\pi^{\mu\nu} \}$. For instance, as mentioned above Landau \cite{LandauLifshitzFluids} defined the flow and the energy density out of equilibrium in such a way that $\mathcal{Q}^\mu$ and $\mathcal{A}$ vanish. 

To proceed, one must decide whether the fields
$\{\mathcal{A},\mathcal{Q}^\mu,\pi^{\mu\nu}\}$
that are absent in the ideal fluid limit are to be treated as independent dynamical variables or are 
fully specified by the original hydrodynamic fields $\{\epsilon,u^\mu\}$. The former implies that 5 extra equations of motion must be given, in addition to the conservation law of energy and momentum. This idea is pursued in the aforementioned IS theories and
more generally in extended irreversible thermodynamics theories \cite{JouetallBook}. In this case, it is natural to employ Landau's definition to define conformal hydrodynamics, with $\pi^{\mu\nu}$ being defined by its own set of equations of motion. The degree of deviation from local equilibrium helps determine the equation of motion for $\pi^{\mu\nu}$ \cite{MIS-2}.

Another option consists in assuming that the set $\{\mathcal{A},\mathcal{Q}^\mu,\pi^{\mu\nu}\}$ is constructed using derivatives of the hydrodynamic fields $\{\epsilon,u^\mu\}$, as in a gradient expansion \cite{Baier:2007ix}. In the standard gradient expansion approach, dissipative effects are taken into account in the energy-momentum tensor via the inclusion of terms 
containing higher order derivatives of the hydrodynamic variables \cite{Baier:2007ix}, which are (formally) assumed to be small corrections around local equilibrium. To a given order in the expansion, one includes in the energy-momentum tensor all the possible terms compatible with the symmetries (e.g., conformal invariance), and this procedure was carried out to second order in \cite{Baier:2007ix} assuming Landau's definition of the hydrodynamic fields (i.e., the Landau frame),
and to third order in \cite{GrozdanovKaplisThirdOrder}.

In a gradient expansion, to first order in derivatives, there is only one choice for $\pi^{\mu\nu}$, namely,
$\pi^{\mu\nu} = -2\eta \sigma^{\mu\nu}$, where $\sigma_{\mu\nu} = \left(\nabla_{\langle\mu\rangle} u_\nu+\nabla_{\langle\nu\rangle} u_\mu\right)/2 - \frac{1}{3}\Delta_{\mu\nu}\nabla_\alpha u^\alpha$ is the shear tensor, $\nabla_{\langle\mu\rangle} = \Delta_\mu^\nu \nabla_\nu$ is the transverse covariant derivative, and $\eta$ is the shear viscosity transport coefficient (for a conformal fluid $\eta \propto s \propto T^3$, with $s=4\ep/(3T)$ being the entropy density). Using the Landau frame and keeping terms up to first order in gradients, one finds the conformal Navier-Stokes energy-momentum tensor $T^{NS}_{\mu\nu} = \epsilon \left(u_\mu u_\nu +\frac{\Delta_{\mu\nu}}{3}\right)-2\eta \sigma_{\mu\nu}$ \cite{Baier:2007ix,Bhattacharyya:2008jc}. At the linear level, this theory accurately describes the long wavelength behavior of sound and shear hydrodynamic disturbances around hydrostatic equilibrium: $\omega_{sound}(\mathbf{k}) = \frac{1}{\sqrt{3}}|\mathbf{k}| - i\frac{2}{3 T}\frac{\eta}{s}\mathbf{k}^2 + \mathcal{O}(k^3)$, $\omega_{shear}(\mathbf{k}) = - i \frac{\eta}{s}\frac{\mathbf{k}^2}{T}+\mathcal{O}(k^4)$ \cite{Romatschke:2009im}, in the sense that these dispersion relations can be directly matched to microscopic calculations, a procedure that may be used to determine the value of $\eta$ in a given system. However, as discussed above, the relativistic NS equations 
are plagued with instabilities and acausal behavior that severely limit their application in fluid dynamic calculations. 

Since the gradient expansion is used to derive the relativistic NS equations, it is believed that such an approach is generally responsible for the aforementioned problems displayed by these equations, as the underlying microscopic theory is widely expected to be free from pathologies. Therefore, one may be tempted to conclude that the particular type of coarse-graining procedure defined by the gradient expansion is inherently incompatible with causality and stability. However, one may also argue that the general reasoning behind the gradient expansion should be the most natural way to construct effective theories that describe the hydrodynamic regime of a fluid that is sufficiently near local equilibrium. In this paper we show that causality and stability are indeed compatible with the gradient expansion as long as one abandons the usual definition of hydrodynamic variables in relativistic viscous fluid dynamics put forward by Landau and Eckart.

\section{New conformal tensor\label{section_new_tensor}} 

Here we investigate causality and stability 
in relativistic viscous hydrodynamics using only the usual hydrodynamic fields
in $T_{\mu\nu}$, thus without introducing new dynamical degrees of freedom
as in IS-like theories. In this section we limit ourselves to introducing our new
conformal tensor and discuss some of its properties. 
Its derivation will be given in section \ref{section_kinetic_theory} using the relativistic
Boltzmann equation and a suitable perturbative expansion in spacetime gradients.

Our new tensor corresponds to \eqref{mostgeneral}
with the choices $\cA = 3\chi \frac{\cD T}{T}$ 
and
$\cQ_\mu = \la \frac{\cD_{\langle\mu\rangle}  T}{T}$,
where $\cD_\mu$ is the Weyl derivative \cite{Loganayagam:2008is} and $\mathcal{D}^{\langle\mu\rangle} = \Delta^\mu_\nu \mathcal{D}^\nu$.
Or, since
$\ep \propto T^4$ and $3 P(\ep) = \ep$, 
we can alternatively write   
$\cA = \chi \frac{\cD \ep}{\ep + P}$ and $\cQ_\mu = \la \frac{\cD_{\langle\mu\rangle}  \ep}{\ep + P}$,
where $\cD \epsilon = u^\mu \nabla_\mu \epsilon + (4/3) \epsilon \nabla_\mu u^\mu$
and $\cD_{\langle\mu\rangle} \ep = 4 \ep u^\la \nabla_\la u_\mu + \nabla_{\langle\mu\rangle} \ep$,
which is more convenient for our purposes.
Using these 
expressions for $\cA$ and $\cQ^\mu$ into (\ref{mostgeneral}) yields
\begin{equation}
T^{\mu\nu} 
= 
\left(\epsilon+\frac{3\chi}{4\epsilon} \mathcal{D}\epsilon\right)
\left(u^\mu u^\nu +\frac{\Delta^{\mu\nu}}{3}\right) 
-2\eta \sigma^{\mu\nu} 
 +  \frac{\lambda}{4\epsilon} \left(u^\mu \mathcal{D}^{\langle\nu\rangle} \epsilon+ u^\nu 
\mathcal{D}^{\langle\mu\rangle} \epsilon \right).
\label{tensorfodaotche}
\end{equation}
This is the most general energy-momentum tensor one can write for a conformal fluid to first order in gradients of the hydrodynamic fields $\{\epsilon,u_\mu\}$. The coefficients $\chi/\epsilon$ and $\lambda/\epsilon$ in \eqref{tensorfodaotche} define timescales ($\propto 1/T$) that control the behavior of the theory in the ultraviolet. 
They work as causal regulators because, as we shall see, when $\la$ and $\chi$ are different than zero and appropriately
chosen, the equations of motion are causal, whereas they become acausal when 
$\la=0=\chi$ (since then (\ref{tensorfodaotche}) reduces to NS). Furthermore, when these coefficients are nonzero and causality holds the disturbances in the hydrodynamic fields are resummed in the sense that the dispersion relations for sound and shear channels are not simple polynomial functions of momenta (see the dispersion relations in section \ref{section_linear_stability}). 
We note that conformal invariance implies that $\chi$ and $\la$ are proportional to
$\eta$.

We remark that the dynamical variables of \eqref{tensorfodaotche} are simply $\epsilon$ and the flow $u_\mu$, which obey second order nonlinear partial differential equations determined by $\nabla_\mu T^{\mu\nu}=0$. Also, we note that the hydrodynamic fields $\epsilon$ and $u_\mu$ in this theory do not coincide with those in either Landau's or Eckart's frames. In fact, due to the ambiguities in the definition of local temperature and velocity in the presence of dissipation (see above discussion), the fields $\ep$ (or $T$) and $u^\mu$ in (\ref{tensorfodaotche}) can be thought as auxiliary fields used to parametrize $T_{\mu\nu}$ \cite{Kovtun:2012rj}. We also stress that, regardless of how
we think of the parametrization given by $\ep$ and $u^\mu$, once the equations of motion have been shown
to satisfy desired physical requirements (e.g., causality and stability), then one can solve
them and reconstruct $T_{\mu\nu}$, from which further physical quantities of interest can be derived. 
Also, we note that for sufficiently small gradients the solutions for the fields $\ep$ and $u^\mu$ in \eqref{tensorfodaotche} will be near the corresponding quantities obtained by solving the ideal fluid equations.

The non-relativistic limit of the conformal fluid introduced here may be computed in the same way as in \cite{Fouxon:2008tb} and this yields the incompressible non-relativistic Navier-Stokes equations, with incompressibility being a consequence of the conformal invariance. Following the steps in \cite{Fouxon:2008tb}, we see that terms containing $\lambda$ and $\chi$ vanish in the non-relativistic limit since they are proportional to higher order terms. This shows that the new tensor in \eqref{tensorfodaotche} also has the correct non-relativistic limit. 

The tensor in \eqref{tensorfodaotche} provides a \emph{causal} generalization of NS theory constructed without the introduction of additional dynamical variables beyond those already present in the ideal fluid limit. We rigorously prove below existence, uniqueness, and causality of solutions to this viscous theory (in the full nonlinear regime) both in a Minkowski background and also when the fluid is dynamically coupled to Einstein's equations.
In a later section, we establish the stability of the solutions to the equations of motion in the linearized regime, and we show how \eqref{tensorfodaotche} can be ultimately derived from kinetic theory. Moreover, we develop applications in important known test-cases. This is the first time that such nontrivial statements can be rigorously made about viscous fluid dynamics in the relativistic regime since Eckart's first proposal in 1940.

\section{Well-posedness and causality\label{section_well_posedness_causality}}
 
In this section we consider Einstein's equations
$R_{\mu\nu}-\dfrac{1}{2}Rg_{\mu\nu}+\Lambda g_{\mu\nu} = 8\pi G T_{\mu\nu}$, with energy-momentum tensor given by \eqref{tensorfodaotche}.
It is assumed that the equation $u^\mu u_\mu = -1$ is also part of the system.
An initial data set $\cI$ for this system consists of the usual initial conditions
$(\Si,g_0,\kappa)$ for Einstein's equations ($\Si$ a three-dimensional manifold
endowed with a Riemannian metric $g_0$ and a symmetric two tensor $\kappa$), two scalar functions
$\ep_0$ and $\ep_1$ on $\Si$ (energy density and its time-derivative at the initial time), and
two  vector fields $v_0$ and $v_1$ on $\Si$ (the initial values for the velocity and its time-derivative),
such that the constraint equations are satisfied \cite[Chapter 10]{WaldBookGR1984}.
For a conformal theory all transport
coefficients are $\propto T^3$ so we can assume
$\chi = a_1 \eta$, $\lambda = a_2 \eta$,  with $a_{1,2}$ constants. 
 The meaning of ``sufficiently regular" stated in the theorem is explained below.
\begin{theorem}
Let $\cI = (\Si, g_0, \kappa, \ep_0, \ep_1, v_0, v_1)$ be a sufficiently regular
initial data set for 
Einstein's equations coupled to (\ref{tensorfodaotche}).
Suppose that $\Si$ is compact with no boundary, $\ep_0 > 0$, and that 
$\eta: (0,\infty) \rar (0,\infty)$ is analytic. Finally, assume  
that $a_1 \geq 4$ and $a_2 \geq \frac{3a_1}{a_1 -1}$. Then: 

(A) There exists a globally hyperbolic development $M$ of $\cI$.

(B) Let $(g,\ep,u)$ be a solution of Einstein's equations provided by the 
globally hyperbolic development $M$. For any $x \in M$ in the future of $\Si$, $(g(x), u(x), \epsilon(x))$ depends only on $\left. \cI \right|_{i(\Si) \cap J^-(x)}$, where $J^{-}(x)$ is the causal past of $x$ and
 $i: \Si \rar M$ is the embedding associated with the globally hyperbolic development $M$.
\label{main_theorem}
\end{theorem}

Statement (A) means the Einstein's equations admit existence and uniqueness of solutions 
(uniqueness up to a diffeomorphism, as usual in general relativity). Statement (B) says that the 
system is causal. $\Si$ is assumed compact and with no boundary for simplicity, as otherwise 
asymptotic and/or boundary conditions would have to be prescribed. The assumption
$\ep_0 >0$ guarantees that the equations of motion are not degenerate
(see section \ref{section_energy_conditions} for more details).
Above, sufficiently regular means that the initial data belongs to appropriate Gevrey spaces
(which are subspaces of the space of smooth functions, see \cite{RodinoGevreyBook} for a definition).
It is crucial to point out, however, that the \emph{causality} of the equations does not depend
on the use of Gevrey spaces, and it will automatically hold in any space of functions
where existence and uniqueness can be established.

\begin{theorem}
Under assumptions $a_1 \geq 4$ and $a_2 \geq \frac{3a_1}{a_1 -1}$ as above, 
a statement similar to Theorem \ref{main_theorem}, i.e., existence, uniqueness, and causality holds
for solutions of $\nabla_\mu T^{\mu\nu}=0$, with $T^{\mu\nu}$ given by \eqref{tensorfodaotche}, in Minkowski background.
\label{Minkowski_theorem}
\end{theorem}

The conditions $a_1 \geq 4$ and $a_2 \geq \frac{3a_1}{a_1 -1}$
 in Theorems \ref{main_theorem}
and \ref{Minkowski_theorem} are technical, but  they provide
a wide range of values for applications in different
situations of interest. Note that these are sufficient conditions, i.e., 
we are not saying (and we do not know) whether causality is lost if one of these
two conditions is not satisfied. Moreover, these conditions are easily accommodated
with those determined by kinetic theory for the coefficients 
$\chi$ and $\la$ (see section \ref{section_kinetic_theory}) and the 
stability conditions of section \ref{section_linear_stability}.

The proofs of Theorems \ref{main_theorem} and \ref{Minkowski_theorem} will be an application of the combined theorems of Leray and Ohya \cite[\S 6, sec. 27]{LerayOhyaNonlinear} and Choquet-Bruhat \cite[p. 381]{CB_diagonal}. A statement of the result as needed here appears in \cite[p. 624]{ChoquetBruhatGRBook} and it can be summarized as follows. Suppose that the characteristic determinant 
\cite[VI, \S 3.2]{Courant_and_Hilbert_book_2} of the system $P^I_K \varphi^K  = 0$ is a product of hyperbolic polynomials whose highest order is at least the order of the equations (all equations in the system are assumed of the same order). Assume that the characteristic cones determined by the hyperbolic polynomials are all contained in the light-cone in coordinate space and their intersection
has non-empty interior. Then, the system admits a unique causal solution in appropriate Gevrey spaces. We recall that a polynomial $p(\xi_0,\dots,\xi_n)$ of order $m$ is called hyperbolic if for every $(\xi_0,\dots,\xi_n) \neq 0$, the equation $p(\xi_0,\dots,\xi_n) = 0$ admits $m$ real distinct solutions $\xi_0 = \xi_0(\xi_1,\dots,\xi_n)$ \cite[VI, \S 3.7]{Courant_and_Hilbert_book_2}. For brevity, our proof will omit certain technicalities 
that might be of interest for more mathematically minded readers but would obfuscate the main
ideas. Those interested in such technical aspects can consult
\cite{DisconziFollowupBemficaNoronha}, where proofs of Theorems
\ref{main_theorem} and \ref{Minkowski_theorem} are given for an audience of mathematically
inclined readers\footnote{In \cite{DisconziFollowupBemficaNoronha}, for simplicity, only the case
$a_1 = 4$, $a_2 \geq \frac{3a_1}{a_1 - 1} =4$ is treated. The arguments there presented, however, are essentially the same to cover the remaining cases. In fact, the only substantial
difference for other values of $a_1$ is the computation of the characteristic determinant, which
is presented in detail here.}.

\vskip 0.1cm
\noindent \emph{Proof of Theorem \ref{main_theorem}.} As usual in general relativity, 
we embed $\Si$ into
$\RR \times \Si$ and work in local coordinates in the neighborhood of a point $p \in \Si$.
We can assume that $g(p)$ is the Minkowski metric.
We consider Einstein's equations written in wave gauge, 
$\nabla_\mu T^{\mu \nu} = 0$, and 
\begin{gather}
u_\la u^\al u^\mu \nabla_\mu \nabla_\al u^\la + u^\al \nabla_\al u_\la u^\mu \nabla_\mu u^\la = 0,
\label{pde_u}
\end{gather}
which follows from $u_\mu u^\mu =-1$ after twice differentiating
and contracting with $u$ (hence, 15 equations for the 15 unknowns $g_{\al\be}$, $u_\al$, $\epsilon$).
The characteristic determinant of the system equals
$ p_1(\xi) 
p_2(\xi)  p_3(\xi) 
 p_4(\xi) $
where
$p_1(\xi) = \frac{1}{12 \ep} \eta^4 (u^\mu \xi_\mu)^2$, $ p_4(\xi) = 
(\xi^\mu \xi_\mu)^{10}$,
\begin{align}
\begin{split}
 p_2(\xi) & =  [ (a_2-1) ( (u^0)^2 \xi_0^2 + (u^1)^2 \xi_1^2 +(u^2)^2 \xi_2^2
\\
&
+ (u^3)^2 \xi_3^2) - \xi^\mu \xi_\mu  
 +2(a_2-1)( u^1u^2 \xi_1 \xi_2 
 \\
 & +u^1 u^3 \xi_1 \xi_3 +u^2u^3 \xi_2 \xi_3 ) 
+ 2(a_2 -1 )u^0 \xi_0 u^i \xi_i ]^2,
\end{split}
\nonumber
\end{align}
and
\begin{align}
\begin{split}
p_3(\xi) & = [ 4a_1(a_2 - 3) - 4 a_2 ] (u^\mu \xi_\mu)^4 
\\
& -4 [ 2a_2 + a_1 (3+a_2) ] (u^\mu \xi_\mu)^2 \xi^\nu \xi_\nu
\\
& - (a_1 - 4) a_2 (\xi^\mu \xi_\mu)^2.
\end{split}
\nonumber
\end{align}
Here, $\xi = (\xi_0,\dots, \xi_3)$ is an arbitrary element of the cotangent bundle at a fixed point in the spacetime manifold (i.e., $\xi$ are coordinates in momentum space), in accordance to the prescription to compute the characteristic 
determinant \cite[VI, \S 3.2]{Courant_and_Hilbert_book_2}.
The polynomials $u^\mu \xi_\mu$ and $\xi^\mu \xi_\mu$ are
hyperbolic polynomials if $g$ is a Lorentzian metric
and $u$ is time-like. 
Thus, $p_1(\xi)$ is the product of two hyperbolic polynomials 
(recall that $\ep > 0$ and $\eta(\ep)>0$) and $p_4(\xi)$ is
the product of ten hyperbolic polynomials which stem from the principal part of Einstein's equations.

To analyze $p_2(\xi)$ we write 
$p_2(\xi) = (\widetilde{p}_2(\xi))^2$, where $\widetilde{p}_2(\xi)$ is
the polynomial between brackets in the definition of $p_2(\xi)$. Note that the 
assumptions on $a_1$ and $a_2$ imply that $a_2 \geq 3$.

Let us investigate the roots $\xi_0 = \xi_0(\xi_1, \xi_2, \xi_3)$ of
the equation $\widetilde{p}_2(\xi) = 0$. Consider first the case where
$\widetilde{p}_2(\xi)$ is evaluated at the origin, in which case $g$ is the Minkowski metric.
Then the roots are
$\xi_{0,\pm} = 
-\frac{1}{1 + (a_2 - 1)(1+\underline{u}^2)}
( (a_2 -1 )\underline{u}\cdot \underline{\xi} \sqrt{ 1 + \underline{u}^2}
\pm \sqrt{ (a_2 + (a_2 -1 ) \underline{u}^2 )\underline{\xi}^2  - (a_2 -1 )(\underline{u}\cdot \underline{\xi} )^2 }
)$,
where $\underline{u}=(u^1,u^2,u^3)$, 
$\underline{u}^2 = (u^1)^2+(u^2)^2+
(u^3)^2$, $\underline{\xi} = (\xi_1,\xi_2, \xi_3)$,
$\underline{\xi}^2 = \xi_1^2 + \xi_2^2 + \xi_3^2$,
 and $\cdot$
is the Euclidean inner product. We see that if  $\underline{\xi} = 0$,  then $\xi_{0,\pm} =0$  and hence
$\xi=0$. Thus, we can assume $\underline{\xi} \neq 0$.
The Cauchy-Schwarz inequality gives 
$\underline{u}^2  \underline{\xi}^2 - (\underline{u} \cdot \underline{\xi} )^2 \geq 0$, hence
 $\xi_{0,+}$ and $\xi_{0,-}$
are real and distinct for $a_2 \geq 3$. We conclude that $\widetilde{p}_2(\xi)$ is a hyperbolic polynomial at the origin. 
Since the roots of a polynomial vary continuously with the polynomial coefficients, $\widetilde{p}_2(\xi)$ will have two distinct
real roots at any point near the origin, hence on the entire coordinate chart (shrinking the chart
if necessary). Therefore, $p_2(\xi)$ is the product of two hyperbolic polynomials. 

We now move to analyze $p_3(\xi)$. First consider $a_1=4$, in which
case $a_2 \geq 4$. Then $p_3(\xi)$ reduces to
$ p_3(\xi) = [
12( -4 +a_2 )(u^\mu \xi_\mu)^2 
- 24(2+a_2) \xi^\mu \xi_\mu] (u^\nu \xi_\nu)^2$. The term $(u^\nu \xi_\nu)^2$ can be grouped
with $p_1(\xi)$, whereas the term between brackets can be analyzed similarly to $p_2(\xi)$ above
and we conclude, using $a_2 \geq 4$, that it is a hyperbolic polynomial of degree two.

Consider now $a_1 >4$ and $a_2 = \frac{3a_1}{a_1-1}$. Then the term coefficient
of $(u^\mu \xi_\mu)^4$ in $p_3(\xi)$ vanishes. We can then factor $(u^\mu \xi_\mu)^2$ and
a direct algebraic computation, as above, reveals that the remaining polynomial is hyperbolic
of degree two for $a_1 > 4$.

It remains to analyze the case $a_1 > 4$ and $a_2 > \frac{3a_1}{a_1 - 1}$. Note that in this case
the coefficients of both $(u^\mu\xi_\mu)^4$ and $(\xi^\mu \xi_\mu)^4$ in $p_3(\xi)$ are positive,
while the middle coefficient, $-4 [ 2a_2 + a_1 (3+a_2) ]$, is negative. Under these circumstances
we can factor $p_3(\xi)$ as
\begin{align}
\begin{split}
p_3(\xi) & = X( \xi^\mu \xi_\mu - Y u^\mu \xi_\mu )( \xi^\nu \xi_\nu - Z u^\nu \xi_\nu )
\end{split}
\nonumber
\end{align}
with $X, Y, Z > 0$. But for any $W > 0$, the polynomial 
$\xi^\mu \xi_\mu - W u^\mu \xi_\mu$ is a hyperbolic polynomial of degree two, as it can be seen
by a direct computation. Alternatively, we can note that if $W > 0$ then 
$\xi^\mu \xi_\mu - W u^\mu \xi_\mu$ is a non-zero multiple of the characteristic polynomial of
the acoustical metric \cite{ChristodoulouShocks} with sound speed equal to $1/(1+W)$.

Using the above explicit expression for the roots of $\widetilde{p}_2(\xi) = 0$,
it is not difficult to verify that the cone defined by
 $p_2(\xi) = 0$  
contains\footnote{By definition of the characteristic determinant,
the polynomials $p_i(\xi)$ are defined in the cotangent bundle or, equivalently, 
in momentum space. By duality, the characteristic cones associated with
$p_i(\xi) = 0$ in coordinate space will be inside the light-cone $g_{\mu\nu} v^\mu v^\nu = 0$, hence causal, if they are outside the light-cone
in momentum space.} the light-cone $g^{\mu\nu} \xi_\mu \xi_\nu = 0$. For 
$p_1(\xi)$ and $p_4(\xi)$ this condition is straightforward. Finally, 
the same
is true for $p_3(\xi)$ under all the above conditions (using again the acoustical metric
as a shortcut,  note that
$\xi^\mu \xi_\mu - W u^\mu \xi_\mu = 0$ defines a cone that contains the light cone if
$W>0$ in that the sound speed in this case satisfies $0 < 1/(1+W) < 1$).
Moreover, the intersection of these cones has non-empty interior.

Since the equations are of second order and the highest degree among the above hyperbolic polynomials is two, we have
verified all the conditions in \cite[p. 624]{ChoquetBruhatGRBook}. We conclude that Einstein's equations in wave gauge admit a unique and causal solution in a neighborhood of $x$. This gives a solution to Einstein's equations
in arbitrary coordinates (near $x$) because, by assumption, the initial data satisfies the Einstein constraint equations. 
Equation (\ref{pde_u}) 
implies that $u^\mu$ remains normalized if it is normalized at time zero.
A standard gluing argument \cite[p. 263]{WaldBookGR1984} now produces a solution defined on the entire manifold. This completes the proof. \hfill \cqd

\noindent \emph{Proof of Theorem \ref{Minkowski_theorem}.} This is exactly as the proof of Theorem \ref{main_theorem}, except
that now the polynomial $p_4(\xi)$, which comes from Einstein's equations, does not figure in the characteristic determinant. \hfill \cqd

\section{Linear stability analysis\label{section_linear_stability}} 

We follow Hiscock and Lindblom \cite{Hiscock_Lindblom_instability_1985} and consider the linearized version of the equations of motion for the theory defined by \eqref{tensorfodaotche}. We perform linear perturbations $\Psi\to \Psi^{(0)}+\delta\Psi$
 around thermodynamical equilibrium characterized by a constant flow $u_\mu^{(0)}$ and equilibrium energy density $\epsilon_0$ (i.e., $\nabla_\mu u^{(0)}_\nu=0=\nabla_\mu \epsilon_0$), where $\Psi=u^\mu,\,\epsilon,\,\eta,\,\chi,\,\lambda$. 
Our background is the Minkowski metric, which remains undisturbed, i.e., 
we work in the Cowling approximation \cite{CowlingApproximation} where $\delta g_{\mu\nu} = 0$.
As in \cite{Hiscock_Lindblom_instability_1985}, we consider only the plane-wave solutions to the perturbation equations $\delta\Psi(x)\to \delta\Psi(k)e^{i k_\mu x^\mu}$ with $k^\mu=(\omega,k,0,0)$. We begin our analysis in the fluid's rest frame so that $u_\mu^{(0)}=(-1,0,0,0)$. The equations of motion separate into two independent channels, the so-called sound and shear channels, whose modes are defined by the solutions of the following equations:
\begin{align}
\begin{split}
\text{sound:} &\, A_0+A_1\Gamma+A_2\Gamma^2+A_3\Gamma^3+A_4\Gamma^4=0,
 \\
\text{shear:} &\,  \bar{\lambda}\Gamma^2+\Gamma+\bar{\eta}k^2=0,
\end{split}
\nonumber
\end{align}
where $\Gamma = -i\omega$ \cite{Hiscock_Lindblom_instability_1985}, $A_0=3 k^2 + k^4 \bar{\lambda} (\bar{\chi}-4\bar{\eta})$, $A_1=3 k^2 (4 \bar{\eta} + \bar{\chi} + \bar{\lambda})$,
$A_2=9 + 6 k^2 (2 \bar{\eta} + \bar{\lambda}) \bar{\chi}$, 
$A_3=9 (\bar{\lambda} + \bar{\chi})$, 
$A_4=9 \bar{\lambda}\bar{\chi}$, momenta are rescaled by the background temperature $T_0$, $\bar{\eta}=\eta/s$, $\bar{\lambda}=\lambda/s$, and $\bar{\chi} = \chi/s$. 
The modes are stable if their solutions are such that $\mathrm{Re}\,\Gamma(k) \leq 0$ \cite{Hiscock_Lindblom_instability_1985} and, in the rest frame, this occurs when $\eta$, $\lambda$, $\chi>0$ and $\chi\geq 4\eta$. 

Tighter constraints appear by analyzing the stability in a boosted fluid where $u_\mu^{(0)} = (-\gamma,\gamma \mathbf{v})$, with $0\leq |\mathbf{v}|<1$ constant \cite{Hiscock_Lindblom_instability_1985,Denicol:2008ha,Pu:2009fj}. 
For the shear channel, the previous rest frame conditions are sufficient to guarantee stability also in a boosted frame. For the sound channel, stability in a boosted frame requires that $\eta>0$, $\chi = a_1\, \eta$, $\lambda \geq 3\eta\, a_1/(a_1-1)$ with $a_1 \geq 4$. We note that these are precisely the same conditions needed in Theorems \ref{main_theorem} and \ref{Minkowski_theorem}, guaranteeing
 causality, existence and uniqueness of solutions, and linear stability around equilibrium. This is the first time that such general statement can be made rigorous in  relativistic viscous hydrodynamics\footnote{At the linearized level, a similar statement was made for Israel-Stewart theory \cite{Hiscock_Lindblom_stability_1983} - see also \cite{Denicol:2008ha,Pu:2009fj} for related work.}. We note that our theory has non-hydrodynamic modes\footnote{These are modes in the linearized theory with dispersion relations such that $\lim_{\mathbf{k}\to 0} \omega(\mathbf{k})\neq 0$. In our theory, these non-hydrodynamic modes are purely imaginary at zero spatial momentum.} in both channels even in the rest frame (as does IS theory) and such modes are stable if $\lambda$ and $\chi$ obey the conditions mentioned above. 

The linearized problem studied here shows that
our conditions ensuring causality imply
some type of resummation. In fact, consider the dispersion relations of a theory (in the rest frame) that has the correct NS limit at small momenta, i.e., $\omega_{sound}(\mathbf{k}) = \frac{1}{\sqrt{3}}|\mathbf{k}| - i\frac{2}{3 T_0}\frac{\eta}{s}\mathbf{k}^2 + \mathcal{O}(k^3)$ and $\omega_{shear}(\mathbf{k}) = - i \frac{\eta}{s}\frac{\mathbf{k}^2}{T_0}+\mathcal{O}(k^4)$, such as the theory defined by \eqref{tensorfodaotche} or IS theory. This should be the case in any theory of viscous hydrodynamics, as it follows directly from gravitational Ward identities \cite{Czajka:2017bod}. On the other hand, the limit $|\mathbf{k}| \rar \infty$ provides a simple test that
suggests causality, namely, 
$\omega_{sound}(\mathbf{k})$ and $\omega_{shear}(\mathbf{k})$ cannot grow faster than $|\mathbf{k}|$
\cite{Pu:2009fj}, which implies that the dispersion relations for sound and shear channels cannot be simple polynomial functions of $|\mathbf{k}|$ -- hence, causality implies a resummation of spatial derivatives. This is obviously true when the full microscopic dynamics is taken into account but here we remark that such a statement must also hold after coarse graining if the fluid dynamic theory respects causality, as is the case for the theory defined by Eq.\ \eqref{tensorfodaotche}.       
Indeed, from the above dispersion relations, we find that
$|\omega_{sound}(\mathbf{k})| \leq | \textbf{k} |$ and 
$|\omega_{shear}(\mathbf{k})| \leq | \textbf{k} |$ for $|\mathbf{k}| \gg T_0$.

\section{Derivation from kinetic theory\label{section_kinetic_theory}}

Since the seminal work of Israel and Stewart \cite{MIS-2,MIS-3,MIS-4,MIS-5,MIS-6}, the relativistic Boltzmann equation has been considered a good starting point to understand the emergence of fluid dynamic behavior in relativistic systems.  As usual in such treatments \cite{degroot},
we consider the Boltzmann equation in flat spacetime. By general covariance, the same form of the 
energy-momentum tensor can then be obtained in curved spacetimes 
\cite[Chapter 5.4]{Weinberg_GR_book}.

Often in kinetic theory one derives the fluid dynamic equations under simplifying assumptions
that allow explicit calculations to be carried out in a perturbative regime.
In our case, we will consider a conformal gas. 
Nevertheless, since this can be viewed as a limiting case of more complex scenarios,
general features, such as  rough bounds
on the transport coefficients or the functional form of the energy-momentum tensor,
are expected to hold for other types of fluids (provided the general features of the derivation, 
such as the validity of a perturbative expansion, still hold). In fact, as we point out further below,
much of what follows is more general and the assumption of a single species conformal
gas is only a useful simplification of the analysis. Before presenting our kinetic theory derivation of \eqref{tensorfodaotche}, we begin the next section reviewing some basic aspects of relativistic kinetic theory that will be needed in this paper.

\subsection{Relativistic kinetic theory}

The Boltzmann equation for a dilute, single species relativistic gas of particles with constant mass 
$\mass$ (in flat spacetime) can be written as \cite{kremer}  
\beq
k^\mu \nabla_\mu f_k = \mathcal{C}[f_k,f_k]
\label{BE}
\eeq
where $\mathcal{C}[f_k,f_k]$ is the collision kernel, $f_k(x)=f(x,k)$ is the distribution function in phase space, which is a (dimensionless) Lorentz scalar that depends on the spacetime coordinates $x^\mu$ and the \emph{on-shell} momenta $k^\mu$ (i.e., $f_k$ may depend on 7 variables altogether).
From $f_k$ we can construct coarse-grained quantities such as the particle current 
\beq
J^\mu(x) = \int_p p^\mu f_p(x)
\nonumber
\eeq
and the energy-momentum tensor
\beq
T^{\mu\nu}(x) =  \int_p p^\mu p^\nu f_p(x),
\nonumber
\eeq
where $\int_p=\int \frac{d^3 \vec{p}}{(2\pi)^3\,p^0} = \int\int\int \frac{dp_1 dp_2 dp_3}{(2\pi)^3\,p^0}$ 
 and $\frac{dp_1 dp_2 dp_3}{(2\pi)^3\,p^0}$ is the Lorentz invariant measure \cite{kremer},  with $p^0 = \sqrt{\vec{p}^{\,2}+\mass^2}$ due to the on-shell mass condition $p^\mu p_\mu = -\mass^2$. 

The collision kernel of the Boltzmann equation encodes the nonlinear behavior of this integro-differential equation. In the limit of classical statistics, the collision kernel is given by \cite{kremer} 
\beq
\mathcal{C}[f_k,f_k] = \int_{k'pp'}W(kk'|pp')(f_p f_{p'}-f_k f_{k'})
\nonumber
\eeq
where 
\beq
W(kk'|pp') =  \frac{1}{2}|M|^2\,\delta^{(4)}(k_\mu+k'_\mu-p_\mu-p'_\mu)
\label{transitionW}
\eeq
and $M$ is the transition amplitude for particle scattering. For instance, for particles interacting with a constant total cross section one finds $|M|^2 \sim  s$ \cite{Denicol:2012cn}, where $s = -(k_\mu+k'_\mu)(k^\mu + k'^\mu)$ is the Mandelstam variable.

We assume the collision kernel to be such that 
\beq
\nabla_\mu J^\mu = \int_k k^\mu \nabla_\mu f_k = \int_k \mathcal{C}[f_k,f_k] = 0,
\label{property1}
\eeq
and also
\beq
\nabla_{\mu}T^{\mu\nu} = \int_k k^\nu k^\mu \nabla_\mu f_k  = \int_k k^\nu \mathcal{C}[f_k,f_k]=0.
\label{property2}
\eeq
Eq.\ \eqref{property1} defines the conservation of the particle current $J^\mu$ while \eqref{property2} implies that the energy-momentum tensor, constructed using the solution of the Boltzmann equation, is covariantly conserved. Also, we note that the so-called equilibrium distribution 
\beq
f^{eq}_k = e^{u_\mu k^\mu/T + \mu/T}
\label{definefeq} 
\eeq
is a zero of the collision kernel, i.e., $\mathcal{C}[f_k^{eq},f_k^{eq}]=0$. This occurs regardless the values assumed for the flow velocity $u_\mu(x)$, chemical potential $\mu(x)$, and temperature $T(x)$ that describe local equilibrium. We remark, however, that $f_k^{eq}(x)$ is only a solution of the Boltzmann equation when $u_\mu/T$ is a Killing vector of the underlying spacetime \cite{kremer}. 

We assume that the collision kernel obeys the standard conditions necessary for the H-theorem to be valid as in \cite{degroot}, i.e., we assume that the interactions are such that the general expression for the entropy current (which is also valid out of equilibrium)
\beq
\mathcal{S}^\mu(x) = -\int_k k^\mu\,f_k(x) \left(\ln f_k(x) -1\right)
\label{defineentropy}
\eeq
obeys the second law of thermodynamics 
\beq
\nabla_\mu \mathcal{S}^\mu = -\int_k \mathcal{C}[f_k,f_k]\ln f_k \geq 0,
\label{2ndlaw}
\eeq
where the equality only holds in equilibrium. In fact, in equilibrium one finds that $\mathcal{S}_{eq}^\mu = \eqent u^\mu$, with $\eqent$ being the equilibrium entropy density obtained from the first law of thermodynamics $T \eqent = \epsilon + P - \mu \nn$, while the equilibrium energy density and number density are given by
\beq
\epsilon(x) =\int_k E_k^2 f_{k}^{eq}(x)\qquad \textrm{and}\qquad \nn(x) = \int_k E_k f_{k}^{eq}(x),
\label{defineepsilonn}
\eeq
respectively. The equilibrium pressure is given by
\beq
P(x) =\frac{1}{3} \int_k \Delta_{\mu\nu}k^\mu k^\nu f_k^{eq}(x) = \frac{1}{3} \int_k k^{\langle\mu\rangle} k_{\langle\mu\rangle} f_k^{eq}(x),
\nonumber
\eeq
where $k^{\langle\mu\rangle} = \Delta^\mu_\nu k^\nu$, $E_k = - u_\mu k^\mu$, and $k^\mu = E_k u^\mu + k^{\langle\mu\rangle}$. From these definitions one can write down the corresponding expressions for the  equilibrium energy-momentum tensor $T_{\mu\nu}^{ideal} = \int_k k_\mu k_\nu f_k^{eq}=\epsilon u_\mu u_\nu + P \Delta_{\mu\nu}$ and particle current $J_\mu^{ideal} = \nn u_\mu$ \cite{kremer}.

\subsection{Conformal kinetic theory dynamics}

Here we will only consider the case of conformal kinetic dynamics, first discussed in Ref.\ \cite{Baier:2007ix} and  later explored in \cite{Denicol:2014xca,Denicol:2014tha}, which emerges in the case of a massless gas $p^\mu p_\mu = 0$ when the collision kernel changes homogeneously under Weyl transformations \cite{Baier:2007ix}, i.e., $g_{\mu\nu} \to e^{-2\Omega}g_{\mu\nu}$,  
\beq
p^\mu \nabla_\mu f(x,p) = \mathcal{C}[f_p,f_p] \to e^{2\Omega}\left(p^\mu \nabla_\mu f(x,p) =\mathcal{C}[f_p,f_p]   \right).
\label{conformalBE}
\eeq
This is the case of a massless gas of scalar bosons with quartic interactions computed at tree level. Also, an even simpler conformal kinetic theory can be constructed in the case of a massless gas with cross section $\sim 1/T^2$. The collision term in this conformal theory (still assuming classical statistics) may be written as
\beq
\mathcal{C}[f_k,f_k] = \frac{\sigma_0}{2T^2}\int_{k'pp'}s(2\pi)^5\,\delta^{(4)}(k+k'-p-p')\,(f_p f_{p'}-f_k f_{k'}),
\label{definemycollision}
\eeq
where $\sigma_0$ is a dimensionless constant that describes the magnitude of the interactions at fixed temperature. We will use this particular conformal theory when explicit calculations become necessary later in this paper.

\subsection{Perturbative expansion}
\label{BDN}

The Boltzmann equation \eqref{BE} is a nonlinear integro-differential equation for $f_k$ and, as such, exact solutions are very rare \cite{Cercignani}\footnote{For instance, the first analytical solution of the Boltzmann equation for an expanding gas was only found recently, see Refs.\ \cite{Bazow:2015dha,Bazow:2016oky}.}. Perturbative methods have been pursued over the years exploring different limits of its dynamics, as reviewed in \cite{Cercignani}. The hydrodynamical regime is of particular interest due to its simplicity as it describes the dynamics of small disturbances near local equilibrium. In this regard, the two most famous perturbative methods are the Hilbert series \cite{Cercignani} and the Chapman-Enskog expansion \cite{ChapmanCowling}, whose relativistic generalization are also known (see, for instance, Ref.\ \cite{kremer}). The Hilbert series does not lead to the usual equations of viscous fluid dynamics \cite{Cercignani} though the 
Chapman-Enskog expansion, when truncated to first order in deviations from local equilibrium, leads to the Navier-Stokes equations. A similar statement holds in the relativistic regime \cite{kremer,degroot}, but in this case the corresponding relativistic Navier-Stokes equations are problematic because of their lack of causality and stability, as mentioned in Sec.\ \ref{section_introduction}. In this paper we perform a different type of perturbative expansion that yields equations of motion for the hydrodynamic fields that describe a viscous relativistic fluid with causal and stable dynamics. This method is based on the technique developed in Ref.\ \cite{Denicol:2011fa}, with the important distinction that here we are only focused on the hydrodynamic regime.    

We start from the Boltzmann equation \eqref{BE} for a conformal fluid and use the fact that it is always possible to write its solution as 
\beq
f_k(x) = f_k^{eq}(x) + \delta f_k(x),
\label{BDN1}
\eeq
where $f_k^{eq}$ is a (fictitious) local equilibrium distribution \eqref{definefeq}, assumed to be the starting point of the perturbative expansion soon to be developed, and $\delta f_k$ represents the deviations from equilibrium. The arbitrariness in the definition of the local hydrodynamic fields $\{T,u_\mu,\mu\}$ in $f_k^{eq}$ must be fixed by imposing conditions on $\delta f_k$ \cite{kremer}, which may be generally written as
\beq
\int_k E_k^n \delta f_k = 0, \qquad \int_k E_k^m \delta f_k = 0, \qquad \textrm{and} \qquad \int_k E_k^r k^{\langle\mu\rangle}\delta f_k = 0
\label{generalmatching}
\eeq
where $n$, $m$, $r$, are non-negative integers. In the literature, the most common choices for these numbers are the Landau conditions $n=2$, $m=r=1$ and Eckart's where $n=2$, $m=1$, and $r=0$ \cite{kremer}. In terms of moments, Landau's conditions are simply $-J^\mu u_\mu = \nn$ and $u_\mu T^{\mu\nu} = -\epsilon u^\nu$, with $\nn$ and $\epsilon$ defined by their equilibrium values in \eqref{defineepsilonn}. Since for both Landau and Eckart the number density and the energy density are matched to their equilibrium expressions, the scalar conditions above are sometimes refereed as matching conditions, with the vector equation in \eqref{generalmatching} being used to the define the so-called ``frame" (i.e., $r=1$ is the Landau frame and $r=0$ is the Eckart frame) \cite{RischkeCollisions}. However, such choices are certainly not unique (as they reflect our choice in the definition of the hydrodynamic fields in $f_k^{eq}$) and other conditions may be used in perturbative expansions \cite{Stewart:1972hg,Tsumura:2006hn}. Thus, a choice of $n$, $m$, and $r$ determines a choice of local temperature, flow velocity, and
chemical potential. In other words, a choice of 
$n$, $m$, and $r$ in \eqref{generalmatching} \emph{corresponds precisely to a choice of frame as discussed in section \ref{section_conformal}}.
The role played by such a choice in the perturbative expansion is discussed below. 

We remark that the \emph{full} solution of the Boltzmann equation does not depend on the choice of the hydrodynamic fields in $f_k^{eq}$ as different choices can always be accounted for in $\delta f_k$. However, the moments of $f_k$ do change with the frame when one employs the truncated solution for $f_k$ in the calculation of these quantities (as we shall do in the following). In fact, it is well-known that $T^{\mu\nu}$ and $J^\mu$ change when going from the Landau to the Eckart frame in the usual Chapman-Enskog expansion truncated at first order \cite{degroot}. Therefore, one may use this freedom in the definition of the hydrodynamic fields in \eqref{generalmatching} to determine which choice is more suited in practice to study the hydrodynamic regime of the Boltzmann equation. As a matter of fact, frames different than Landau's and Eckart's have been already discussed and pursued in the literature, see Refs.\ \cite{Tsumura:2006hn,Tsumura:2007wu,Van:2007pw,Van:2011yn,Tsumura:2011cj,Tsumura:2012ss,
Monnai:2018rgs}. 


We substitute \eqref{BDN1} in the Boltzmann equation to find 
\bea
k^\mu \nabla_\mu f_k^{eq} + k^\mu \nabla_\mu \delta f_k  &=& \int_{k'pp'}W(pp'|kk') f_k^{eq} f_{k'}^{eq}\left( \frac{\delta f_p}{f_p^{eq}}+\frac{\delta f_{p'}}{f_{p'}^{eq}} - \frac{\delta f_k}{f_k^{eq}}  -\frac{\delta f_{k'}}{f_{k'}^{eq}} \right) \nonumber \\ &+&  \int_{k'pp'}W(pp'|kk') \,\left(\delta f_p \delta f_{p'}-\delta f_k \delta f_{k'}\right).
\label{exactboltz1}
\eea
Up to this point, no approximations in the Boltzmann dynamics were made. However, now we assume that the deviations from equilibrium are small \cite{Denicol:2011fa} and linearize the equation above by neglecting\footnote{This perturbative solution can be performed systematically as follows. First, we introduce a book keeping parameter $\alpha$ on the nonlinear term in \eqref{exactboltz1} and then assume a power series behavior for $\delta f_k = \sum_{n=0}^\infty \alpha^n \delta f_k^{(n)}$. The lowest order term in this expansion gives Eq.\ \eqref{BDN1storder}.} the contribution from terms that are quadratic in $\delta f_k$. This gives 
\beq
k^\mu \nabla_\mu f_k^{eq}+k^\mu \nabla_\mu \left(f_k^{eq}\phi_k\right) - f_k^{eq} \mathcal{L}[\phi_k] =0,
\label{BDN1storder}
\eeq
where we defined $\phi_k = \delta f_k/f_k^{eq}$ and $\mathcal{L}$ is the linearized collision operator
\beq
\mathcal{L}[\phi_k] = \int_{k'pp'}W(pp'|kk') f_{k'}^{eq}\left(\phi_p + \phi_{p'}-\phi_k -\phi_{k'}\right).
\nonumber
\eeq
It will be useful for our analysis to know that  the functions $\{1,k^\mu\}$ span the kernel of this operator, i.e., $\mathcal{L}[1]=\mathcal{L}[k^\mu]=0$ and that this operator is self-adjoint in the sense that 
\beq
\int_k f_k^{eq} h_k \mathcal{L}[z_k] = \int_k f_k^{eq} z_k \mathcal{L}[h_k] 
\nonumber
\eeq
with $h_k(x)$ and $z_k(x)$ being arbitrary functions\footnote{These functions are assumed to be such that $\int_k f_{k}^{eq} h_k$ and $\int_k f_k^{eq} z_k$ are finite.}. Also, this operator is non-positive
\beq
\int_k f_k^{eq} \phi_k \mathcal{L}[\phi_k]  \leq 0,
\nonumber
\eeq
with the equality corresponding to the case where $\phi_k=\{1,k^\mu\}$. A more detailed discussion of the mathematical properties of $\mathcal{L}$ can be found, for instance, in Ref.\ \cite{Cercignani}.

Now our task is to solve Eq.\ \eqref{BDN1storder} subject to the conditions \eqref{generalmatching}. This problem can be solved \cite{kremer} by considering integral moments of \eqref{BDN1storder} with respect to the tensorial basis $k^{\mu_1 }\dots k^{\mu_j}$ ($j\geq 0$), i.e., 
\beq
\int_k k^{\mu_1 }\dots k^{\mu_j}\, \left\{k^\mu \nabla_\mu f_k^{eq}+k^\mu \nabla_\mu \left(f_k^{eq}\phi_k\right) - f_k^{eq} \mathcal{L}[\phi_k]\right\} =0,
\nonumber
\eeq
which leads to an infinite set of (coupled) differential equations for the moments of the non-equilibrium correction determined by $\phi_k$. In this paper we truncate this set of equations and consider only the cases where $j=0,1,2$. This type of truncation is commonly used in the derivation of hydrodynamics from the Boltzmann equation \cite{MIS-6,kremer}. For a more systematic approach that includes the contribution from higher order moments, we refer the reader to Ref.\ \cite{Denicol:2012cn}.  

Using that $\mathcal{L}$ is a self-adjoint operator, and the functions $\{1,k^\mu\}$ are in its kernel, one can see that $j=0$ simply gives the conservation law of particle number, $\nabla_\mu J^\mu = 0$, while $j=1$ implies the conservation of energy and momentum, $\nabla_\mu T^{\mu\nu}=0$, with both $J^\mu$ and $T^{\mu\nu}$ being constructed using $f_k = f_k^{eq}\left(1 +\phi_k\right)$. We now use the decomposition 
\beq
k^\mu k^\nu = \left(u^\mu u^\nu + \frac{\Delta^{\mu\nu}}{3}\right)E_k^2 + E_k u^\mu k^{\langle\nu\rangle}+E_k u^\nu k^{\langle\mu\rangle} + k^{\langle\mu}k^{\nu\rangle}
\nonumber
\eeq
to show that the $j=2$ term can be divided into three separate equations
\beq
\int_k E_k^2\, \left\{f_k^{eq} E_k^2 \frac{\mathcal{D}T}{T^2}+k^\mu \nabla_\mu \left(f_k^{eq}\phi_k\right) - f_k^{eq} \mathcal{L}[\phi_k]\right\} =0,
\label{definescalarEq}
\eeq
\beq
\int_k E_k k^{\langle\mu\rangle}\, \left\{f_k^{eq} E_k \frac{k^{\langle\nu\rangle}\mathcal{D}_{\langle\nu\rangle}T}{T^2} +k^\mu \nabla_\mu \left(f_k^{eq}\phi_k\right) - f_k^{eq} \mathcal{L}[\phi_k]\right\} =0,
\label{definevectorEq}
\eeq
and
\beq
\int_k k^{\langle\alpha}k^{\beta\rangle}\, \left\{f_k^{eq} \frac{k^{\langle\mu}k^{\nu\rangle}\sigma_{\mu\nu}}{T}+k^\mu \nabla_\mu \left(f_k^{eq}\phi_k\right) - f_k^{eq} \mathcal{L}[\phi_k]\right\} =0
\label{definetensorEq}
\eeq
where the shear tensor is $\sigma^{\mu\nu} = \Delta^{\mu\nu}_{\alpha\beta}\nabla^\alpha u^\beta$ and $\Delta^{\mu\nu}_{\alpha\beta} = \left(\Delta^\mu_\alpha \Delta^\nu_\beta +\Delta^\mu_\beta \Delta^\nu_\alpha \right)/2 - \Delta^{\mu\nu}\Delta_{\alpha\beta}/3$ is the projection tensor \cite{degroot}, and $k^{\langle\mu}k^{\nu\rangle} = \Delta^{\mu\nu}_{\alpha\beta}k^\alpha k^\beta$. For convenience, we have introduced the Weyl derivative notation \cite{Loganayagam:2008is} $\mathcal{D}T = DT + \theta T/3$ and $\mathcal{D}_{\langle\mu\rangle}T = T Du_\mu + \nabla_{\langle\mu\rangle}T$, with $D=u^\mu \nabla_\mu$, $\nabla^{\langle\mu\rangle}=\Delta^\mu_\alpha\nabla^\alpha$, and the expansion rate $\theta = \nabla_\mu u^\mu$. Furthermore, since $\mu/T$ is constant in a conformal fluid, no gradients of this quantity appear when computing $k^\mu \nabla_\mu f_k^{eq}$ in the equations above. Therefore, for convenience we set the chemical potential $\mu=0$ in the following.   

We are primarily interested in the case where the hydrodynamic fields are sufficiently slowly varying functions of space and time, since this is the situation when the hydrodynamic limit
is expected to be a good approximation to the underlying kinetic theory, i.e., when 
a gradient expansion provides a good representation of the dynamics of the system. Thus, only an approximate solution for $\phi_k$ valid in this limit will be pursued. Since the source terms in the equations above are already of first order in derivatives of the hydrodynamic fields, $\phi_k$ must be of first order in gradients at lowest order in a derivative expansion. But the term $k^\mu \nabla_\mu \left(f_k^{eq}\phi_k\right)$ only contributes at 2nd order\footnote{Recall that, in a gradient expansion for a field $\psi$, both
$\nabla^2 \psi$ and $(\nabla \psi)^2$ count as second order terms.}. Therefore, 
$k^\mu \nabla_\mu \left(f_k^{eq}\phi_k\right)$ can be omitted in Eqs.\ \eqref{definescalarEq}, \eqref{definevectorEq}, and \eqref{definetensorEq} when determining $\phi_k$ to first order in gradients. In this case, the general solution for $\phi_k$, valid at first order in the derivative expansion, can be written as follows 
\beq
\phi_k = \phi_A \frac{k^{\langle\mu}k^{\nu\rangle}\sigma_{\mu\nu}}{T^3} + \phi_B \frac{E_k^2\mathcal{D}T}{T^4} + \phi_C \frac{E_k k^{\langle\nu\rangle}\mathcal{D}_{\langle\nu\rangle}T}{T^4} + \xi + v \frac{E_k}{T} + v_{\langle\mu\rangle}\frac{k^{\langle\mu\rangle}}{T},
\label{phikgeral}
\eeq
where $\xi$, $v$ and $v_{\langle\mu\rangle}$ parametrize the kernel of the collision
operator. Using Eqs.\ \eqref{definescalarEq}, \eqref{definevectorEq}, and \eqref{definetensorEq} one can show that $\phi_A$, $\phi_B$, and $\phi_C$ are determined by the equations
\beq
\phi_A\left(\frac{1}{T^8} \int_k f_k^{eq} k^{\langle\alpha}k^{\beta\rangle}\mathcal{L}[k^{\langle\mu}k^{\nu\rangle}]  \right)\sigma_{\mu\nu} = \frac{8}{\pi^2}\sigma^{\alpha\beta},
\label{definephiA}
\eeq
\beq
\phi_B\left(\frac{1}{T^8} \int_k f_k^{eq} E_k^2  \mathcal{L}[E_k^2]\right)   = \frac{60}{\pi^2},
\label{definephiB}
\eeq
and 
\beq
\phi_C\left(\frac{1}{T^8}\int_k f_k^{eq} E_k k^{\langle\mu\rangle}\mathcal{L}[E_k k^{\langle\nu\rangle}]\right)\mathcal{D}_{\langle\nu\rangle}T = \frac{20}{\pi^2} \mathcal{D}^{\langle\mu\rangle}T,
\label{definephiC}
\eeq
where we used that
\beq
\int_k E_k^n f_k^{eq} =T^{n+2} \frac{(n+1)! }{2\pi^2} \qquad \textrm{and} \qquad \int_k f_k^{eq} k^{\langle\mu}k^{\nu\rangle} k^{\langle\alpha}k^{\beta\rangle} = \frac{8T^6}{\pi^2}\Delta^{\mu\nu\alpha\beta}.
\nonumber
\eeq
Since $\mathcal{L}$ is non-positive, the quantities $\phi_A$, $\phi_B$, and $\phi_C$ are negative and their specific values only depend on the properties of the collision kernel. On the other hand, the coefficients $\{\xi,v,v_{\langle\mu\rangle}\}$ are fixed by our definition of the hydrodynamic fields via the constraints in Eq.\ \eqref{generalmatching} (and the corresponding results for $\phi_A$, $\phi_B$, and $\phi_C$). Using \eqref{phikgeral} in \eqref{generalmatching} we find 
\beq
\xi = \phi_B \frac{\mathcal{D}T}{T^2}(m+2)(n+2), \qquad v = -\phi_B\frac{\mathcal{D}T}{T^2}(m+n+5), \qquad \textrm{and} \qquad v^{\langle\mu\rangle} = -\phi_C \frac{\mathcal{D}^{\langle\mu\rangle}T}{T^2}(r+4).
\nonumber
\eeq
One can see that these coefficients are nonzero for any choice of frame.

Now let us determine the energy-momentum tensor of the gas. A simple calculation reveals that
\beq
T^{\mu\nu} =\int_k k^\mu k^\nu f_k^{eq}\left(1 + \phi_k\right)= \left(\epsilon + \mathcal{A}\right)\left(u^\mu u^\nu + \frac{\Delta^{\mu\nu}}{3}\right) + \pi^{\mu\nu} + \mathcal{Q}^{\langle\mu\rangle}u^\nu + \mathcal{Q}^{\langle\nu\rangle}u^\mu.
\label{newtensorkinetic1}
\eeq
The non-equilibrium correction to the energy density is
\beq
\mathcal{A} = \int_k E_k^2 f_k^{eq} \phi_k= 3 \chi \frac{\mathcal{D}T}{T}
\nonumber
\eeq
with 
\beq
\chi = \phi_B \frac{T^3}{\pi^2} (n-2)(m-2).
\nonumber
\eeq
On the other hand, the heat flow in \eqref{newtensorkinetic1} is given by
\beq
\mathcal{Q}^{\langle\mu\rangle} =\int_k E_k k^{\langle\mu\rangle}f_k^{eq}\phi_k = \lambda \frac{\mathcal{D}^{\langle\mu\rangle}T}{T}
\nonumber
\eeq
with
\beq
\lambda = \phi_C\frac{4T^3}{\pi^2}(1-r).
\nonumber
\eeq
Finally, the shear stress tensor is given by 
\bea
\pi^{\mu\nu} &=&\int_k k^{\langle \mu}k^{\nu\rangle}f_k^{eq} \phi_k= -2\eta \sigma^{\mu\nu}
\nonumber
\eea
where
\beq
\eta = -\phi_A\frac{4T^3}{\pi^2}
\label{defineeta}
\eeq
is the shear viscosity transport coefficient \cite{Denicol:2011fa}. It is interesting to notice that while $\eta$ does not depend on our choice of frame, the new coefficients $\chi$ and $\lambda$ that appear in our perturbative expansion certainly do. In fact, even their sign can change as different choices in \eqref{generalmatching} are made. For instance, in the Landau frame $n=2$, $m=r=1$ and, thus, $\chi=\lambda=0$. On the other hand, for Eckart's $\chi=0$ but $\lambda\neq 0$ and there is nonzero heat flow. 

Another property of the system that can be easily computed is the entropy production. Using the general expression \eqref{phikgeral} in Eq.\ \eqref{2ndlaw}, we keep the lowest order terms in the expansion to find
\beq
\nabla_\mu \mathcal{S}^\mu = - \int_k f_k^{eq} \phi_k \mathcal{L}[\phi_k] = \frac{2\eta}{T}\sigma_{\mu\nu}\sigma^{\mu\nu} -\phi_B\frac{60}{\pi^2} \left(\mathcal{D}T\right)^2 - \phi_C \frac{20}{\pi^2} \mathcal{D}^{\langle\mu\rangle}T \,\mathcal{D}_{\langle\mu\rangle}T,
\nonumber
\eeq
which is non-negative since $\phi_B$ and $\phi_C$ are negative. We note that the production of entropy does not depend on \eqref{generalmatching}, being thus independent on the choice of 
frame.

Now we have to specify the interactions in the collision kernel to determine $\phi_A$, $\phi_B$, and $\phi_C$. For simplicity, in this paper we use the simple conformal gas defined in Eq.\ \eqref{definemycollision}. Using the results from \cite{Denicol:2011fa,Bazow:2015dha,Bazow:2016oky} in Eqs.\ \eqref{definephiA}, \eqref{definephiB}, and \eqref{definephiC}, standard calculations give
\beq
\phi_A = -\frac{3\pi^2}{10\sigma_0}, \qquad \phi_B = -\frac{15\pi^2}{2\sigma_0}, \qquad \textrm{and} \qquad \phi_C = -\frac{5\pi^2}{2\sigma_0}.
\nonumber
\eeq
One can show that for this gas $\eta=6T^3/(5\sigma_0)$,  $\lambda/\eta=25(r-1)/3$, and $\chi/\eta=-25(n-2)(m-2)/4$. Given that positive values for these coefficients are preferred according to the well-posedness, causality, and stability results derived in Sections \ref{section_well_posedness_causality} and \ref{section_linear_stability}, one can see that  frames where $r>1$ and $n<2$, $m>2$ provide a suitable definition of the hydrodynamic fields since in this case the propagation of energy and momentum is causal and stable. A possible choice of frame would be, for instance, $n=0$, $m=3$, and $r=2$ which gives values for $\lambda$ and $\chi$ that satisfy the conditions established for causality and stability of sections 
\ref{section_well_posedness_causality} and \ref{section_linear_stability}. In terms of the following moments of $f_k$
\beq
\rho^{\mu_1\ldots \mu_j} = \int_k k^{\mu_1} \ldots k^{\mu_j} (f_k^{eq}+\delta f_k)
\nonumber
\eeq
the choice frame mentioned above corresponds to setting
\beq
\rho = \rho^{eq}\qquad \textrm{and}\qquad  \rho_{\mu\nu\lambda}u^\nu u^\lambda = \rho^{eq}_{\mu\nu\lambda}u^\nu u^\lambda,
\nonumber
\eeq
where $\rho^{eq}$ and $\rho^{eq}_{\mu\nu\lambda}$ are computed using the equilibrium distribution. Therefore, we see that \emph{causality and stability can be obtained in viscous relativistic hydrodynamics from a derivative expansion} as long as a judicious choice of frame involving the definition of hydrodynamic fields in $f_k^{eq}$ is made.

\section{Applications\label{section_applications}}

In this section we initiate an investigation of the immediate applications of the theory discussed in this paper. We focus on problems of relevance to high energy nuclear physics, more specifically the space-time evolution of the quark-gluon plasma formed in heavy ion collisions, where the conformal fluid approximation has been already used \cite{Baier:2007ix,Luzum:2008cw}. We solve our equations of motion for fluids undergoing Bjorken and Gubser flows in \ref{section_Bjorken} and \ref{section_Gubser}, respectively, where the flow velocity is completely determined by symmetry arguments. A discussion about how to set up the initial value problem in more general situations is presented in \ref{section_initial_conditions}.  

\subsection{Hydrodynamic attractor in Bjorken flow}
\label{section_Bjorken}

Motivated by the hydrodynamic studies of the quark-gluon plasma formed in heavy ion collisions, we first consider the case of the Bjorken flow \cite{Bjorken:1982qr} where $u^\mu = (1,0,0,0)$ in Milne coordinates defined as $x^\mu = (\tau,x,y,\varsigma)$, with $\tau = \sqrt{t^2-z^2}$ and $\varsigma = \tanh^{-1}(z/t)$. This configuration corresponds to a fluid rapidly expanding in the longitudinal $z$ direction (being homogeneous in the $xy$ plane) and the only unknown is $\epsilon=\epsilon(\tau)$
or equivalently $T = T(\tau)$, which is the solution of $u_\nu \nabla_\mu T^{\mu\nu}=0$ (we note that the term with $\lambda$ in \eqref{tensorfodaotche} does not contribute to Bjorken flow dynamics). 

\begin{figure}[th]
\includegraphics[width=0.6\textwidth]{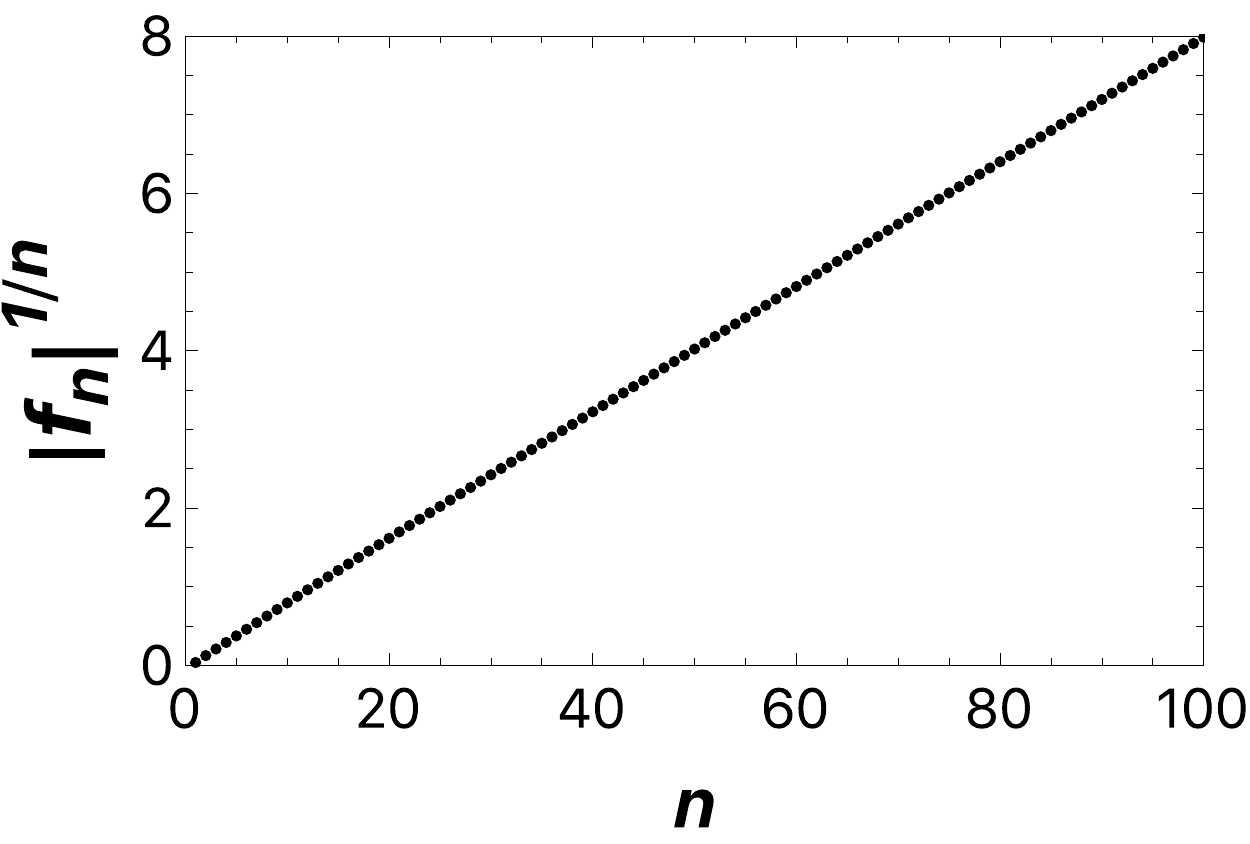}
\caption{$ | f_n |^{1/n}$ as a function of $n$, computed using \eqref{largefn}, for $\eta/s=0.08$ and $\chi=4\eta$.}
\label{fig2}
\end{figure}

The equation for the temperature $T(\tau)$ is 
\begin{align}
\begin{split}
\tau\bar{\chi}\frac{\partial_\tau^2 T}{T}
+2\tau\bar{\chi}\left(\frac{\partial_\tau T}{T}\right)^2 + \frac{7}{3}\bar{\chi} \,\frac{\partial_\tau T}{T}+\frac{\bar{\chi}}{9\tau}  
+ \tau \partial_\tau T- \frac{4}{9}\frac{\bar{\eta}}{\tau} +\frac{T}{3}=0,
\end{split}
\label{eqBjorken}
\end{align}
where we used that $\nabla_\mu u^\mu=1/\tau$ and $\sigma_{\mu\nu}\sigma^{\mu\nu} = 2/(3\tau^2)$. If $\chi=0$, \eqref{eqBjorken} describes the well-known NS equations for Bjorken flow \cite{Danielewicz:1984ww}. Eq.\ \eqref{eqBjorken} can be rewritten in a clearer form by defining the variables $w=\tau T$ and $f=1+\tau \partial_\tau T/T$ \cite{Heller:2011ju,Heller:2015dha}, 
leading to
\begin{align}
\begin{split}
\bar{\chi}w f(w)\frac{df(w)}{dw}+ 3\bar{\chi} f(w)^2 + f(w)\left(w-\frac{14}{3}\bar{\chi}\right)
  +\frac{16\bar{\chi}}{9}-\frac{4\bar{\eta}}{9}-\frac{2w}{3}=0.
\end{split}
\label{eqattractorBj}
\end{align}
This equation is very similar to the one found in the case of Israel-Stewart theory first reported in \cite{Heller:2015dha} and, as such, it shares the same qualitative features. In this system, the Knudsen number (i.e., the ratio between micro and macroscopic length scales) $K_N \propto 1/w$ and, thus, the NS limit should be recovered in the large $w$ (i.e., large $\tau$) regime. This can be seen by considering the formal large $w$ series solution $f(w) = \sum_{n=0}^\infty f_n w^{-n}$ for \eqref{eqattractorBj}, which describes the gradient expansion series around equilibrium, and leads to the following equation for its coefficients
\begin{equation}
f_{n+1} = \bar{\chi}\sum_{m=0}^n (n-m-3)f_{n-m} f_m+\frac{14}{3}\bar{\chi}f_n
\label{largefn}
\end{equation}  
for $n>1$, while $f_0=2/3$ and $f_1 = 4\bar{\eta}/9$. The exact result for NS corresponds to truncating the infinite series as $f_0+f_1/w$. However, such a truncation leads to acausal equations, which can only be resolved by properly resumming the series. In fact, the series coefficients diverge as $f_n \sim n!$ for large $n$, as shown in Fig.\ \ref{fig2}. We remark that the divergence of the large $\tau$ expansion in Bjorken flow was previously found in holography \cite{Heller:2013fn,Buchel:2016cbj}, kinetic theory \cite{Denicol:2016bjh,Heller:2016rtz}, as well as in hydrodynamic theories involving extended variables \cite{Heller:2015dha,Basar:2015ava,Aniceto:2015mto,Florkowski:2016zsi} (for a review see \cite{Florkowski:2017olj}). Connections with resurgence theory have been investigated in \cite{Heller:2015dha,Basar:2015ava,Aniceto:2015mto,Buchel:2016cbj}. 

\begin{figure}[th]
\includegraphics[width=0.6\textwidth]{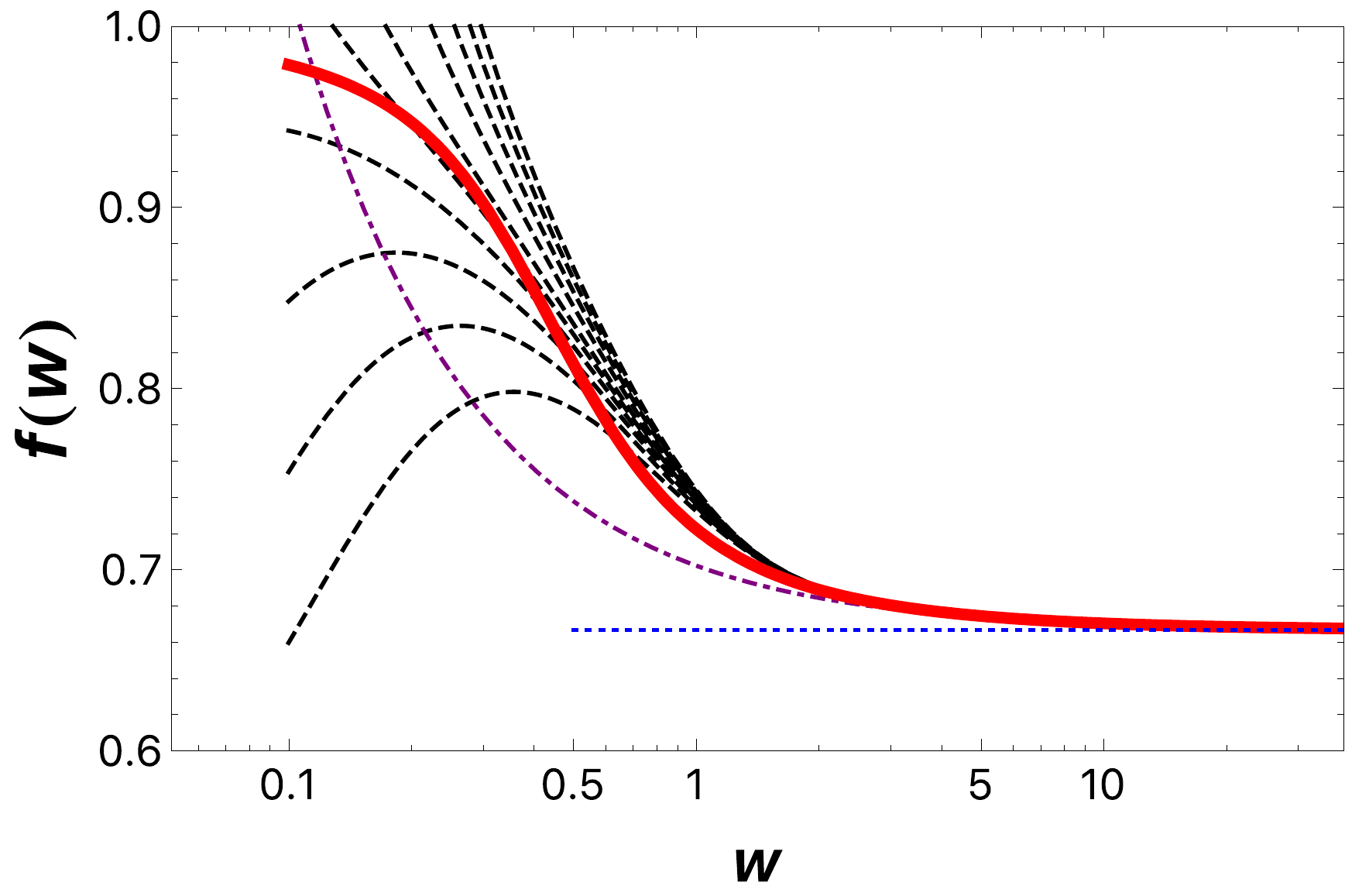}
\caption{(Color online) Hydrodynamic attractor solution for the causal tensor \eqref{tensorfodaotche} with $\eta/s=0.08$ and $\chi=4\eta$. The black dashed lines represent solutions of \eqref{eqattractorBj} with different initial conditions and the solid red line corresponds to the attractor solution. The NS solution is given by the purple dotted-dashed curve while the dotted blue line denotes the equilibrium limit.}
\label{fig3}
\end{figure}

In general, one expects that such a result indicates that new properties of the solutions of the equations of motion, which do not appear at any finite order in the series expansion, may emerge after resummation. 
As a matter of fact, linear disturbances around the series solution decay exponentially at large times \cite{Heller:2015dha} on a time scale controlled by the non-hydrodynamic mode that appears when $\bar{\chi}\neq 0$, which indicates the presence of a non-equilibrium structure called the hydrodynamic attractor. This is confirmed numerically in Fig.\ \ref{fig3} by investigating the behavior of the solutions of \eqref{eqattractorBj} generated using different initial conditions for $f(w)$. As noticed in \cite{Heller:2015dha}, the attractor solution can be determined using the analogous of the slow-roll expansion in cosmology \cite{Liddle:1994dx}, which here corresponds to setting $df/dw\to 0$ (the red line in Fig.\ \ref{fig3} shows the result of this procedure taking into account first order corrections). One can see that already at very short times the system rapidly ``erases" its memory of the initial conditions and converges to the hydrodynamic attractor (in solid red) before it reaches equilibrium (where $f\to 2/3$). We also show the NS solution where $f(w)= 2/3 + 4\bar{\eta}/(9w)$ for comparison. 

We stress that in our case the only dynamic variables of the system are the original hydrodynamic fields and, thus, the presence of a hydrodynamic attractor, even in this case, suggests that this may be a generic feature of (causal) viscous relativistic fluids (at least in the case of Bjorken flow). 

\subsection{Gubser flow}
\label{section_Gubser}

Another important type of (conformal) hydrodynamic flow employed in the study of simple models of heavy-ion collisions is the so-called Gubser flow \cite{Gubser:2010ze}. In this case, the flow is invariant under $SO(3)\otimes SO(1,1) \otimes Z_2$, with the $SO(3)$ being a particular sub-group of the $SO(4,2)$ conformal group which includes the symmetry of the solutions under rotations around the $z$ axis and two operations constructed using special conformal transformations that replace the translation invariance in the $xy$ plane present in Bjorken flow.  Full analytical solutions for the ideal fluid and NS approximations were derived in \cite{Gubser:2010ze} and the geometrical interpretation of $SO(3)\otimes SO(1,1) \otimes Z_2$ symmetry after Weyl rescaling was explained in \cite{Gubser:2010ui}.

Ref.\ \cite{Marrochio:2013wla} went beyond the NS limit and studied the case of a fluid described by the conformal IS equations undergoing Gubser flow. The semi-analytical solutions obtained in \cite{Marrochio:2013wla} have since then become the standard test of the accuracy of numerical schemes used in the large scale codes that realistically model the hydrodynamic evolution of the quark-gluon plasma \cite{Shen:2014vra}. Moreover, they have also motivated a series of studies on the emergence of hydrodynamic behavior in rapidly expanding fluids described by kinetic theory models \cite{Denicol:2014xca,Denicol:2014tha,Nopoush:2014qba,Noronha:2015jia,Bazow:2015dha}. 

One interesting aspect of the NS solution for Gubser flow is that there are regions in space-time where the temperature becomes negative as long as $\eta/s > 0$. Ref.\ \cite{Gubser:2010ze} argued that in these regions the gradients are so large that the NS equations do not apply and it was observed in \cite{Marrochio:2013wla} that the higher order resummed dynamics included in IS theory (and kinetic theory \cite{Denicol:2014xca,Denicol:2014tha}) resolved this issue guaranteeing that $T$ remained positive-definite. In this section we show that the same occurs in the new theory in \eqref{tensorfodaotche}, which provides a powerful consistency test of the formalism developed here.

The symmetry pattern that defines Gubser flow exactly determines \cite{Gubser:2010ze} the flow velocity to be 
\begin{equation}
u_\mu = \left(u_\tau(\tau,r),u_r(\tau,r),0,0\right)
\nonumber
\end{equation}
with
\begin{eqnarray}
u_\tau&=& -\cosh\left[\tanh^{-1}\left(\frac{2\tau r q}{1+q^2 \tau^2 + q^2 r^2}\right)\right]\nonumber \\
u_r &=& \sinh\left[\tanh^{-1}\left(\frac{2\tau r q}{1+q^2 \tau^2 + q^2 r^2}\right)\right],
\nonumber
\end{eqnarray}    
where we used Milne coordinates $x^\mu = (\tau,r,\phi,\varsigma)$, with $r=\sqrt{x^2+y^2}$ and $\phi = \tan^{-1}(y/x)$. Above, $q$ is an arbitrary energy scale that describes the spatial extent of the solutions in the $xy$ plane (we note that the Bjorken solution is recovered in limit $q\to 0$). Without loss of generality, we set $q=1 \,\mathrm{fm}^{-1}$ \cite{Marrochio:2013wla}. Just as it happened in the Bjorken flow case, since $u_\mu$ is already known (and the momentum part of the conservation laws is automatically satisfied) the only quantity left to characterize the hydrodynamic solution in our theory is $T=T(\tau,r)$, which is obtained as a solution of $u_\nu \nabla_\mu T^{\mu\nu}=0$. In this case, the nonlinear 2nd order partial differential equation for $T$ depends on $(\tau,r)$, which makes the problem considerably more complicated than the Bjorken flow case.   

However, the underlying conformal invariance of the equations of motion allows one to perform a Weyl transformation of the metric and rephrase this complicated flow pattern in terms of a locally static flow in the curved space-time $dS_3\otimes \mathbb{R}$ \cite{Gubser:2010ui}, where $dS_3$ denotes the 3-dimensional de Sitter space \cite{Weinberg_GR_book}. In fact, starting with the line element written in Milne coordinates $ds^2 = -d\tau^2 + dr^2 +r^2 d\phi^2 + \tau^2 d\varsigma^2$, one may rescale the flat space-time metric $ds^2 \to ds^2/\tau^2$ to obtain a metric of $dS_3\otimes \mathbb{R}$, which may be written in global coordinates as $d\hat{s}^2 = -d\rho^2+\cosh^2 \rho \,d\theta^2 + \sin^2\theta\cosh^2\rho\, d\phi^2+d\varsigma^2$, where
\begin{eqnarray}
\sinh\rho = -\frac{1-\tau^2+r^2}{2\tau},\qquad 
\tan\theta = \frac{2r}{1+\tau^2-r^2}.\nonumber
\end{eqnarray}
After this procedure, the fluid is at rest $\hat{u}_\mu = (-1,0,0,0)$ and the temperature $\hat{T}=\hat{T}(\rho)$, which now obeys a 2nd order nonlinear ODE that can be easily solved. Once that is done, a simple Weyl transformation gives $T(\tau,r) = \hat{T}(\rho(\tau,r))/\tau$ \cite{Gubser:2010ui}. This approach to solve the conformal hydrodynamic equations was applied in \cite{Hatta:2014gqa} to obtain axisymmetric exact solutions with nonzero vorticity, while other flow profiles were systematically developed in \cite{Hatta:2014gga}. 

\begin{figure}[th]
\includegraphics[width=0.6\textwidth]{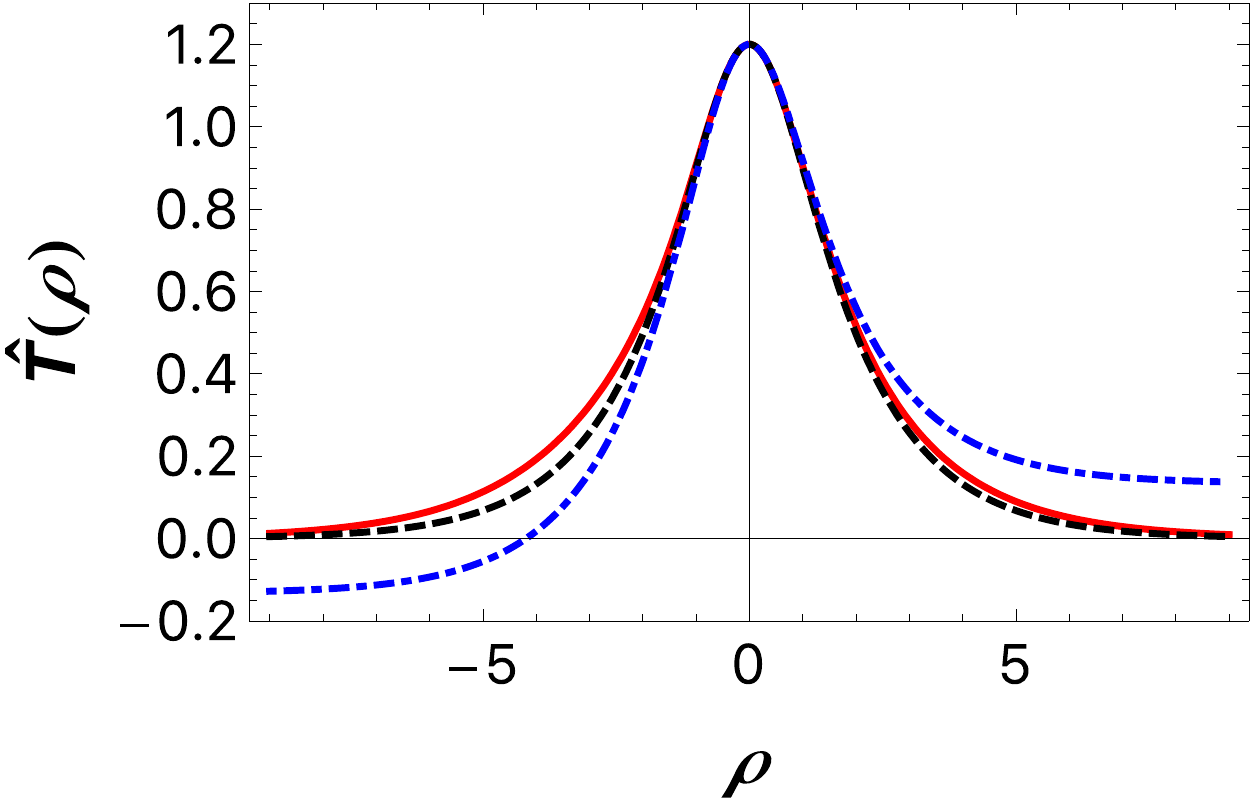}
\caption{(Color online) Temperature profile in $dS_3\otimes \mathbb{R}$ as a function of de Sitter time $\rho$. The red line is the solution of \eqref{EOMGubser}, the blue dotted-dashed line denotes the NS solution \eqref{NSsolution}, and the ideal fluid case is shown in black (dashed). In this plot, $\bar{\eta}=0.2$, $\bar{\chi}=4\eta$, and $T_0 = 1.2$.}
\label{fig4}
\end{figure}

\begin{figure}[th]
\includegraphics[width=0.6\textwidth]{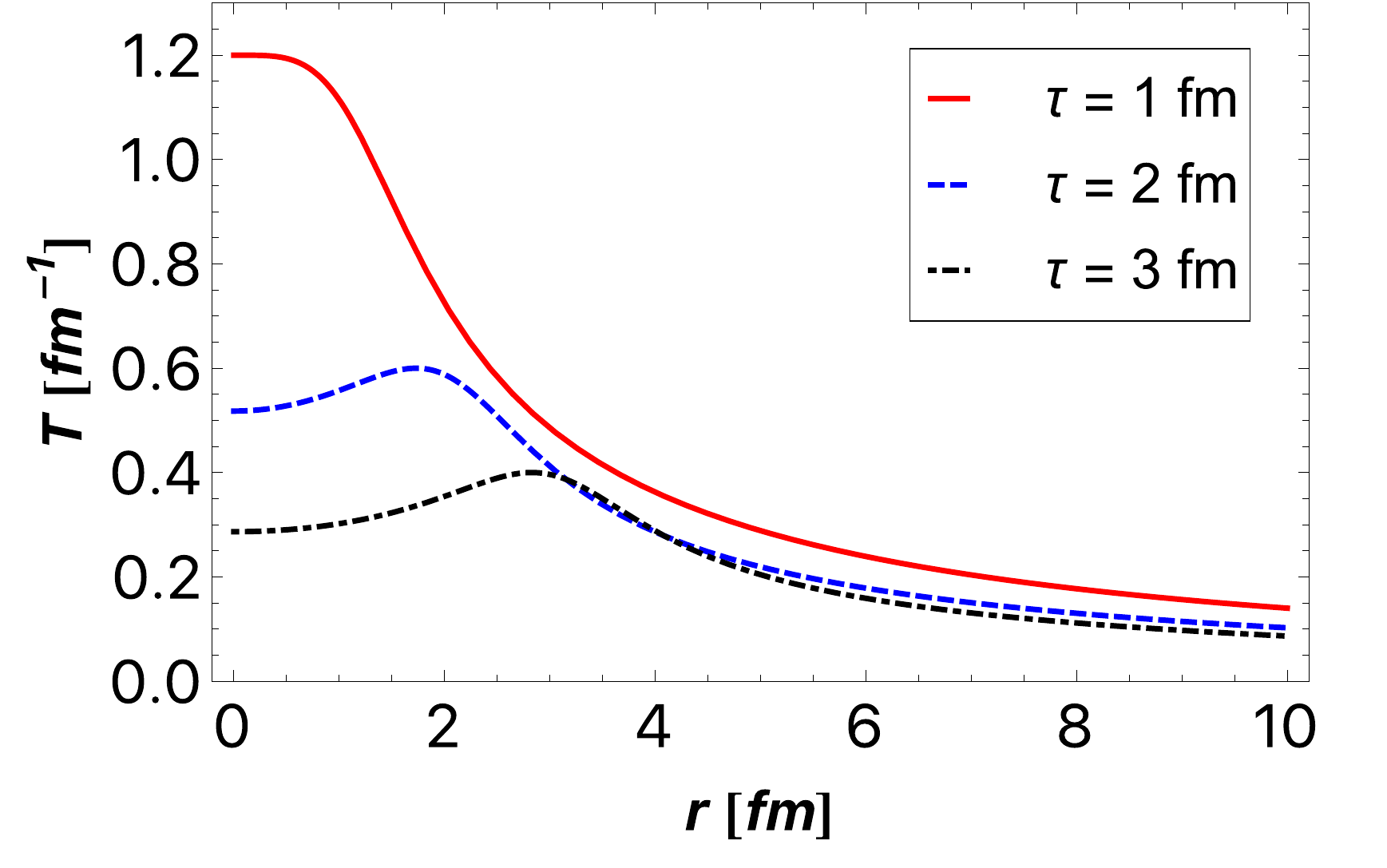}
\caption{(Color online) Dependence of the temperature with the transverse radius $r=\sqrt{x^2+y^2}$, evaluated at different Milne times $\tau=1,2,3$ fm, for the viscous fluid defined by \eqref{tensorfodaotche} undergoing Gubser flow. The fluid also expands in the $z$ direction (not shown). In this plot, $\bar{\eta}=0.2$, $\chi=4\eta$, and $T_0 = 1.2$.}
\label{fig5}
\end{figure}

In order to compare to the IS case studied in \cite{Marrochio:2013wla}, it is convenient to write our 2nd order equation of motion for $\hat{T}$ in $dS_3\otimes \mathbb{R}$ as a set of coupled 1st order differential equations
\begin{align}
\begin{split}
&\frac{1}{\hat{T}}\frac{d\hat{T}}{d\rho}+\frac{2}{3}\tanh\rho=\hat{\mathcal{F}}(\rho), \\
&\bar{\chi} \frac{d\hat{\mathcal{F}}}{d\rho}+3\bar{\chi} \hat{\mathcal{F}}^2 + \frac{2}{3}\bar{\chi} \hat{\mathcal{F}}\tanh\rho+\hat{T} \hat{\mathcal{F}} 
 - \frac{4}{9}\bar{\eta}(\tanh\rho)^2=0,
\end{split}
\label{EOMGubser}
\end{align}
where we used that in $dS_3\otimes \mathbb{R}$ the expansion rate is $\nabla_\mu\hat{u}^\mu=2\tanh\rho$, $\hat{u}^\lambda \nabla_\lambda \hat{u}^\mu = 0$, and the only nonzero components of the shear tensor are $\hat{\sigma}^\varsigma_\varsigma = -2\tanh\rho/3$, $\hat{\sigma}^\phi_\phi = \hat{\sigma}^\theta_\theta = \tanh\rho/3$. One can appreciate the similarity between \eqref{EOMGubser} and the IS equations (11) and (12) obtained in \cite{Marrochio:2013wla}. The NS limit ($\bar{\chi}=0$) gives the exact solution \cite{Gubser:2010ze}
\begin{align}
\begin{split}
\hat{T}_{NS}(\rho) = \frac{T_0}{\cosh^{2/3}\rho} + \frac{4}{27}\bar{\eta} \frac{\sinh^3\rho}{\cosh^{2/3}\rho}
 \, _2F_1\left(\frac{3}{2};\frac{7}{6};\frac{5}{2};-\sinh^2\rho\right),
\end{split}
\label{NSsolution}
\end{align} 
where $_2F_1$ is a hypergeometric function and $T_0$ is a constant that characterizes the solution at $\rho=0$. The equation above shows that $\hat{T}$ is positive-definite in the ideal fluid limit ($\bar{\eta}=0$) but for NS $\lim_{\rho\to \pm \infty} \hat{T}_{NS}(\rho) = \pm 2\bar{\eta}/3$ \cite{Gubser:2010ze}, which implies that for any time $\tau$ there is an $r$ for which the temperature turns negative (for any value of $\eta>0$). This pathology of NS does  not occur in the solution obtained from Eq.\ \eqref{EOMGubser}, as illustrated in Fig.\ \ref{fig4}. In this plot we used $\bar{\eta}=0.2$, $\chi=4\eta$, $T_0=1.2$ and $\hat{\mathcal{F}}(0)=0$, to facilitate the comparison with the results obtained for IS theory in \cite{Marrochio:2013wla}. The red line denotes our numerical solution, the black dashed line corresponds to the ideal fluid limit and the NS solution, which becomes negative at sufficiently large negative $\rho$, is shown in blue. One can see that our solution for the temperature is positive-definite, taking values strikingly similar to the IS solution reported in Fig.\ 1 of Ref.\ \cite{Marrochio:2013wla} (in which $\bar{\eta}=0.2$ was also used).  
We show in Fig.\ \ref{fig5} the time evolution of the new solution as a function of the transverse radius $r$, using the same parameters employed in Fig. \ref{fig4}. The fluid rapidly expands in the transverse direction while also expanding in the longitudinal $z$ direction, similarly to 
what occurs with the IS solution \cite{Marrochio:2013wla}. We remark that hydrodynamic attractor solutions can also be investigated in Gubser flow, as shown in Refs.\ \cite{Behtash:2017wqg,Denicol:2018pak}. Since our equations of motion \eqref{EOMGubser} are very similar to the IS equations for Gubser flow \cite{Marrochio:2013wla}, we expect that hydrodynamic attractor behavior will also be present in our case. We leave such a study for future work. 

\subsection{Initial conditions\label{section_initial_conditions}}

The equations of motion derived from \eqref{tensorfodaotche} are second order 
evolution equations. Their initial value formulation, therefore, requires a
complete knowledge of eight quantities
\begin{align}
\begin{split}
\left. u^i \right|_{t=0}, \,\left. \ep \right|_{t=0},\,
\left. \nabla_0 u^i \right|_{t=0},\, \text{ and } \, \left. \nabla_0 \ep \right|_{t=0},
\end{split}
\label{data_at_time_zero}
\end{align}
where we assume to be working locally in coordinates $(x^0,x^i) = (t,x^i)$ such that
initial data is given on the hypersurface $\{t=0\}$. Note that the $u^0$ component and its time derivative at $t=0$ are obtained from the normalization $u^\mu u_\mu=-1$. For a comparison, we note that the relativistic NS equations require the knowledge of  the same quantities as above, with the exception of $\nabla_0 \ep |_{t=0}$ which is only needed in our formulation. 

Clearly, the choice of initial data depends on the particular problem at hand. In the
study of the quark-gluon plasma, one is  often given an energy-momentum tensor $\cT_{\mu\nu}$ at some initial time\footnote{As above, here we denote the initial time as $t=0$. However, we note that in heavy-ion applications one initializes hydrodynamics at some nonzero initial Bjorken time $\tau_0$.}, computed for instance using the IP-Glasma model \cite{Schenke:2012wb}, that is expected to be matched to the corresponding energy-momentum tensor of the fluid. In principle, this matching can always be performed in IS-like approaches since the ten independent quantities in $\cT_{\mu\nu}$ can be directly mapped into the ten dynamical degrees of freedom of IS theory defined, e.g., in the Landau frame. However, it is important to remark that this is not free from problems. For instance, the initial state physics contained in $\cT_{\mu\nu}$ in the case of the quark-gluon plasma, which for instance involves solving the classical Yang-Mills equations in the case of IP-Glasma, may be such that the solution of the Landau condition $u_\mu \cT^{\mu\nu} = -\mathcal{E} \,u^\nu$ at the initial time gives regions in space where $\mathcal{E}$ is not positive-definite \cite{Arnold:2014jva}. Also, even if that is not the case, the deviations from local equilibrium at the initial time may be so large that the extracted $\pi^{\mu\nu}$ of the fluid is larger than the equilibrium pressure, which implies that the system is already outside the regime of applicability of the IS equations \cite{MIS-6} and higher order corrections become necessary. Furthermore, in extremely rapidly expanding systems depending on the size of the bulk viscosity, it is possible that the bulk scalar $\Pi$ in IS theory is such that the local total pressure changes sign, which leads to phenomenon of relativistic cavitation \cite{Torrieri:2008ip,Rajagopal:2009yw}. This possibility was already found in realistic simulations of heavy-ion collisions in \cite{Denicol:2015bpa}. Moreover, since there is no proof of causality and well-posedness for these equations in the nonlinear regime, it is not known if the spacetime evolution of the fluid described by these equations is always well behaved in the case highly inhomogeneous initial conditions. Therefore, even for the case of IS-like equations, it is not guaranteed that their use makes sense in the extreme conditions that may occur in some heavy-ion collision events (such as in the case of small collision systems formed in proton-nucleus or even proton-proton collisions). Thus, we limit our discussion to consider the case where the system is close to local equilibrium and such issues do not appear.

If the system is close to equilibrium (though still nonlinear), it becomes then again meaningful to match the initial $\cT_{\mu\nu}$ to a fluid dynamic description. In this case, one can approach this problem considering different levels of approximation. Assuming that the eigenvalue problem $w_\mu \cT^{\mu\nu} = -\alpha\, w^{\nu}$, with $\alpha > 0$ and $w_\mu w^\mu = -1$ can be solved at the initial time, the first approximation consists in assuming that the dynamics of the system can be described by the ideal fluid equations of motion with initial conditions given by $\epsilon |_{t=0} = \alpha$ and $u^i |_{t=0} = w^i$. This approximation may be locally improved by assuming that the system evolves according to viscous fluid dynamics. However, besides the problems with causality and stability, the NS equations contain less information than the general initial $\cT^{\mu\nu}$ and, thus, information about the initial state is necessarily lost when setting up the initial conditions for $\epsilon$, $u^i$, and $\nabla_0 u^i$. On the other hand, as remarked above the initial values of the fields in the conformal IS equations can be directly matched to the initial (traceless) $\cT^{\mu\nu}$, though in this case there is no way to know a priori if causality violations (and other theoretical issues) may appear in the nonlinear regime.

In comparison to the NS equations, the new tensor derived in this paper can in principle better detail the system in the initial state since the number of input variables  equals 8 in comparison to the 9 present in the most general $\cT^{\mu\nu}$ (assuming conformal invariance). However, differently than IS theory, in our case  some amount of information about the initial condition is necessarily lost though causality and well-posedness have been proven in the nonlinear regime. Given that currently the majority of the simulations of the quark-gluon plasma employing IS theory do not fully take into account all the possible information in the initial $\cT^{\mu\nu}$ computed from quantum chromodynamics, we believe that it is important to investigate which properties are more important for the specific problems at hand by comparing the results for the evolution of $T^{\mu\nu}$ obtained using IS and the new theory proposed in this paper. 

Such a comparison could be meaningfully performed as follows. Let us first assume that $\cT^{\mu\nu}$ models the initial energy-momentum tensor of the quark-gluon plasma but this system is not very far from equilibrium, being thus close to the NS regime\footnote{This statement can be made more formal in the sense of Geroch's work in Ref.\ \cite{Geroch-RelativisticDissipative}. We note that both IS equations and ours can have the NS equations as a limit. For the former, this limit is well understood \cite{Baier:2007ix} while in the case of our tensor \eqref{tensorfodaotche} this occurs when the contribution from the terms $\mathcal{A}=\chi \mathcal{D}\epsilon/(\epsilon+P)$ and $\mathcal{Q}^\mu = \lambda \mathcal{D}^{\langle\mu\rangle} \epsilon/(\epsilon+P)$ is neglected  in  the equations of motion.}. One solves $w_\mu \cT^{\mu\nu} = -\alpha\, w^{\nu}$ and uses these quantities to set $\epsilon |_{t=0} = \alpha$ and $u^i |_{t=0} = w^i$ in our theory and, correspondingly, 
$\mathcal{E} |_{t=0} = \alpha$ and $u^i |_{t=0} = w^i$ in conformal IS theory 
 (where $\mathcal{E}$ is as in section \ref{section_conformal}). Using NS theory as guidance, we set $\mathcal{D}_\mu\epsilon|_{t=0}=0$ in our equations. Projecting this condition onto the flow one finds
\begin{equation}
\nabla_0\epsilon|_{t=0}=\frac{4\ep\,u^0}{3+2\bar{u}^2}\left (\frac{u^lu^m\nabla_lu_m}{1+\bar{u}^2}-\nabla_lu^l-\frac{u^l\nabla_l\ep}{2\ep}\right ),
\nonumber
\end{equation}
where $\bar{u}^2=(u_1)^2+(u_2)^2+(u_3)^2$. The remaining conditions then give
\begin{eqnarray}
u^0\nabla_0u_j|_{t=0}=\left (u_0^2\nabla_l u^l-u^lu^m\nabla_l u_m-\frac{u^l\nabla_l\ep}{4\ep}\right )
\frac{u_j}{3+2\bar{u}^2}-u^l\nabla_l u_j-\frac{\nabla_j\ep}{4\ep}.
\nonumber
\end{eqnarray}
This sets up the initial value problem for our tensor. At the same time, we also use this last equation to provide the remaining initial condition needed for NS. On the other hand, one can use the equations above to determine all the components of $\sigma_{\mu\nu}|_{t=0}$, which can then be used to set $\pi^{\mu\nu}|_{t=0}$ in IS theory. Therefore, in this case all the different descriptions, i.e., NS, IS, and ours, 
would have the same initial $T^{\mu\nu}|_{t=0}$. 
One could then compare the solutions of the equations of motion for these different theories under heavy-ion like conditions. Unfortunately, such a study requires solving the equations of motion in situations that are significantly more complex than those presented in \ref{section_Bjorken} and \ref{section_Gubser}. We intend to investigate this interesting problem in our future work.

\section{Limitations, open questions, and discussions\label{section_discussion}}

Given the novelty of (\ref{tensorfodaotche}), it is natural that many questions remain
open. In this section we will briefly discuss some of them, tying the discussion with limitations
and potential shortcomings of the theory here presented.


\subsection{Generalizations of Theorems \ref{main_theorem} and \ref{Minkowski_theorem} and other fluid theories\label{section_generalizations}}

Theorems \ref{main_theorem} and \ref{Minkowski_theorem} establish well-posedness of
Einstein's equations coupled to \eqref{tensorfodaotche} in Gevrey spaces. Such spaces
are commonly used in the study of fluid dynamic equations (see, e.g., \cite{TadmorBesovGevrey,  TitiGevreyNavier, TitiGevreyParabolic, TemamGevrey, RodinoGevreyBook} and references therein),
and they have been used in the study of Einstein's equations before
\cite{ChoquetBruhatGRBook, Lich_MHD_paper, Lichnerowicz_MHD_book}.
In fact, 
in some circumstances, Einstein's equations coupled to ideal magneto-hydrodynamics appear 
to  have been shown to be well-posed only in the 
Gevrey spaces \cite{ChoquetBruhatGRBook,FriRenCauchy}\footnote{Although probably
the formulation of \cite{AnilePennisiMHD} would carry over to the 
coupling
with Einstein's equations. A proof of this statement, however, does not
seem to be available in the literature.}. Nevertheless, it would be important to establish
a well-posedness result in larger function spaces,  not only for the sake of generality but
also because many important questions, such as those concerning the long-time dynamics
(see section \ref{section_energy_conditions}) are better posed in other function spaces
such as Sobolev spaces (see \cite{ChristodoulouBlackHoles} for an example in the context
of Einstein's equations). The main difficulty to generalize Theorems \ref{main_theorem}
and \ref{Minkowski_theorem} to Sobolev spaces is that the equations derived 
from \eqref{tensorfodaotche} are only weakly hyperbolic (for $a_1 \geq 4$ $a_2 \geq \frac{3a_1}{a_1-1}$; it is not clear whether the equations are hyperbolic in any sense if these conditions
do not hold). This is manifest by the presence of repeated roots in the characteristic determinant.
Absent further structural properties, weakly hyperbolic systems are not, in general, well-posed
in Sobolev spaces \cite{MizohataCauchyProblem}. A more refined analysis, therefore,
has to be carried out in order to generalize our Theorems to Sobolev spaces.
This will be presented in a future work, since the proof is quite technical (relying on delicate resolvent
estimates and an in-depth study of the regularity properties of $C^0$-semi-groups in Banach spaces),
and thus is beyond the scope of this work. 

Generalizations of Theorems \ref{main_theorem} and \ref{Minkowski_theorem} notwithstanding,
one should contrast our results with what is currently known about the IS and resummed
BRSSS theories, for which no analogues of such theorems are available, even in spaces more
restrictive than Gevrey spaces such as the space of analytic functions. It is also 
interesting to note the IS and resummed BRSSS theories posses multiple characteristics
\cite{Hiscock_Lindblom_stability_1983}, which 
would render the equations of motion weakly hyperbolic at best (unless, of course,
the equations are rewritten in terms of new ``better" variables). Thus, it is likely that for such theories
results beyond Gevrey spaces, if available at all, will be difficult to be obtained.

The comments of the previous paragraph should by
 no means minimize the importance of the IS and ressumed BRSSS theories, given their wide
 use in the study of relativistic fluids with viscosity. It remains an extremely
important open problem to find reasonable conditions that guarantee that these
theories are well-posed and causal in the non-linear
regime, both in Minkowski space and  when dynamically coupled to gravity. Unfortunately,
these are very difficult questions. The characteristic determinant of the IS theory
seems more complicated than the one we have here, and we were so far unable to find
any structure that would allow an application of techniques similar to the ones employed here.
The situation becomes even more complicated if the 2nd order terms proposed by \cite[Eq.\ (3.11)]{Baier:2007ix} are included, as these terms turn the causality and stability analysis significantly more complicated.
The first reason for this is computational: the equations will be of third order in derivatives,
increasing the complexity of the system's characteristics.
The second reason is structural. The coefficients of the principal part 
will now depend on derivatives of the fields. In our case, the coefficients
of the principal part depend on the fields but not on its derivatives (e.g., 
terms of the form
$\eta(\ep) g^{\mu\nu} \partial^2_{\mu\nu} u^\al$ and $u^\mu u^\nu \partial^2_{\mu\nu} \ep$),
where $\partial_\mu$ are coordinate partial derivatives,
thus the system characteristics can be understood solely in terms of the intrinsic properties
of the fields, i.e., the facts that $g$ is a Lorentzian metric, $u$ is time-like, and $\ep > 0$.
With a few exceptions (e.g., the property that the acceleration is orthogonal to the velocity), this 
is no longer the case when the coefficients depend on derivatives of the fields,
and the geometric and physical meaning of the system's characteristics become much more obscure.
Therefore, a systematic investigation of well-posedness and causality
(in the full nonlinear sense meant here) is extremely challenging 
when the 2nd order gradient terms proposed by \cite[Eq.\ (3.11)]{Baier:2007ix} are included.

We also mention that an important problem for applications in astrophysics is to understand the linear stability of fluid theories
beyond the Cowling approximation, not only for \eqref{tensorfodaotche}
but for other fluid theories as well.

\subsection{Energy conditions and positivity of $\ep$\label{section_energy_conditions}}
Because in our tensor $T^{\mu\nu} u_\mu u_\nu = \ep +\chi \nabla_\mu u^\mu +\frac{3\chi}{4\ep} u^\mu \nabla_\mu\ep = \ep +\cA$, the weak energy condition 
 $u_\mu u_\nu T^{\mu\nu}  \geq 0$ \cite{WaldBookGR1984}
can be violated for sufficiently large dissipative contributions.
This is not so much a limitation of (\ref{tensorfodaotche}) but rather  
a consequence of the assumptions of the 
gradient expansion employed to derive \eqref{tensorfodaotche} in section \ref{section_kinetic_theory}, as the theory is not supposed to be valid
for very large deviations from local equilibrium. In fact, in applications (including numerical
simulations) keeping track of the positivity of
$T^{\mu\nu} u_\mu u_\nu$ may provide a criterion to determine when the limit 
of validity of the theory has been crossed. This can be useful because,
while we know the theory to be valid only for small gradients, in practice it is 
not always evident when its regime of applicability has been reached. 
Note that by continuity, we know that
$T^{\mu\nu} u_\mu u_\nu$ will remain positive for some time interval  if positive initially.
Therefore, whether or not the weak energy condition is in fact violated 
depends on the long time behavior of the system. The latter, in turn,
depends on particular features of specific models, such as the values of $\chi/\eta$ and
$\lambda/\eta$ or the initial conditions chosen for the system. 

Another question tied to the long term dynamics is that of the positivity of $\ep$.
Again by continuity, $\ep$ will be strictly greater than zero if so initially. But the equations of motion degenerate if $\ep=0$, in which case our causality and well-posedness results no longer apply. 
Such difficulties caused by $\ep=0$, however, are no different than what happens already in other fluid theories. For the non-relativistic Euler equations, one has
$\partial_t v^i +v^j \nabla_j v^i + \frac{1}{\rho} \nabla^i p = 0$, where $v^i$, $\rho$, and
$p$ are the fluid's velocity, density, and pressure, respectively. We see that $\rho > 0$ is needed;
and if one writes the equations as $\rho(\partial_t v^i +v^j \nabla_j v^i )+  \nabla^i p = 0$, the situation
is hardly better, since well-posedness, among other traditional results \cite{MajdaCompressibleFlow}, no longer applies when
$\rho=0$ because the equation degenerates. 
The same problem also arises in the
non-relativistic Navier-Stokes equations when the density vanishes, and, in fact,
in the relativistic NS, IS, and resummed BRSSS theories as well. To see this, note
that $u^\mu \nabla_\nu T^\nu_\mu = 0$ can be written, for all such theories and ours, as
\begin{align}
\begin{split}
u^\mu \nabla_\mu \epsilon + \frac{4}{3}\epsilon \nabla_\mu u^\mu + \mathcal{V} = 0,
\end{split}
\label{generic_epsilon_evolution}
\end{align}
where $\mathcal{V}$ represents the viscous contributions and we used $P(\epsilon) = \frac{1}{3} 
\epsilon$. Assume that we know $\epsilon$ to be positive at certain time that we can 
take as $t=0$. By continuity\footnote{Assuming, say, that a solution exists
and is continuous.} $\epsilon$ will be positive for some time interval $[0,\mathsf{T})$.
The question of whether $\epsilon$ remains positive after $\mathsf{T}$ can be reduced to determine
whether 
\begin{gather}
\lim_{t\rightarrow \mathsf{T}^-} \epsilon(t, x) \equiv \epsilon_\mathsf{T}(x) > 0
\label{limit_epsilon_positive}
\end{gather}
for all $x$. For, if this is the case,
we can then take $\epsilon_\mathsf{T}$ as initial data for the equations on $t=\mathsf{T}$. Solving the 
corresponding initial value problem\footnote{Assuming a well-posedness result to be available.
Thus, even to discuss whether $\epsilon$ remains positive, we see that a well-posedness
theorem seems to be needed.} with $\ep_\mathsf{T}$ as initial condition, we then obtain that the solution
now exists on a larger interval $[0,\mathsf{T} + \mathsf{T}^\prime)$, $\mathsf{T}^\prime >0$; again by continuity
(and shrinking $\mathsf{T}^\prime$ a bit if necessary) we conclude that $\epsilon$ is positive 
on $[0,\mathsf{T} + \mathsf{T}^\prime)$. We can now repeat the argument to obtain
positivity after $\mathsf{T}+\mathsf{T}^\prime$ and so on.

Thus, we need to obtain \eqref{limit_epsilon_positive} to show that $\epsilon$ will remain
positive. Given $(t,x)$, $t < \mathsf{T}$, we can integrate \eqref{generic_epsilon_evolution}
along an integral curve $\gamma$ of $u^\mu$ connecting $(t,x)$ to some $(0,x_0)$, 
yielding
\begin{align}
\begin{split}
\epsilon(t,x) = \epsilon(0,x_0)e^{-\frac{4}{3} \int_\gamma \nabla_\mu u^\mu - \int_\gamma
\frac{\mathcal{V}}{\epsilon} }.
\end{split}
\label{generic_epsilon_evolution_integrated}
\end{align}
In producing this identity we had to use that $\epsilon > 0$,
which is the case for $t< \mathsf{T}$. Consider first the case without viscosity, i.e., $\mathcal{V} = 0$.
Then \eqref{limit_epsilon_positive} clearly holds unless
$\lim_{t \rightarrow \mathsf{T}^-} \int_\gamma \nabla_\mu u^\mu = \infty$, i.e., unless
$\nabla_\mu u^\mu$ becomes singular. However, the same argument does not work when $\mathcal{V} 
\neq 0$ due to the presence of $\epsilon$ on the RHS of \eqref{generic_epsilon_evolution_integrated}. Indeed, in order to conclude
\eqref{limit_epsilon_positive} we need 
$\lim_{t \rightarrow \mathsf{T}^-} \int_\gamma \frac{\mathcal{V}}{\ep}$ to remain finite. This limit
depends on the form of $\mathcal{V}$. In particular, it will involve (for a conformal fluid)
terms in $\frac{\eta}{\epsilon} \propto \epsilon^{-\frac{1}{4}}$. Thus, 
for $\lim_{t \rightarrow \mathsf{T}^-} \int_\gamma \frac{\mathcal{V}}{\ep}$ to be finite we need
$\epsilon$ to remain positive in the limit $t \rightarrow \mathsf{T}^-$, which is  
what we are trying to prove to begin with.

The above shows that the mechanism that enforces $\epsilon$ to remain positive in 
an ideal fluid no longer holds when viscosity is present. Moreover, if $\epsilon$
reaches the value zero, there is no a priori reason why it could not become negative
(assuming that we can guarantee solutions to still exist if $\epsilon=0$, see below),
thus violating the weak energy condition. Note that, as stressed, this 
is a potential issue in the NS, IS, and BRSSS theories alike.

Upon closer inspection, it is not surprising that many difficulties arise when the fluid 
energy density vanishes since zero energy/matter-density corresponds to a vacuum region. Thus, 
$\ep=0$ marks an interface
where the fluid is separated from the vacuum. A typical scenario where one has such an interface
is in the study of stars, where the star is modeled as a fluid body and $\ep=0$ 
corresponds to the boundary of the star. The main difficulty
in this case is that the interface $\ep=0$ is not prescribed but rather it is dynamic, i.e., it changes with
the motion of the fluid. Unfortunately, establishing well-posedness and causality in such cases
is extremely difficult. Even for the non-relativistic Euler equations the problem has been solved
only over the last decade or so \cite{CoutandShkollerFreeBoundary,CoutandHoleShkollerLimit,DisconziEbinFreeBoundary2d,
DisconziEbinFreeBoundary3d,LindbladFreeBoundary,Lindblad-FreeBoundaryCompressbile},
and it remains largely open for the equations of relativistic ideal 
fluids\footnote{When $\epsilon$ is allowed to vanish, the relativistic Euler equations
degenerate and standard well-posedness results \cite{AnileBook} no longer apply.
As just showed, for an ideal fluid $\epsilon$ will remain positive if initially so (absent
singularities). Thus, for ideal fluids we only have $\epsilon=0$ if the initial data
is chosen with regions of zero energy density. But, as remarked, this situation
is important in the study of stellar evolution and hence needs to be addressed.}
\cite{JangLeFlochMasmoudi,HadzicShkollerSpeck,RendallFluidBodies}.

It is, of course, possible that $\ep$ remains strictly greater than zero if initially so, in which case
the issues of previous paragraphs do not arise. But, as mentioned, whether or not
$\ep$ remains positive requires understanding the long term dynamics. The takeaway
of this discussion is that to answer whether or not certain features (positivity of $\ep$,
energy conditions, etc.) persist for longer times we need to go beyond well-posedness
results and understand problems such as the potential formation of singularities or degeneracy
of the equations, how large can the interval of existence be, and so on. Such questions,
albeit very important, are typically very challenging for non-linear equations (in fact, they are
intimately tied to the problem of global existence briefly mentioned in section 
\ref{section_introduction}) and are beyond the scope of this work.

The difficulties discussed above essentially boil down to the question of whether properties
that hold initially persist for long times, e.g., whether $\ep > 0$ or the weak energy condition remains
valid for a long time interval (beyond what is valid by a simple continuity argument).
The challenges in answering these questions reflect more on the difficulties common to the analysis of non-linear partial differential equations than limitations of our model per se. In fact, as discussed above, similar difficulties are present in both the IS and BRSSS theories.

\subsection{The non-conformal case}

Another important question is whether it is possible to generalize the ideas used to derive
(\ref{tensorfodaotche}) to  construct more general causal and stable
energy-momentum tensors, including theories with derivatives higher than
second order, theories with more conserved charges, 
and the non-conformal case. The short answer to most of these questions is yes. 
The strategy leading to (\ref{tensorfodaotche}), namely, start with kinetic theory
but leave the choice of frame (i.e., the choice of $m$, $n$, and $r$ in
\eqref{generalmatching}) general,
 can be reproduced for other types of gases. The main difficulty now is that we will have a larger number of transport coefficients  and 
a more complicated equation of state. Finding conditions for well-posedness, causality, and stability
will then require determining substantially more complex relations among these quantities.
Moreover, such relations must be compatible with the choices of hydrodynamic frames allowed
by different values of $m$, $n$, and $r$. 

Going beyond kinetic theory, it would be interesting to investigate how \eqref{tensorfodaotche} may be derived using holographic techniques. In fact, it is known how to obtain the BRSSS equations (and the corresponding non-conformal generalization) from the fluid/gravity duality \cite{Bhattacharyya:2008jc} and it is possible that modifications of this approach can be devised to obtain \eqref{tensorfodaotche}. 

\subsection{Choice of frames}

A crucial element in the fluid theory introduced here that was essential for causality 
and stability (and also for the possibility of extending our results to non-conformal 
theories as just mentioned)
is the fact that we have not adopted
either the Landau or Eckart frame. 
In essence, our philosophy is that the fundamental principle of causality should determine what frames are physically meaningful, 
and not the other way around. 

Even if in practice causality and stability are determined a posteriori, i.e., one 
establishes conditions guaranteeing these properties and then verify that they are compatible
with the choices given by \eqref{generalmatching},
this would not have been possible had we imposed Landau or Eckart's frames at the beginning (without introducing new dynamic degrees of freedom).
In this regard, it is interesting to notice that the causal theories
of \cite{TempleViscous,TempleViscous2,TempleViscous3} do not use Landau or Eckart's frames
either.
In fact, any pre-determined choice of frame at the beginning would probably prevent
us from establishing causality for the full nonlinear system of equations (i.e., fluid + Einstein's) considered here.

We showed in this paper that the theory in \eqref{tensorfodaotche} provides a causal generalization of conformal NS theory. As an effective theory, our construction is rigorously well defined in the sense that it is causal and stable, though admittedly not accurate in the ultraviolet (as it must be the case in any effective theory at sufficiently large energy scales). While we considered the more general case where $\lambda$ and $\chi$ are distinct, in practical applications it may be more convenient to assume these quantities to be the same. For instance, the choice $\lambda=\chi=4\eta$ would satisfy our causality, well-posedness, and stability conditions. In this case, the only free parameter needed to determine dissipative effects would be the value of $\eta/s$, just as in conformal NS theory.  

\section{Conclusions\label{section_conclusions}}

In this manuscript, we have presented what is, to the best of our knowledge,
the first example of a viscous relativistic fluid that is causal, stable, well-posed
(in the non-linear regime with or without dynamic coupling to gravity), that is derivable from kinetic theory and as such obeys the second law of thermodynamics, and at the same time producing meaningful physical results in widely used test models. The equations of motion involve only the hydrodynamic fields and are simpler than those from extended irreversible thermodynamics, including IS theory.
We have solved numerically the equations of motion for the case of Bjorken flow and found the presence of an out-of-equilibrium hydrodynamic attractor. Causality was identified here as the root behind the resummation present in the dispersion relations obtained from the linear stability analysis and also in the hydrodynamic attractor of the (fully nonlinear) Bjorken flow solution. We also investigated the case of Gubser flow, where our approach was shown to also lead to meaningful results by resolving the negative temperature problem found in NS equations in this case. Further properties were also discussed together with some of the limitations and open questions surrounding this theory,
and we briefly pointed out how the general principles here employed can be used
to construct causal and stable theories beyond the conformal case, which may be later used in numerical simulations of astrophysical phenomena such as binary neutron-star mergers. 

Our work emphasizes the importance of critically analyzing 
the most basic assumptions involved in current theories of relativistic fluid dynamics. As mentioned, a key element in our
causality and stability results was the avoidance of the Landau and Eckart frames.
These seemingly harmless assumptions have been almost universally employed
for nearly 75 years, even when it is known that they are not necessary
conditions for the study of viscous hydrodynamics
\cite{Tsumura:2006hn,RezzollaZanottiBookRelHydro}. We hope our work will lead to new insights in the study of the quark-gluon plasma formed in heavy ion collisions and also in astrophysics applications where viscous fluid dynamics is dynamically coupled to Einstein's equations.

\section*{Acknowledgements} We thank G.\ S.\ Denicol, M.\ Luzum, J.\ Noronha-Hostler, and R.\ Rougemont for discussions. MMD is partially supported by 
a Sloan Research Fellowship provided by the Alfred P. Sloan foundation, 
a NSF grant DMS-1812826, and a Discovery Grant. 
JN thanks Conselho Nacional de Desenvolvimento Cient\'{\i}fico e Tecnol\'{o}gico (CNPq) and Funda\c c\~ao de Amparo \`{a} Pesquisa do Estado de S\~{a}o Paulo (FAPESP) under grant 2017/05685-2 for financial support.

\bibliography{References.bib}

\def\cprime{$'$}
\begin{thebibliography}{173}
\expandafter\ifx\csname natexlab\endcsname\relax\def\natexlab#1{#1}\fi
\expandafter\ifx\csname bibnamefont\endcsname\relax
  \def\bibnamefont#1{#1}\fi
\expandafter\ifx\csname bibfnamefont\endcsname\relax
  \def\bibfnamefont#1{#1}\fi
\expandafter\ifx\csname citenamefont\endcsname\relax
  \def\citenamefont#1{#1}\fi
\expandafter\ifx\csname url\endcsname\relax
  \def\url#1{\texttt{#1}}\fi
\expandafter\ifx\csname urlprefix\endcsname\relax\def\urlprefix{URL }\fi
\providecommand{\bibinfo}[2]{#2}
\providecommand{\eprint}[2][]{\url{#2}}

\bibitem[{\citenamefont{Heinz and Snellings}(2013)}]{Heinz:2013th}
\bibinfo{author}{\bibfnamefont{U.}~\bibnamefont{Heinz}} \bibnamefont{and}
  \bibinfo{author}{\bibfnamefont{R.}~\bibnamefont{Snellings}},
  \bibinfo{journal}{Ann. Rev. Nucl. Part. Sci.} \textbf{\bibinfo{volume}{63}},
  \bibinfo{pages}{123} (\bibinfo{year}{2013}), \eprint{1301.2826}.

\bibitem[{\citenamefont{Weinberg}(2008)}]{WeinbergCosmology}
\bibinfo{author}{\bibfnamefont{S.}~\bibnamefont{Weinberg}},
  \emph{\bibinfo{title}{Cosmology}} (\bibinfo{publisher}{Oxford University
  Press; 1 edition}, \bibinfo{year}{2008}).

\bibitem[{\citenamefont{Rezzolla and
  Zanotti}(2013)}]{RezzollaZanottiBookRelHydro}
\bibinfo{author}{\bibfnamefont{L.}~\bibnamefont{Rezzolla}} \bibnamefont{and}
  \bibinfo{author}{\bibfnamefont{O.}~\bibnamefont{Zanotti}},
  \emph{\bibinfo{title}{Relativistic Hydrodynamics}}
  (\bibinfo{publisher}{Oxford University Press}, \bibinfo{address}{New York},
  \bibinfo{year}{2013}).

\bibitem[{\citenamefont{Gyulassy and McLerran}(2005)}]{Gyulassy:2004zy}
\bibinfo{author}{\bibfnamefont{M.}~\bibnamefont{Gyulassy}} \bibnamefont{and}
  \bibinfo{author}{\bibfnamefont{L.}~\bibnamefont{McLerran}},
  \bibinfo{journal}{Nucl. Phys.} \textbf{\bibinfo{volume}{A750}},
  \bibinfo{pages}{30} (\bibinfo{year}{2005}), \eprint{nucl-th/0405013}.

\bibitem[{\citenamefont{Baiotti and Rezzolla}(2017)}]{Baiotti:2016qnr}
\bibinfo{author}{\bibfnamefont{L.}~\bibnamefont{Baiotti}} \bibnamefont{and}
  \bibinfo{author}{\bibfnamefont{L.}~\bibnamefont{Rezzolla}},
  \bibinfo{journal}{Rept. Prog. Phys.} \textbf{\bibinfo{volume}{80}},
  \bibinfo{pages}{096901} (\bibinfo{year}{2017}), \eprint{1607.03540}.

\bibitem[{\citenamefont{Baier et~al.}(2008)\citenamefont{Baier, Romatschke,
  Son, Starinets, and Stephanov}}]{Baier:2007ix}
\bibinfo{author}{\bibfnamefont{R.}~\bibnamefont{Baier}},
  \bibinfo{author}{\bibfnamefont{P.}~\bibnamefont{Romatschke}},
  \bibinfo{author}{\bibfnamefont{D.~T.} \bibnamefont{Son}},
  \bibinfo{author}{\bibfnamefont{A.~O.} \bibnamefont{Starinets}},
  \bibnamefont{and} \bibinfo{author}{\bibfnamefont{M.~A.}
  \bibnamefont{Stephanov}}, \bibinfo{journal}{JHEP}
  \textbf{\bibinfo{volume}{04}}, \bibinfo{pages}{100} (\bibinfo{year}{2008}),
  \eprint{0712.2451}.

\bibitem[{\citenamefont{Anile}(1990)}]{AnileBook}
\bibinfo{author}{\bibfnamefont{A.~M.} \bibnamefont{Anile}},
  \emph{\bibinfo{title}{Relativistic Fluids and Magneto-fluids: With
  Applications in Astrophysics and Plasma Physics (Cambridge Monographs on
  Mathematical Physics)}} (\bibinfo{publisher}{Cambridge University Press; 1
  edition}, \bibinfo{year}{1990}).

\bibitem[{\citenamefont{Four\`es-Bruhat}(1958)}]{Choquet-BruhatFluidsExistence}
\bibinfo{author}{\bibfnamefont{Y.}~\bibnamefont{Four\`es-Bruhat}},
  \bibinfo{journal}{Bull. Soc. Math. France} \textbf{\bibinfo{volume}{86}},
  \bibinfo{pages}{155} (\bibinfo{year}{1958}), ISSN \bibinfo{issn}{0037-9484},
  \urlprefix\url{http://www.numdam.org/item?id=BSMF_1958__86__155_0}.

\bibitem[{\citenamefont{Disconzi}(2015)}]{DisconziRemarksEinsteinEuler}
\bibinfo{author}{\bibfnamefont{M.~M.} \bibnamefont{Disconzi}},
  \bibinfo{journal}{Reviews in Mathematical Physics}
  \textbf{\bibinfo{volume}{27}}, \bibinfo{pages}{1550014}
  (\bibinfo{year}{2015}), \bibinfo{note}{45 pages}.

\bibitem[{\citenamefont{Choquet-Bruhat}(2009)}]{ChoquetBruhatGRBook}
\bibinfo{author}{\bibfnamefont{Y.}~\bibnamefont{Choquet-Bruhat}},
  \emph{\bibinfo{title}{General Relativity and the Einstein Equations}}
  (\bibinfo{publisher}{Oxford University Press}, \bibinfo{address}{New York},
  \bibinfo{year}{2009}).

\bibitem[{\citenamefont{Wald}(2010)}]{WaldBookGR1984}
\bibinfo{author}{\bibfnamefont{R.~M.} \bibnamefont{Wald}},
  \emph{\bibinfo{title}{General relativity}} (\bibinfo{publisher}{University of
  Chicago press}, \bibinfo{year}{2010}).

\bibitem[{\citenamefont{Geroch}(1995)}]{Geroch-RelativisticDissipative}
\bibinfo{author}{\bibfnamefont{R.}~\bibnamefont{Geroch}}, \bibinfo{journal}{J.
  Math. Phys.} \textbf{\bibinfo{volume}{36}}, \bibinfo{pages}{4226}
  (\bibinfo{year}{1995}), ISSN \bibinfo{issn}{0022-2488},
  \urlprefix\url{http://dx.doi.org/10.1063/1.530958}.

\bibitem[{\citenamefont{Herrera and Pavon}(2001)}]{Herrera:2001if}
\bibinfo{author}{\bibfnamefont{L.}~\bibnamefont{Herrera}} \bibnamefont{and}
  \bibinfo{author}{\bibfnamefont{D.}~\bibnamefont{Pavon}},
  \bibinfo{journal}{Phys. Rev.} \textbf{\bibinfo{volume}{D64}},
  \bibinfo{pages}{088503} (\bibinfo{year}{2001}), \eprint{gr-qc/0102026}.

\bibitem[{\citenamefont{Kostadt and Liu}(2001)}]{Kostadt:2001rr}
\bibinfo{author}{\bibfnamefont{P.}~\bibnamefont{Kostadt}} \bibnamefont{and}
  \bibinfo{author}{\bibfnamefont{M.}~\bibnamefont{Liu}},
  \bibinfo{journal}{Phys. Rev.} \textbf{\bibinfo{volume}{D64}},
  \bibinfo{pages}{088504} (\bibinfo{year}{2001}).

\bibitem[{\citenamefont{Kostadt and Liu}(2000)}]{Kostadt:2000ty}
\bibinfo{author}{\bibfnamefont{P.}~\bibnamefont{Kostadt}} \bibnamefont{and}
  \bibinfo{author}{\bibfnamefont{M.}~\bibnamefont{Liu}},
  \bibinfo{journal}{Phys. Rev.} \textbf{\bibinfo{volume}{D62}},
  \bibinfo{pages}{023003} (\bibinfo{year}{2000}).

\bibitem[{\citenamefont{Lindblom}(1996)}]{LindblomRelaxationEffect}
\bibinfo{author}{\bibfnamefont{L.}~\bibnamefont{Lindblom}},
  \bibinfo{journal}{Annals of Physics} \textbf{\bibinfo{volume}{247}},
  \bibinfo{pages}{1} (\bibinfo{year}{1996}), \eprint{9508058}.

\bibitem[{\citenamefont{Eckart}(1940)}]{EckartViscous}
\bibinfo{author}{\bibfnamefont{C.}~\bibnamefont{Eckart}},
  \bibinfo{journal}{Physical Review} \textbf{\bibinfo{volume}{58}},
  \bibinfo{pages}{919} (\bibinfo{year}{1940}).

\bibitem[{\citenamefont{Landau and Lifshitz}(1987)}]{LandauLifshitzFluids}
\bibinfo{author}{\bibfnamefont{L.~D.} \bibnamefont{Landau}} \bibnamefont{and}
  \bibinfo{author}{\bibfnamefont{E.~M.} \bibnamefont{Lifshitz}},
  \emph{\bibinfo{title}{Fluid Mechanics - Volume 6 (Corse of Theoretical
  Physics)}} (\bibinfo{publisher}{Pergamon Press}, \bibinfo{year}{1987}).

\bibitem[{\citenamefont{Hiscock and
  Lindblom}(1985)}]{Hiscock_Lindblom_instability_1985}
\bibinfo{author}{\bibfnamefont{W.~A.} \bibnamefont{Hiscock}} \bibnamefont{and}
  \bibinfo{author}{\bibfnamefont{L.}~\bibnamefont{Lindblom}},
  \bibinfo{journal}{Phys. Rev. D} \textbf{\bibinfo{volume}{31}},
  \bibinfo{pages}{725} (\bibinfo{year}{1985}).

\bibitem[{\citenamefont{Pichon}(1965)}]{PichonViscous}
\bibinfo{author}{\bibfnamefont{G.}~\bibnamefont{Pichon}},
  \bibinfo{journal}{Ann. Inst. H. Poincar\'e Sect. A (N.S.)}
  \textbf{\bibinfo{volume}{2}}, \bibinfo{pages}{21} (\bibinfo{year}{1965}).

\bibitem[{\citenamefont{Israel}(1976)}]{MIS-2}
\bibinfo{author}{\bibfnamefont{W.}~\bibnamefont{Israel}},
  \bibinfo{journal}{Ann. Phys.} \textbf{\bibinfo{volume}{100}},
  \bibinfo{pages}{310} (\bibinfo{year}{1976}).

\bibitem[{\citenamefont{Israel and Stewart}(1979{\natexlab{a}})}]{MIS-6}
\bibinfo{author}{\bibfnamefont{W.}~\bibnamefont{Israel}} \bibnamefont{and}
  \bibinfo{author}{\bibfnamefont{J.~M.} \bibnamefont{Stewart}},
  \bibinfo{journal}{Ann. Phys.} \textbf{\bibinfo{volume}{118}},
  \bibinfo{pages}{341} (\bibinfo{year}{1979}{\natexlab{a}}).

\bibitem[{\citenamefont{Denicol et~al.}(2012)\citenamefont{Denicol, Niemi,
  Molnar, and Rischke}}]{Denicol:2012cn}
\bibinfo{author}{\bibfnamefont{G.~S.} \bibnamefont{Denicol}},
  \bibinfo{author}{\bibfnamefont{H.}~\bibnamefont{Niemi}},
  \bibinfo{author}{\bibfnamefont{E.}~\bibnamefont{Molnar}}, \bibnamefont{and}
  \bibinfo{author}{\bibfnamefont{D.~H.} \bibnamefont{Rischke}},
  \bibinfo{journal}{Phys. Rev.} \textbf{\bibinfo{volume}{D85}},
  \bibinfo{pages}{114047} (\bibinfo{year}{2012}), \bibinfo{note}{[Erratum:
  Phys. Rev.D91,no.3,039902(2015)]}, \eprint{1202.4551}.

\bibitem[{\citenamefont{Hiscock and
  Lindblom}(1983)}]{Hiscock_Lindblom_stability_1983}
\bibinfo{author}{\bibfnamefont{W.~A.} \bibnamefont{Hiscock}} \bibnamefont{and}
  \bibinfo{author}{\bibfnamefont{L.}~\bibnamefont{Lindblom}},
  \bibinfo{journal}{Annals of Physics} \textbf{\bibinfo{volume}{151}},
  \bibinfo{pages}{466} (\bibinfo{year}{1983}).

\bibitem[{\citenamefont{Hiscock and
  Lindblom}(1988)}]{Hiscock_Lindblom_pathologies_1988}
\bibinfo{author}{\bibfnamefont{W.~A.} \bibnamefont{Hiscock}} \bibnamefont{and}
  \bibinfo{author}{\bibfnamefont{L.}~\bibnamefont{Lindblom}},
  \bibinfo{journal}{Physics Letters A} \textbf{\bibinfo{volume}{131}},
  \bibinfo{pages}{509} (\bibinfo{year}{1988}).

\bibitem[{\citenamefont{Maartens}(1995)}]{Maartens:1995wt}
\bibinfo{author}{\bibfnamefont{R.}~\bibnamefont{Maartens}},
  \bibinfo{journal}{Class. Quant. Grav.} \textbf{\bibinfo{volume}{12}},
  \bibinfo{pages}{1455} (\bibinfo{year}{1995}).

\bibitem[{\citenamefont{Jeon and Heinz}(2015)}]{Jeon:2015dfa}
\bibinfo{author}{\bibfnamefont{S.}~\bibnamefont{Jeon}} \bibnamefont{and}
  \bibinfo{author}{\bibfnamefont{U.}~\bibnamefont{Heinz}},
  \bibinfo{journal}{Int. J. Mod. Phys.} \textbf{\bibinfo{volume}{E24}},
  \bibinfo{pages}{1530010} (\bibinfo{year}{2015}), \eprint{1503.03931}.

\bibitem[{\citenamefont{Romatschke and Romatschke}(2007)}]{Romatschke:2007mq}
\bibinfo{author}{\bibfnamefont{P.}~\bibnamefont{Romatschke}} \bibnamefont{and}
  \bibinfo{author}{\bibfnamefont{U.}~\bibnamefont{Romatschke}},
  \bibinfo{journal}{Phys. Rev. Lett.} \textbf{\bibinfo{volume}{99}},
  \bibinfo{pages}{172301} (\bibinfo{year}{2007}), \eprint{0706.1522}.

\bibitem[{\citenamefont{Song and Heinz}(2008)}]{Song:2007ux}
\bibinfo{author}{\bibfnamefont{H.}~\bibnamefont{Song}} \bibnamefont{and}
  \bibinfo{author}{\bibfnamefont{U.~W.} \bibnamefont{Heinz}},
  \bibinfo{journal}{Phys. Rev.} \textbf{\bibinfo{volume}{C77}},
  \bibinfo{pages}{064901} (\bibinfo{year}{2008}), \eprint{0712.3715}.

\bibitem[{\citenamefont{Luzum and Romatschke}(2008)}]{Luzum:2008cw}
\bibinfo{author}{\bibfnamefont{M.}~\bibnamefont{Luzum}} \bibnamefont{and}
  \bibinfo{author}{\bibfnamefont{P.}~\bibnamefont{Romatschke}},
  \bibinfo{journal}{Phys. Rev.} \textbf{\bibinfo{volume}{C78}},
  \bibinfo{pages}{034915} (\bibinfo{year}{2008}), \bibinfo{note}{[Erratum:
  Phys. Rev.C79,039903(2009)]}, \eprint{0804.4015}.

\bibitem[{\citenamefont{Schenke et~al.}(2011)\citenamefont{Schenke, Jeon, and
  Gale}}]{Schenke:2010rr}
\bibinfo{author}{\bibfnamefont{B.}~\bibnamefont{Schenke}},
  \bibinfo{author}{\bibfnamefont{S.}~\bibnamefont{Jeon}}, \bibnamefont{and}
  \bibinfo{author}{\bibfnamefont{C.}~\bibnamefont{Gale}},
  \bibinfo{journal}{Phys. Rev. Lett.} \textbf{\bibinfo{volume}{106}},
  \bibinfo{pages}{042301} (\bibinfo{year}{2011}), \eprint{1009.3244}.

\bibitem[{\citenamefont{Niemi et~al.}(2011)\citenamefont{Niemi, Denicol,
  Huovinen, Molnar, and Rischke}}]{Niemi:2011ix}
\bibinfo{author}{\bibfnamefont{H.}~\bibnamefont{Niemi}},
  \bibinfo{author}{\bibfnamefont{G.~S.} \bibnamefont{Denicol}},
  \bibinfo{author}{\bibfnamefont{P.}~\bibnamefont{Huovinen}},
  \bibinfo{author}{\bibfnamefont{E.}~\bibnamefont{Molnar}}, \bibnamefont{and}
  \bibinfo{author}{\bibfnamefont{D.~H.} \bibnamefont{Rischke}},
  \bibinfo{journal}{Phys. Rev. Lett.} \textbf{\bibinfo{volume}{106}},
  \bibinfo{pages}{212302} (\bibinfo{year}{2011}), \eprint{1101.2442}.

\bibitem[{\citenamefont{Noronha-Hostler
  et~al.}(2013)\citenamefont{Noronha-Hostler, Denicol, Noronha, Andrade, and
  Grassi}}]{Noronha-Hostler:2013gga}
\bibinfo{author}{\bibfnamefont{J.}~\bibnamefont{Noronha-Hostler}},
  \bibinfo{author}{\bibfnamefont{G.~S.} \bibnamefont{Denicol}},
  \bibinfo{author}{\bibfnamefont{J.}~\bibnamefont{Noronha}},
  \bibinfo{author}{\bibfnamefont{R.~P.~G.} \bibnamefont{Andrade}},
  \bibnamefont{and} \bibinfo{author}{\bibfnamefont{F.}~\bibnamefont{Grassi}},
  \bibinfo{journal}{Phys. Rev.} \textbf{\bibinfo{volume}{C88}},
  \bibinfo{pages}{044916} (\bibinfo{year}{2013}), \eprint{1305.1981}.

\bibitem[{\citenamefont{Del~Zanna et~al.}(2013)\citenamefont{Del~Zanna,
  Chandra, Inghirami, Rolando, Beraudo, De~Pace, Pagliara, Drago, and
  Becattini}}]{DelZanna:2013eua}
\bibinfo{author}{\bibfnamefont{L.}~\bibnamefont{Del~Zanna}},
  \bibinfo{author}{\bibfnamefont{V.}~\bibnamefont{Chandra}},
  \bibinfo{author}{\bibfnamefont{G.}~\bibnamefont{Inghirami}},
  \bibinfo{author}{\bibfnamefont{V.}~\bibnamefont{Rolando}},
  \bibinfo{author}{\bibfnamefont{A.}~\bibnamefont{Beraudo}},
  \bibinfo{author}{\bibfnamefont{A.}~\bibnamefont{De~Pace}},
  \bibinfo{author}{\bibfnamefont{G.}~\bibnamefont{Pagliara}},
  \bibinfo{author}{\bibfnamefont{A.}~\bibnamefont{Drago}}, \bibnamefont{and}
  \bibinfo{author}{\bibfnamefont{F.}~\bibnamefont{Becattini}},
  \bibinfo{journal}{Eur. Phys. J.} \textbf{\bibinfo{volume}{C73}},
  \bibinfo{pages}{2524} (\bibinfo{year}{2013}), \eprint{1305.7052}.

\bibitem[{\citenamefont{Noronha-Hostler
  et~al.}(2014)\citenamefont{Noronha-Hostler, Noronha, and
  Grassi}}]{Noronha-Hostler:2014dqa}
\bibinfo{author}{\bibfnamefont{J.}~\bibnamefont{Noronha-Hostler}},
  \bibinfo{author}{\bibfnamefont{J.}~\bibnamefont{Noronha}}, \bibnamefont{and}
  \bibinfo{author}{\bibfnamefont{F.}~\bibnamefont{Grassi}},
  \bibinfo{journal}{Phys. Rev.} \textbf{\bibinfo{volume}{C90}},
  \bibinfo{pages}{034907} (\bibinfo{year}{2014}), \eprint{1406.3333}.

\bibitem[{\citenamefont{Habich et~al.}(2015)\citenamefont{Habich, Nagle, and
  Romatschke}}]{Habich:2014jna}
\bibinfo{author}{\bibfnamefont{M.}~\bibnamefont{Habich}},
  \bibinfo{author}{\bibfnamefont{J.~L.} \bibnamefont{Nagle}}, \bibnamefont{and}
  \bibinfo{author}{\bibfnamefont{P.}~\bibnamefont{Romatschke}},
  \bibinfo{journal}{Eur. Phys. J.} \textbf{\bibinfo{volume}{C75}},
  \bibinfo{pages}{15} (\bibinfo{year}{2015}), \eprint{1409.0040}.

\bibitem[{\citenamefont{Shen et~al.}(2016)\citenamefont{Shen, Qiu, Song,
  Bernhard, Bass, and Heinz}}]{Shen:2014vra}
\bibinfo{author}{\bibfnamefont{C.}~\bibnamefont{Shen}},
  \bibinfo{author}{\bibfnamefont{Z.}~\bibnamefont{Qiu}},
  \bibinfo{author}{\bibfnamefont{H.}~\bibnamefont{Song}},
  \bibinfo{author}{\bibfnamefont{J.}~\bibnamefont{Bernhard}},
  \bibinfo{author}{\bibfnamefont{S.}~\bibnamefont{Bass}}, \bibnamefont{and}
  \bibinfo{author}{\bibfnamefont{U.}~\bibnamefont{Heinz}},
  \bibinfo{journal}{Comput. Phys. Commun.} \textbf{\bibinfo{volume}{199}},
  \bibinfo{pages}{61} (\bibinfo{year}{2016}), \eprint{1409.8164}.

\bibitem[{\citenamefont{Romatschke}(2015)}]{Romatschke:2015gxa}
\bibinfo{author}{\bibfnamefont{P.}~\bibnamefont{Romatschke}},
  \bibinfo{journal}{Eur. Phys. J.} \textbf{\bibinfo{volume}{C75}},
  \bibinfo{pages}{305} (\bibinfo{year}{2015}), \eprint{1502.04745}.

\bibitem[{\citenamefont{Ryu et~al.}(2015)\citenamefont{Ryu, Paquet, Shen,
  Denicol, Schenke, Jeon, and Gale}}]{Ryu:2015vwa}
\bibinfo{author}{\bibfnamefont{S.}~\bibnamefont{Ryu}},
  \bibinfo{author}{\bibfnamefont{J.~F.} \bibnamefont{Paquet}},
  \bibinfo{author}{\bibfnamefont{C.}~\bibnamefont{Shen}},
  \bibinfo{author}{\bibfnamefont{G.~S.} \bibnamefont{Denicol}},
  \bibinfo{author}{\bibfnamefont{B.}~\bibnamefont{Schenke}},
  \bibinfo{author}{\bibfnamefont{S.}~\bibnamefont{Jeon}}, \bibnamefont{and}
  \bibinfo{author}{\bibfnamefont{C.}~\bibnamefont{Gale}},
  \bibinfo{journal}{Phys. Rev. Lett.} \textbf{\bibinfo{volume}{115}},
  \bibinfo{pages}{132301} (\bibinfo{year}{2015}), \eprint{1502.01675}.

\bibitem[{\citenamefont{Niemi et~al.}(2016)\citenamefont{Niemi, Eskola, and
  Paatelainen}}]{Niemi:2015qia}
\bibinfo{author}{\bibfnamefont{H.}~\bibnamefont{Niemi}},
  \bibinfo{author}{\bibfnamefont{K.~J.} \bibnamefont{Eskola}},
  \bibnamefont{and}
  \bibinfo{author}{\bibfnamefont{R.}~\bibnamefont{Paatelainen}},
  \bibinfo{journal}{Phys. Rev.} \textbf{\bibinfo{volume}{C93}},
  \bibinfo{pages}{024907} (\bibinfo{year}{2016}), \eprint{1505.02677}.

\bibitem[{\citenamefont{Bazow et~al.}(2016{\natexlab{a}})\citenamefont{Bazow,
  Heinz, and Strickland}}]{Bazow:2016yra}
\bibinfo{author}{\bibfnamefont{D.}~\bibnamefont{Bazow}},
  \bibinfo{author}{\bibfnamefont{U.~W.} \bibnamefont{Heinz}}, \bibnamefont{and}
  \bibinfo{author}{\bibfnamefont{M.}~\bibnamefont{Strickland}}
  (\bibinfo{year}{2016}{\natexlab{a}}), \eprint{1608.06577}.

\bibitem[{\citenamefont{Okamoto and Nonaka}(2017)}]{Okamoto:2017ukz}
\bibinfo{author}{\bibfnamefont{K.}~\bibnamefont{Okamoto}} \bibnamefont{and}
  \bibinfo{author}{\bibfnamefont{C.}~\bibnamefont{Nonaka}},
  \bibinfo{journal}{Eur. Phys. J.} \textbf{\bibinfo{volume}{C77}},
  \bibinfo{pages}{383} (\bibinfo{year}{2017}), \eprint{1703.01473}.

\bibitem[{\citenamefont{Duez et~al.}(2004)\citenamefont{Duez, Liu, Shapiro, and
  Stephens}}]{DuezetallEinsteinNavierStokes}
\bibinfo{author}{\bibfnamefont{M.~D.} \bibnamefont{Duez}},
  \bibinfo{author}{\bibfnamefont{Y.~T.} \bibnamefont{Liu}},
  \bibinfo{author}{\bibfnamefont{S.~L.} \bibnamefont{Shapiro}},
  \bibnamefont{and} \bibinfo{author}{\bibfnamefont{B.~C.}
  \bibnamefont{Stephens}}, \bibinfo{journal}{Physical Review D}
  \textbf{\bibinfo{volume}{69}}, \bibinfo{pages}{104030}
  (\bibinfo{year}{2004}).

\bibitem[{\citenamefont{Herrera et~al.}(2014)\citenamefont{Herrera, Prisco,
  Ib\'a{\~n}ez, and Ospino}}]{Herr_axially}
\bibinfo{author}{\bibfnamefont{L.}~\bibnamefont{Herrera}},
  \bibinfo{author}{\bibfnamefont{A.~D.} \bibnamefont{Prisco}},
  \bibinfo{author}{\bibfnamefont{J.}~\bibnamefont{Ib\'a{\~n}ez}},
  \bibnamefont{and} \bibinfo{author}{\bibfnamefont{J.}~\bibnamefont{Ospino}},
  \bibinfo{journal}{Physical Review D} \textbf{\bibinfo{volume}{89}},
  \bibinfo{pages}{084034} (\bibinfo{year}{2014}).

\bibitem[{\citenamefont{Herrera et~al.}(2009)\citenamefont{Herrera, Prisco,
  Fuenmayor, and Troconis}}]{HerreraFullCausal}
\bibinfo{author}{\bibfnamefont{L.}~\bibnamefont{Herrera}},
  \bibinfo{author}{\bibfnamefont{A.~D.} \bibnamefont{Prisco}},
  \bibinfo{author}{\bibfnamefont{E.}~\bibnamefont{Fuenmayor}},
  \bibnamefont{and} \bibinfo{author}{\bibfnamefont{O.}~\bibnamefont{Troconis}},
  \bibinfo{journal}{Int. J. Mod. Phys. D} \textbf{\bibinfo{volume}{18}},
  \bibinfo{pages}{129} (\bibinfo{year}{2009}), \eprint{0804.3584}.

\bibitem[{\citenamefont{Alford et~al.}(2017)\citenamefont{Alford, Bovard,
  Hanauske, Rezzolla, and Schwenzer}}]{Alford:2017rxf}
\bibinfo{author}{\bibfnamefont{M.~G.} \bibnamefont{Alford}},
  \bibinfo{author}{\bibfnamefont{L.}~\bibnamefont{Bovard}},
  \bibinfo{author}{\bibfnamefont{M.}~\bibnamefont{Hanauske}},
  \bibinfo{author}{\bibfnamefont{L.}~\bibnamefont{Rezzolla}}, \bibnamefont{and}
  \bibinfo{author}{\bibfnamefont{K.}~\bibnamefont{Schwenzer}}
  (\bibinfo{year}{2017}), \eprint{1707.09475}.

\bibitem[{\citenamefont{Bancel}(1973)}]{Bancel}
\bibinfo{author}{\bibfnamefont{D.}~\bibnamefont{Bancel}},
  \bibinfo{journal}{Ann. Inst. H. Poincare Sect. A (N.S.)}
  \textbf{\bibinfo{volume}{18}}, \bibinfo{pages}{263} (\bibinfo{year}{1973}).

\bibitem[{\citenamefont{Bancel and Choquet-Bruhat}(1973)}]{BancelBruhat}
\bibinfo{author}{\bibfnamefont{D.}~\bibnamefont{Bancel}} \bibnamefont{and}
  \bibinfo{author}{\bibfnamefont{Y.}~\bibnamefont{Choquet-Bruhat}},
  \bibinfo{journal}{Commun. Math. Phys.} \textbf{\bibinfo{volume}{33}},
  \bibinfo{pages}{83} (\bibinfo{year}{1973}).

\bibitem[{\citenamefont{Groot}(1980)}]{degroot}
\bibinfo{author}{\bibfnamefont{S.~R.~D.} \bibnamefont{Groot}},
  \emph{\bibinfo{title}{Relativistic Kinetic Theory. Principles and
  Applications}} (\bibinfo{publisher}{Amsterdam, Netherlands: North-holland (
  1980) 417p}, \bibinfo{year}{1980}).

\bibitem[{\citenamefont{Freist\"uhler and Temple}(2014)}]{TempleViscous}
\bibinfo{author}{\bibfnamefont{H.}~\bibnamefont{Freist\"uhler}}
  \bibnamefont{and} \bibinfo{author}{\bibfnamefont{B.}~\bibnamefont{Temple}},
  \bibinfo{journal}{Proc. R. Soc. A} \textbf{\bibinfo{volume}{470}},
  \bibinfo{pages}{20140055} (\bibinfo{year}{2014}).

\bibitem[{\citenamefont{Freist\"uhler and Temple}(2017)}]{TempleViscous2}
\bibinfo{author}{\bibfnamefont{H.}~\bibnamefont{Freist\"uhler}}
  \bibnamefont{and} \bibinfo{author}{\bibfnamefont{B.}~\bibnamefont{Temple}},
  \bibinfo{journal}{Proc. R. Soc. A} \textbf{\bibinfo{volume}{473}},
  \bibinfo{pages}{20160729} (\bibinfo{year}{2017}).

\bibitem[{\citenamefont{Freist\"uhler and Temple}(2018)}]{TempleViscous3}
\bibinfo{author}{\bibfnamefont{H.}~\bibnamefont{Freist\"uhler}}
  \bibnamefont{and} \bibinfo{author}{\bibfnamefont{B.}~\bibnamefont{Temple}},
  \bibinfo{journal}{Journal of Mathematical Physics}
  \textbf{\bibinfo{volume}{59}}, \bibinfo{pages}{063101}
  (\bibinfo{year}{2018}).

\bibitem[{\citenamefont{Lichnerowicz}(1955)}]{LichnerowiczBookGR}
\bibinfo{author}{\bibfnamefont{A.}~\bibnamefont{Lichnerowicz}},
  \emph{\bibinfo{title}{Th\'eories Relativistes de la Gravitation et de
  l'\'Electromagn\'etism}} (\bibinfo{publisher}{Masson et Cie},
  \bibinfo{address}{Paris}, \bibinfo{year}{1955}).

\bibitem[{\citenamefont{Disconzi}(2014)}]{DisconziViscousFluidsNonlinearity}
\bibinfo{author}{\bibfnamefont{M.~M.} \bibnamefont{Disconzi}},
  \bibinfo{journal}{Nonlinearity} \textbf{\bibinfo{volume}{27}},
  \bibinfo{pages}{1915} (\bibinfo{year}{2014}).

\bibitem[{\citenamefont{Disconzi et~al.}(2017)\citenamefont{Disconzi, Kephart,
  and Scherrer}}]{DisconziKephartScherrerNew}
\bibinfo{author}{\bibfnamefont{M.~M.} \bibnamefont{Disconzi}},
  \bibinfo{author}{\bibfnamefont{T.~W.} \bibnamefont{Kephart}},
  \bibnamefont{and} \bibinfo{author}{\bibfnamefont{R.~J.}
  \bibnamefont{Scherrer}}, \bibinfo{journal}{International Journal of Modern
  Physics D} \textbf{\bibinfo{volume}{26}}, \bibinfo{pages}{1750146 (52 pages)}
  (\bibinfo{year}{2017}).

\bibitem[{\citenamefont{Czubak and Disconzi}(2016)}]{DisconziCzubakNonzero}
\bibinfo{author}{\bibfnamefont{M.}~\bibnamefont{Czubak}} \bibnamefont{and}
  \bibinfo{author}{\bibfnamefont{M.~M.} \bibnamefont{Disconzi}},
  \bibinfo{journal}{Journal of Mathematical Physics}
  \textbf{\bibinfo{volume}{57}}, \bibinfo{pages}{042501}
  (\bibinfo{year}{2016}), \bibinfo{note}{21 pages}.

\bibitem[{\citenamefont{Disconzi et~al.}(2015)\citenamefont{Disconzi, Kephart,
  and Scherrer}}]{Disconzi_Kephart_Scherrer_2015}
\bibinfo{author}{\bibfnamefont{M.~M.} \bibnamefont{Disconzi}},
  \bibinfo{author}{\bibfnamefont{T.~W.} \bibnamefont{Kephart}},
  \bibnamefont{and} \bibinfo{author}{\bibfnamefont{R.~J.}
  \bibnamefont{Scherrer}}, \bibinfo{journal}{Physical Review D}
  \textbf{\bibinfo{volume}{91}}, \bibinfo{pages}{043532 (6 pages)}
  (\bibinfo{year}{2015}).

\bibitem[{\citenamefont{Montani and
  Venanzi}(2017)}]{MontaniLichnerowiczViscosityBianchi}
\bibinfo{author}{\bibfnamefont{G.}~\bibnamefont{Montani}} \bibnamefont{and}
  \bibinfo{author}{\bibfnamefont{M.}~\bibnamefont{Venanzi}},
  \bibinfo{journal}{The European Physical Journal C}
  \textbf{\bibinfo{volume}{77}} (\bibinfo{year}{2017}).

\bibitem[{\citenamefont{Geroch and
  Lindblom}(1990)}]{GerochLindblomDivergenceType}
\bibinfo{author}{\bibfnamefont{R.}~\bibnamefont{Geroch}} \bibnamefont{and}
  \bibinfo{author}{\bibfnamefont{L.}~\bibnamefont{Lindblom}},
  \bibinfo{journal}{Phys. Rev. D (3)} \textbf{\bibinfo{volume}{41}},
  \bibinfo{pages}{1855} (\bibinfo{year}{1990}), ISSN \bibinfo{issn}{0556-2821},
  \urlprefix\url{http://dx.doi.org/10.1103/PhysRevD.41.1855}.

\bibitem[{\citenamefont{Geroch and Lindblom}(1991)}]{GerochLindblomCausal}
\bibinfo{author}{\bibfnamefont{R.}~\bibnamefont{Geroch}} \bibnamefont{and}
  \bibinfo{author}{\bibfnamefont{L.}~\bibnamefont{Lindblom}},
  \bibinfo{journal}{Ann. Physics} \textbf{\bibinfo{volume}{207}},
  \bibinfo{pages}{394} (\bibinfo{year}{1991}), ISSN \bibinfo{issn}{0003-4916},
  \urlprefix\url{http://dx.doi.org/10.1016/0003-4916(91)90063-E}.

\bibitem[{\citenamefont{Liu et~al.}(1986)\citenamefont{Liu, M{\"u}ller, and
  Ruggeri}}]{LiuMullerRuggeri-RelThermoGases}
\bibinfo{author}{\bibfnamefont{I.-S.} \bibnamefont{Liu}},
  \bibinfo{author}{\bibfnamefont{I.}~\bibnamefont{M{\"u}ller}},
  \bibnamefont{and} \bibinfo{author}{\bibfnamefont{T.}~\bibnamefont{Ruggeri}},
  \bibinfo{journal}{Ann. Physics} \textbf{\bibinfo{volume}{169}},
  \bibinfo{pages}{191} (\bibinfo{year}{1986}), ISSN \bibinfo{issn}{0003-4916},
  \urlprefix\url{http://dx.doi.org/10.1016/0003-4916(86)90164-8}.

\bibitem[{\citenamefont{Mueller and Ruggeri}(1998)}]{MuellerRuggeriBook}
\bibinfo{author}{\bibfnamefont{I.}~\bibnamefont{Mueller}} \bibnamefont{and}
  \bibinfo{author}{\bibfnamefont{T.}~\bibnamefont{Ruggeri}},
  \emph{\bibinfo{title}{Rational Extended Thermodynamics}}
  (\bibinfo{publisher}{Springer}, \bibinfo{year}{1998}).

\bibitem[{\citenamefont{Calzetta and Peralta-Ramos}(2009)}]{RamosCalzettaDT1}
\bibinfo{author}{\bibfnamefont{E.}~\bibnamefont{Calzetta}} \bibnamefont{and}
  \bibinfo{author}{\bibfnamefont{J.}~\bibnamefont{Peralta-Ramos}},
  \bibinfo{journal}{Phys. Rev. D} \textbf{\bibinfo{volume}{80}},
  \bibinfo{pages}{126002} (\bibinfo{year}{2009}), \eprint{0908.2646}.

\bibitem[{\citenamefont{Calzetta and Peralta-Ramos}(2010)}]{RamosCalzettaDT2}
\bibinfo{author}{\bibfnamefont{E.}~\bibnamefont{Calzetta}} \bibnamefont{and}
  \bibinfo{author}{\bibfnamefont{J.}~\bibnamefont{Peralta-Ramos}},
  \bibinfo{journal}{Phys. Rev. C} \textbf{\bibinfo{volume}{82}},
  \bibinfo{pages}{054905} (\bibinfo{year}{2010}), \eprint{1003.1091}.

\bibitem[{\citenamefont{Lehner et~al.}(2018)\citenamefont{Lehner, Reula, and
  Rubio}}]{Lehner:2017yes}
\bibinfo{author}{\bibfnamefont{L.}~\bibnamefont{Lehner}},
  \bibinfo{author}{\bibfnamefont{O.~A.} \bibnamefont{Reula}}, \bibnamefont{and}
  \bibinfo{author}{\bibfnamefont{M.~E.} \bibnamefont{Rubio}},
  \bibinfo{journal}{Phys. Rev.} \textbf{\bibinfo{volume}{D97}},
  \bibinfo{pages}{024013} (\bibinfo{year}{2018}), \eprint{1710.08033}.

\bibitem[{\citenamefont{Nagy et~al.}(1994)\citenamefont{Nagy, Ortiz, and
  Reula}}]{Nagy_et_all-Hyperbolic_parabolic_limit}
\bibinfo{author}{\bibfnamefont{G.~B.} \bibnamefont{Nagy}},
  \bibinfo{author}{\bibfnamefont{O.~E.} \bibnamefont{Ortiz}}, \bibnamefont{and}
  \bibinfo{author}{\bibfnamefont{O.~A.} \bibnamefont{Reula}},
  \bibinfo{journal}{J. Math. Phys.} \textbf{\bibinfo{volume}{35}},
  \bibinfo{pages}{4334} (\bibinfo{year}{1994}), ISSN \bibinfo{issn}{0022-2488},
  \urlprefix\url{http://dx.doi.org/10.1063/1.530856}.

\bibitem[{\citenamefont{Kreiss et~al.}(1997)\citenamefont{Kreiss, Nagy, Ortiz,
  and Reula}}]{Kreiss_et_al}
\bibinfo{author}{\bibfnamefont{H.-O.} \bibnamefont{Kreiss}},
  \bibinfo{author}{\bibfnamefont{G.~B.} \bibnamefont{Nagy}},
  \bibinfo{author}{\bibfnamefont{O.~E.} \bibnamefont{Ortiz}}, \bibnamefont{and}
  \bibinfo{author}{\bibfnamefont{O.~A.} \bibnamefont{Reula}},
  \bibinfo{journal}{Journal of Mathematical Physics}
  \textbf{\bibinfo{volume}{38}}, \bibinfo{pages}{5272} (\bibinfo{year}{1997}).

\bibitem[{\citenamefont{Reula and Nagy}(1997)}]{Reula_et_al-CausalStatistical}
\bibinfo{author}{\bibfnamefont{O.~A.} \bibnamefont{Reula}} \bibnamefont{and}
  \bibinfo{author}{\bibfnamefont{G.~B.} \bibnamefont{Nagy}},
  \bibinfo{journal}{J. Phys. A} \textbf{\bibinfo{volume}{30}},
  \bibinfo{pages}{1695} (\bibinfo{year}{1997}), ISSN \bibinfo{issn}{0305-4470},
  \urlprefix\url{http://dx.doi.org/10.1088/0305-4470/30/5/030}.

\bibitem[{\citenamefont{V{\'a}n and Bir{\'o}}(2012)}]{VanFirstOrder}
\bibinfo{author}{\bibfnamefont{P.}~\bibnamefont{V{\'a}n}} \bibnamefont{and}
  \bibinfo{author}{\bibfnamefont{T.~S.} \bibnamefont{Bir{\'o}}},
  \bibinfo{journal}{Physics Letters B (2012)} \textbf{\bibinfo{volume}{709}},
  \bibinfo{pages}{106} (\bibinfo{year}{2012}).

\bibitem[{\citenamefont{Tsumura and Kunihiro}(2011)}]{Tsumura:2011cj}
\bibinfo{author}{\bibfnamefont{K.}~\bibnamefont{Tsumura}} \bibnamefont{and}
  \bibinfo{author}{\bibfnamefont{T.}~\bibnamefont{Kunihiro}},
  \bibinfo{journal}{Prog. Theor. Phys.} \textbf{\bibinfo{volume}{126}},
  \bibinfo{pages}{761} (\bibinfo{year}{2011}), \eprint{1108.1519}.

\bibitem[{\citenamefont{Florkowski and Ryblewski}(2011)}]{Florkowski:2010cf}
\bibinfo{author}{\bibfnamefont{W.}~\bibnamefont{Florkowski}} \bibnamefont{and}
  \bibinfo{author}{\bibfnamefont{P.}~\bibnamefont{Ryblewski}},
  \bibinfo{journal}{Phys. Rev.} \textbf{\bibinfo{volume}{C83}},
  \bibinfo{pages}{034907} (\bibinfo{year}{2011}), \eprint{1007.0130}.

\bibitem[{\citenamefont{Martinez and Strickland}(2010)}]{Martinez:2010sc}
\bibinfo{author}{\bibfnamefont{M.}~\bibnamefont{Martinez}} \bibnamefont{and}
  \bibinfo{author}{\bibfnamefont{M.}~\bibnamefont{Strickland}},
  \bibinfo{journal}{Nucl. Phys.} \textbf{\bibinfo{volume}{A848}},
  \bibinfo{pages}{183} (\bibinfo{year}{2010}), \eprint{1007.0889}.

\bibitem[{\citenamefont{Strickland}(2014)}]{Strickland:2014pga}
\bibinfo{author}{\bibfnamefont{M.}~\bibnamefont{Strickland}},
  \bibinfo{journal}{Acta Phys. Polon.} \textbf{\bibinfo{volume}{B45}},
  \bibinfo{pages}{2355} (\bibinfo{year}{2014}), \eprint{1410.5786}.

\bibitem[{\citenamefont{Klainerman and Nicolo}(2003)}]{KlainermanNicoloBook}
\bibinfo{author}{\bibfnamefont{S.}~\bibnamefont{Klainerman}} \bibnamefont{and}
  \bibinfo{author}{\bibfnamefont{F.}~\bibnamefont{Nicolo}},
  \emph{\bibinfo{title}{The Evolution Problem in General Relativity (Progress
  in Mathematical Physics)}} (\bibinfo{publisher}{Birkh\"auser},
  \bibinfo{year}{2003}).

\bibitem[{\citenamefont{ek and Yoneda}(2016)}]{MisiolekYonedaIllPosednessEuler}
\bibinfo{author}{\bibfnamefont{G.~M.} \bibnamefont{ek}} \bibnamefont{and}
  \bibinfo{author}{\bibfnamefont{T.}~\bibnamefont{Yoneda}},
  \bibinfo{journal}{Math. Ann.} \textbf{\bibinfo{volume}{364}},
  \bibinfo{pages}{243} (\bibinfo{year}{2016}), ISSN \bibinfo{issn}{0025-5831}.

\bibitem[{\citenamefont{ek and
  T.~Yoneda}(2012)}]{MisiolekYonedaIllPosednessQuasiGeostrophic}
\bibinfo{author}{\bibfnamefont{G.~M.} \bibnamefont{ek}} \bibnamefont{and}
  \bibinfo{author}{\bibfnamefont{T.}~\bibnamefont{T.~Yoneda}}, in
  \emph{\bibinfo{booktitle}{Analysis, geometry and quantum field theory}}
  (\bibinfo{publisher}{Amer. Math. Soc., Providence, RI},
  \bibinfo{year}{2012}), vol. \bibinfo{volume}{584} of
  \emph{\bibinfo{series}{Contemp. Math.}}, pp. \bibinfo{pages}{251--258},
  \urlprefix\url{https://doi.org/10.1090/conm/584/11589}.

\bibitem[{\citenamefont{Mizohata}(1985)}]{MizohataCauchyProblem}
\bibinfo{author}{\bibfnamefont{S.}~\bibnamefont{Mizohata}},
  \emph{\bibinfo{title}{On the {C}auchy problem}} (\bibinfo{publisher}{Science
  Press and Academic Press, Inc.}, \bibinfo{address}{Hong Kong},
  \bibinfo{year}{1985}).

\bibitem[{\citenamefont{Chan and Czubak}(2013)}]{CC10}
\bibinfo{author}{\bibfnamefont{C.-H.} \bibnamefont{Chan}} \bibnamefont{and}
  \bibinfo{author}{\bibfnamefont{M.}~\bibnamefont{Czubak}},
  \bibinfo{journal}{Dyn. Partial Differ. Equ.} \textbf{\bibinfo{volume}{10}},
  \bibinfo{pages}{43} (\bibinfo{year}{2013}), ISSN \bibinfo{issn}{1548-159X}.

\bibitem[{\citenamefont{Chan and Czubak}(2016)}]{CzubakChanRemarks}
\bibinfo{author}{\bibfnamefont{C.~H.} \bibnamefont{Chan}} \bibnamefont{and}
  \bibinfo{author}{\bibfnamefont{M.}~\bibnamefont{Czubak}},
  \bibinfo{journal}{Ann. Inst. H. Poincar\'e Anal. Non Lin\'eaire}
  \textbf{\bibinfo{volume}{33}}, \bibinfo{pages}{655} (\bibinfo{year}{2016}),
  ISSN \bibinfo{issn}{0294-1449},
  \urlprefix\url{https://doi.org/10.1016/j.anihpc.2015.01.002}.

\bibitem[{\citenamefont{Taylor}(1997)}]{Taylor3}
\bibinfo{author}{\bibfnamefont{M.~E.} \bibnamefont{Taylor}},
  \emph{\bibinfo{title}{Partial differential equations. {III}}}, vol.
  \bibinfo{volume}{117} of \emph{\bibinfo{series}{Applied Mathematical
  Sciences}} (\bibinfo{publisher}{Springer-Verlag}, \bibinfo{address}{New
  York}, \bibinfo{year}{1997}), ISBN \bibinfo{isbn}{0-387-94652-7},
  \bibinfo{note}{nonlinear equations, Corrected reprint of the 1996 original}.

\bibitem[{\citenamefont{Bourgain and Li}(2015)}]{BourgainLiIllPosedEuler}
\bibinfo{author}{\bibfnamefont{J.}~\bibnamefont{Bourgain}} \bibnamefont{and}
  \bibinfo{author}{\bibfnamefont{D.}~\bibnamefont{Li}},
  \bibinfo{journal}{Invent. Math.} \textbf{\bibinfo{volume}{201}},
  \bibinfo{pages}{97} (\bibinfo{year}{2015}), ISSN \bibinfo{issn}{0020-9910}.

\bibitem[{\citenamefont{Christodoulou and
  Klainerman}(1993)}]{ChristodouKlainermanStability}
\bibinfo{author}{\bibfnamefont{D.}~\bibnamefont{Christodoulou}}
  \bibnamefont{and}
  \bibinfo{author}{\bibfnamefont{S.}~\bibnamefont{Klainerman}},
  \emph{\bibinfo{title}{The global nonlinear stability of {M}inkowski space}}
  (\bibinfo{publisher}{Princeton University Press},
  \bibinfo{address}{Princeton, N.J.}, \bibinfo{year}{1993}).

\bibitem[{\citenamefont{Lindblad and
  Rodnianski}(2010)}]{LindbladRodnianskiStabilityMinkowski}
\bibinfo{author}{\bibfnamefont{H.}~\bibnamefont{Lindblad}} \bibnamefont{and}
  \bibinfo{author}{\bibfnamefont{I.}~\bibnamefont{Rodnianski}},
  \bibinfo{journal}{Ann. of Math. (2)} \textbf{\bibinfo{volume}{171}},
  \bibinfo{pages}{1401} (\bibinfo{year}{2010}), ISSN \bibinfo{issn}{0003-486X},
  \urlprefix\url{https://doi.org/10.4007/annals.2010.171.1401}.

\bibitem[{\citenamefont{Rodnianski and Speck}(2013)}]{MR3120746}
\bibinfo{author}{\bibfnamefont{I.}~\bibnamefont{Rodnianski}} \bibnamefont{and}
  \bibinfo{author}{\bibfnamefont{J.}~\bibnamefont{Speck}}, \bibinfo{journal}{J.
  Eur. Math. Soc. (JEMS)} \textbf{\bibinfo{volume}{15}}, \bibinfo{pages}{2369}
  (\bibinfo{year}{2013}), ISSN \bibinfo{issn}{1435-9855},
  \urlprefix\url{http://dx.doi.org/10.4171/JEMS/424}.

\bibitem[{\citenamefont{Speck}(2012)}]{SpeckNonlinearStability}
\bibinfo{author}{\bibfnamefont{J.}~\bibnamefont{Speck}},
  \bibinfo{journal}{Selecta Mathematica} \textbf{\bibinfo{volume}{18}},
  \bibinfo{pages}{633} (\bibinfo{year}{2012}).

\bibitem[{\citenamefont{Speck}(2013)}]{MR3101792}
\bibinfo{author}{\bibfnamefont{J.}~\bibnamefont{Speck}},
  \bibinfo{journal}{Arch. Ration. Mech. Anal.} \textbf{\bibinfo{volume}{210}},
  \bibinfo{pages}{535} (\bibinfo{year}{2013}), ISSN \bibinfo{issn}{0003-9527},
  \urlprefix\url{http://dx.doi.org/10.1007/s00205-013-0655-3}.

\bibitem[{\citenamefont{Hawking and Ellis}(1975)}]{HawkingEllisBook}
\bibinfo{author}{\bibfnamefont{S.~W.} \bibnamefont{Hawking}} \bibnamefont{and}
  \bibinfo{author}{\bibfnamefont{G.~F.~R.} \bibnamefont{Ellis}},
  \emph{\bibinfo{title}{The Large Scale Structure of Space-Time (Cambridge
  Monographs on Mathematical Physics)}} (\bibinfo{publisher}{Cambridge
  University Press}, \bibinfo{year}{1975}).

\bibitem[{\citenamefont{Christodoulou}(2007)}]{ChristodoulouShocks}
\bibinfo{author}{\bibfnamefont{D.}~\bibnamefont{Christodoulou}},
  \emph{\bibinfo{title}{The formation of shocks in 3-dimensional fluids}}, EMS
  Monographs in Mathematics (\bibinfo{publisher}{European Mathematical Society
  (EMS), Z\"urich}, \bibinfo{year}{2007}), ISBN
  \bibinfo{isbn}{978-3-03719-031-9},
  \urlprefix\url{https://doi.org/10.4171/031}.

\bibitem[{\citenamefont{Christodoulou and
  Miao}(2014)}]{ChristodoulouMiaoShocks}
\bibinfo{author}{\bibfnamefont{D.}~\bibnamefont{Christodoulou}}
  \bibnamefont{and} \bibinfo{author}{\bibfnamefont{S.}~\bibnamefont{Miao}},
  \emph{\bibinfo{title}{Compressible flow and {E}uler's equations}},
  vol.~\bibinfo{volume}{9} of \emph{\bibinfo{series}{Surveys of Modern
  Mathematics}} (\bibinfo{publisher}{International Press, Somerville, MA;
  Higher Education Press, Beijing}, \bibinfo{year}{2014}), ISBN
  \bibinfo{isbn}{978-1-57146-297-8}.

\bibitem[{\citenamefont{Speck}(2016)}]{SpeckBook}
\bibinfo{author}{\bibfnamefont{J.}~\bibnamefont{Speck}},
  \textbf{\bibinfo{volume}{214}}, \bibinfo{pages}{xxiii+515}
  (\bibinfo{year}{2016}).

\bibitem[{\citenamefont{Arnold et~al.}(2014)\citenamefont{Arnold, Romatschke,
  and van~der Schee}}]{Arnold:2014jva}
\bibinfo{author}{\bibfnamefont{P.}~\bibnamefont{Arnold}},
  \bibinfo{author}{\bibfnamefont{P.}~\bibnamefont{Romatschke}},
  \bibnamefont{and} \bibinfo{author}{\bibfnamefont{W.}~\bibnamefont{van~der
  Schee}}, \bibinfo{journal}{JHEP} \textbf{\bibinfo{volume}{10}},
  \bibinfo{pages}{110} (\bibinfo{year}{2014}), \eprint{1408.2518}.

\bibitem[{\citenamefont{Kovtun}(2012)}]{Kovtun:2012rj}
\bibinfo{author}{\bibfnamefont{K.}~\bibnamefont{Kovtun}}, \bibinfo{journal}{J.
  Phys.} \textbf{\bibinfo{volume}{A45}}, \bibinfo{pages}{473001}
  (\bibinfo{year}{2012}), \eprint{1205.5040}.

\bibitem[{\citenamefont{Stewart}(1972)}]{Stewart:1972hg}
\bibinfo{author}{\bibfnamefont{J.~M.} \bibnamefont{Stewart}}
  (\bibinfo{year}{1972}).

\bibitem[{\citenamefont{Tsumura et~al.}(2007)\citenamefont{Tsumura, Kunihiro,
  and Ohnishi}}]{Tsumura:2006hn}
\bibinfo{author}{\bibfnamefont{K.}~\bibnamefont{Tsumura}},
  \bibinfo{author}{\bibfnamefont{T.}~\bibnamefont{Kunihiro}}, \bibnamefont{and}
  \bibinfo{author}{\bibfnamefont{K.}~\bibnamefont{Ohnishi}},
  \bibinfo{journal}{Phys. Lett.} \textbf{\bibinfo{volume}{B656}},
  \bibinfo{pages}{274} (\bibinfo{year}{2007}), \bibinfo{note}{[Phys.
  Lett.B646,134(2007)]}, \eprint{hep-ph/0609056}.

\bibitem[{\citenamefont{Monnai}(2018)}]{Monnai:2018rgs}
\bibinfo{author}{\bibfnamefont{A.}~\bibnamefont{Monnai}}
  (\bibinfo{year}{2018}), \eprint{1803.03318}.

\bibitem[{\citenamefont{Bhattacharyyai
  et~al.}(2008)\citenamefont{Bhattacharyyai, Hubeny, Minwalla, and
  Rangamani}}]{Bhattacharyya:2008jc}
\bibinfo{author}{\bibfnamefont{S.}~\bibnamefont{Bhattacharyyai}},
  \bibinfo{author}{\bibfnamefont{V.~E.} \bibnamefont{Hubeny}},
  \bibinfo{author}{\bibfnamefont{S.}~\bibnamefont{Minwalla}}, \bibnamefont{and}
  \bibinfo{author}{\bibfnamefont{M.}~\bibnamefont{Rangamani}},
  \bibinfo{journal}{JHEP} \textbf{\bibinfo{volume}{02}}, \bibinfo{pages}{045}
  (\bibinfo{year}{2008}), \eprint{0712.2456}.

\bibitem[{\citenamefont{Borsanyi et~al.}(2016)}]{Borsanyi:2016ksw}
\bibinfo{author}{\bibfnamefont{S.}~\bibnamefont{Borsanyi}}
  \bibnamefont{et~al.}, \bibinfo{journal}{Nature}
  \textbf{\bibinfo{volume}{539}}, \bibinfo{pages}{69} (\bibinfo{year}{2016}),
  \eprint{1606.07494}.

\bibitem[{\citenamefont{Jou et~al.}(2009)\citenamefont{Jou, Casas-Vazsquez, and
  Lebon}}]{JouetallBook}
\bibinfo{author}{\bibfnamefont{D.}~\bibnamefont{Jou}},
  \bibinfo{author}{\bibfnamefont{J.}~\bibnamefont{Casas-Vazsquez}},
  \bibnamefont{and} \bibinfo{author}{\bibfnamefont{G.}~\bibnamefont{Lebon}},
  \emph{\bibinfo{title}{Extended Irreversible Thermodynamics}}
  (\bibinfo{publisher}{Springer}, \bibinfo{year}{2009}).

\bibitem[{\citenamefont{Grozdanov and
  Kaplis}(2016)}]{GrozdanovKaplisThirdOrder}
\bibinfo{author}{\bibfnamefont{S.}~\bibnamefont{Grozdanov}} \bibnamefont{and}
  \bibinfo{author}{\bibfnamefont{N.}~\bibnamefont{Kaplis}},
  \bibinfo{journal}{Physical Review D} \textbf{\bibinfo{volume}{93}},
  \bibinfo{pages}{066012} (\bibinfo{year}{2016}).

\bibitem[{\citenamefont{Romatschke}(2010)}]{Romatschke:2009im}
\bibinfo{author}{\bibfnamefont{P.}~\bibnamefont{Romatschke}},
  \bibinfo{journal}{Int. J. Mod. Phys.} \textbf{\bibinfo{volume}{E19}},
  \bibinfo{pages}{1} (\bibinfo{year}{2010}), \eprint{0902.3663}.

\bibitem[{\citenamefont{Loganayagam}(2008)}]{Loganayagam:2008is}
\bibinfo{author}{\bibfnamefont{R.}~\bibnamefont{Loganayagam}},
  \bibinfo{journal}{JHEP} \textbf{\bibinfo{volume}{05}}, \bibinfo{pages}{087}
  (\bibinfo{year}{2008}), \eprint{0801.3701}.

\bibitem[{\citenamefont{Fouxon and Oz}(2008)}]{Fouxon:2008tb}
\bibinfo{author}{\bibfnamefont{I.}~\bibnamefont{Fouxon}} \bibnamefont{and}
  \bibinfo{author}{\bibfnamefont{Y.}~\bibnamefont{Oz}}, \bibinfo{journal}{Phys.
  Rev. Lett.} \textbf{\bibinfo{volume}{101}}, \bibinfo{pages}{261602}
  (\bibinfo{year}{2008}), \eprint{0809.4512}.

\bibitem[{\citenamefont{Rodino}(1993)}]{RodinoGevreyBook}
\bibinfo{author}{\bibfnamefont{L.}~\bibnamefont{Rodino}},
  \emph{\bibinfo{title}{Linear partial differential operators in Gevrey
  spaces}} (\bibinfo{publisher}{World Scientific},
  \bibinfo{address}{Singapore}, \bibinfo{year}{1993}).

\bibitem[{\citenamefont{Leray and Ohya}(1967)}]{LerayOhyaNonlinear}
\bibinfo{author}{\bibfnamefont{J.}~\bibnamefont{Leray}} \bibnamefont{and}
  \bibinfo{author}{\bibfnamefont{Y.}~\bibnamefont{Ohya}},
  \bibinfo{journal}{Math. Ann.} \textbf{\bibinfo{volume}{170}},
  \bibinfo{pages}{167} (\bibinfo{year}{1967}), ISSN \bibinfo{issn}{0025-5831}.

\bibitem[{\citenamefont{Choquet-Bruhat}(1966)}]{CB_diagonal}
\bibinfo{author}{\bibfnamefont{Y.}~\bibnamefont{Choquet-Bruhat}},
  \bibinfo{journal}{J. Math. Pures Appl. (9)} \textbf{\bibinfo{volume}{45}},
  \bibinfo{pages}{371} (\bibinfo{year}{1966}), ISSN \bibinfo{issn}{0021-7824}.

\bibitem[{\citenamefont{Courant and
  Hilbert}(1989)}]{Courant_and_Hilbert_book_2}
\bibinfo{author}{\bibfnamefont{C.}~\bibnamefont{Courant}} \bibnamefont{and}
  \bibinfo{author}{\bibfnamefont{D.}~\bibnamefont{Hilbert}},
  \emph{\bibinfo{title}{Methods of Mathematical Physics, Vol. 2.}}
  (\bibinfo{publisher}{Wiley-VCH}, \bibinfo{year}{1989}).

\bibitem[{\citenamefont{Disconzi}(2017)}]{DisconziFollowupBemficaNoronha}
\bibinfo{author}{\bibfnamefont{M.~M.} \bibnamefont{Disconzi}},
  \bibinfo{journal}{arXiv:1708.06572 [math.AP]}  (\bibinfo{year}{2017}),
  \bibinfo{note}{22 pages}.

\bibitem[{\citenamefont{Cowling}(1941)}]{CowlingApproximation}
\bibinfo{author}{\bibfnamefont{T.~G.} \bibnamefont{Cowling}},
  \bibinfo{journal}{Mon. Not. R. Astron. Soc.} \textbf{\bibinfo{volume}{101}},
  \bibinfo{pages}{367} (\bibinfo{year}{1941}).

\bibitem[{\citenamefont{Denicol et~al.}(2008)\citenamefont{Denicol, Kodama,
  Koide, and Mota}}]{Denicol:2008ha}
\bibinfo{author}{\bibfnamefont{G.~S.} \bibnamefont{Denicol}},
  \bibinfo{author}{\bibfnamefont{T.}~\bibnamefont{Kodama}},
  \bibinfo{author}{\bibfnamefont{T.}~\bibnamefont{Koide}}, \bibnamefont{and}
  \bibinfo{author}{\bibfnamefont{P.}~\bibnamefont{Mota}}, \bibinfo{journal}{J.
  Phys.} \textbf{\bibinfo{volume}{G35}}, \bibinfo{pages}{115102}
  (\bibinfo{year}{2008}), \eprint{0807.3120}.

\bibitem[{\citenamefont{Pu et~al.}(2010)\citenamefont{Pu, Koide, and
  Rischke}}]{Pu:2009fj}
\bibinfo{author}{\bibfnamefont{S.}~\bibnamefont{Pu}},
  \bibinfo{author}{\bibfnamefont{T.}~\bibnamefont{Koide}}, \bibnamefont{and}
  \bibinfo{author}{\bibfnamefont{D.~H.} \bibnamefont{Rischke}},
  \bibinfo{journal}{Phys. Rev.} \textbf{\bibinfo{volume}{D81}},
  \bibinfo{pages}{114039} (\bibinfo{year}{2010}), \eprint{0907.3906}.

\bibitem[{\citenamefont{Czajka and Jeon}(2017)}]{Czajka:2017bod}
\bibinfo{author}{\bibfnamefont{A.}~\bibnamefont{Czajka}} \bibnamefont{and}
  \bibinfo{author}{\bibfnamefont{S.}~\bibnamefont{Jeon}},
  \bibinfo{journal}{Phys. Rev.} \textbf{\bibinfo{volume}{C95}},
  \bibinfo{pages}{064906} (\bibinfo{year}{2017}), \eprint{1701.07580}.

\bibitem[{\citenamefont{Israel and Stewart}(1976)}]{MIS-3}
\bibinfo{author}{\bibfnamefont{W.}~\bibnamefont{Israel}} \bibnamefont{and}
  \bibinfo{author}{\bibfnamefont{J.~M.} \bibnamefont{Stewart}},
  \bibinfo{journal}{Phys. Lett. A} \textbf{\bibinfo{volume}{38}},
  \bibinfo{pages}{213} (\bibinfo{year}{1976}).

\bibitem[{\citenamefont{Stewart}(1977)}]{MIS-4}
\bibinfo{author}{\bibfnamefont{J.~M.} \bibnamefont{Stewart}},
  \bibinfo{journal}{Proc. R. Soc. London, Ser. A}
  \textbf{\bibinfo{volume}{357}}, \bibinfo{pages}{59} (\bibinfo{year}{1977}).

\bibitem[{\citenamefont{Israel and Stewart}(1979{\natexlab{b}})}]{MIS-5}
\bibinfo{author}{\bibfnamefont{W.}~\bibnamefont{Israel}} \bibnamefont{and}
  \bibinfo{author}{\bibfnamefont{J.~M.} \bibnamefont{Stewart}},
  \bibinfo{journal}{Proc. R. Soc. London, Ser. A}
  \textbf{\bibinfo{volume}{365}}, \bibinfo{pages}{43}
  (\bibinfo{year}{1979}{\natexlab{b}}).

\bibitem[{\citenamefont{Weinberg}(1972)}]{Weinberg_GR_book}
\bibinfo{author}{\bibfnamefont{S.}~\bibnamefont{Weinberg}},
  \emph{\bibinfo{title}{Gravitation and Cosmology: principles and applications
  of the General Theory of Relativity}} (\bibinfo{publisher}{John Wiley \&
  Sons, Inc.}, \bibinfo{year}{1972}).

\bibitem[{\citenamefont{Cercignani and Kremer}(2002)}]{kremer}
\bibinfo{author}{\bibfnamefont{C.}~\bibnamefont{Cercignani}} \bibnamefont{and}
  \bibinfo{author}{\bibfnamefont{G.~M.} \bibnamefont{Kremer}},
  \emph{\bibinfo{title}{The Relativistic Boltzmann Equation: Theory and
  Applications}} (\bibinfo{publisher}{Birkhauser Verlag},
  \bibinfo{address}{Basel}, \bibinfo{year}{2002}).

\bibitem[{\citenamefont{Denicol
  et~al.}(2014{\natexlab{a}})\citenamefont{Denicol, Heinz, Martinez, Noronha,
  and Strickland}}]{Denicol:2014xca}
\bibinfo{author}{\bibfnamefont{G.~S.} \bibnamefont{Denicol}},
  \bibinfo{author}{\bibfnamefont{U.~W.} \bibnamefont{Heinz}},
  \bibinfo{author}{\bibfnamefont{M.}~\bibnamefont{Martinez}},
  \bibinfo{author}{\bibfnamefont{J.}~\bibnamefont{Noronha}}, \bibnamefont{and}
  \bibinfo{author}{\bibfnamefont{M.}~\bibnamefont{Strickland}},
  \bibinfo{journal}{Phys. Rev. Lett.} \textbf{\bibinfo{volume}{113}},
  \bibinfo{pages}{202301} (\bibinfo{year}{2014}{\natexlab{a}}),
  \eprint{1408.5646}.

\bibitem[{\citenamefont{Denicol
  et~al.}(2014{\natexlab{b}})\citenamefont{Denicol, Heinz, Martinez, , Noronha,
  and Strickland}}]{Denicol:2014tha}
\bibinfo{author}{\bibfnamefont{G.~S.} \bibnamefont{Denicol}},
  \bibinfo{author}{\bibfnamefont{U.~W.} \bibnamefont{Heinz}},
  \bibinfo{author}{\bibfnamefont{M.}~\bibnamefont{Martinez}}, ,
  \bibinfo{author}{\bibfnamefont{J.}~\bibnamefont{Noronha}}, \bibnamefont{and}
  \bibinfo{author}{\bibfnamefont{M.}~\bibnamefont{Strickland}},
  \bibinfo{journal}{Phys. Rev.} \textbf{\bibinfo{volume}{D90}},
  \bibinfo{pages}{125026} (\bibinfo{year}{2014}{\natexlab{b}}),
  \eprint{1408.7048}.

\bibitem[{\citenamefont{Cercignani}(1988)}]{Cercignani}
\bibinfo{author}{\bibfnamefont{C.}~\bibnamefont{Cercignani}},
  \emph{\bibinfo{title}{The Boltzmann Equation and Its Applications}}
  (\bibinfo{publisher}{Springer-Verlag, New York}, \bibinfo{year}{1988}).

\bibitem[{\citenamefont{Bazow et~al.}(2016{\natexlab{b}})\citenamefont{Bazow,
  Denicol, Heinz, Martinez, and Noronha}}]{Bazow:2015dha}
\bibinfo{author}{\bibfnamefont{D.}~\bibnamefont{Bazow}},
  \bibinfo{author}{\bibfnamefont{G.~S.} \bibnamefont{Denicol}},
  \bibinfo{author}{\bibfnamefont{U.}~\bibnamefont{Heinz}},
  \bibinfo{author}{\bibfnamefont{M.}~\bibnamefont{Martinez}}, \bibnamefont{and}
  \bibinfo{author}{\bibfnamefont{J.}~\bibnamefont{Noronha}},
  \bibinfo{journal}{Phys. Rev. Lett.} \textbf{\bibinfo{volume}{116}},
  \bibinfo{pages}{022301} (\bibinfo{year}{2016}{\natexlab{b}}),
  \eprint{1507.07834}.

\bibitem[{\citenamefont{Bazow et~al.}(2016{\natexlab{c}})\citenamefont{Bazow,
  Denicol, Heinz, Martinez, and Noronha}}]{Bazow:2016oky}
\bibinfo{author}{\bibfnamefont{D.}~\bibnamefont{Bazow}},
  \bibinfo{author}{\bibfnamefont{G.~S.} \bibnamefont{Denicol}},
  \bibinfo{author}{\bibfnamefont{U.}~\bibnamefont{Heinz}},
  \bibinfo{author}{\bibfnamefont{M.}~\bibnamefont{Martinez}}, \bibnamefont{and}
  \bibinfo{author}{\bibfnamefont{J.}~\bibnamefont{Noronha}},
  \bibinfo{journal}{Phys. Rev.} \textbf{\bibinfo{volume}{D94}},
  \bibinfo{pages}{125006} (\bibinfo{year}{2016}{\natexlab{c}}),
  \eprint{1607.05245}.

\bibitem[{\citenamefont{Chapman and Cowling}(1970)}]{ChapmanCowling}
\bibinfo{author}{\bibfnamefont{S.}~\bibnamefont{Chapman}} \bibnamefont{and}
  \bibinfo{author}{\bibfnamefont{T.~G.} \bibnamefont{Cowling}},
  \emph{\bibinfo{title}{The mathematical theory of non-uniform gases, 3rd
  edition}} (\bibinfo{publisher}{Cambridge University Press},
  \bibinfo{year}{1970}).

\bibitem[{\citenamefont{Denicol et~al.}(2011)\citenamefont{Denicol, Noronha,
  Niemi, and Rischke}}]{Denicol:2011fa}
\bibinfo{author}{\bibfnamefont{G.~S.} \bibnamefont{Denicol}},
  \bibinfo{author}{\bibfnamefont{J.}~\bibnamefont{Noronha}},
  \bibinfo{author}{\bibfnamefont{H.}~\bibnamefont{Niemi}}, \bibnamefont{and}
  \bibinfo{author}{\bibfnamefont{D.~H.} \bibnamefont{Rischke}},
  \bibinfo{journal}{Phys. Rev.} \textbf{\bibinfo{volume}{D83}},
  \bibinfo{pages}{074019} (\bibinfo{year}{2011}), \eprint{1102.4780}.

\bibitem[{\citenamefont{Rischke}(1999)}]{RischkeCollisions}
\bibinfo{author}{\bibfnamefont{D.}~\bibnamefont{Rischke}}, in
  \emph{\bibinfo{booktitle}{Hadrons in Dense Matter and Hadrosynthesis
  (Cleymans J., Geyer H.B., Scholtz F.G. (eds))}}
  (\bibinfo{publisher}{Springer, Berlin, Heidelberg}, \bibinfo{year}{1999}),
  vol. \bibinfo{volume}{516} of \emph{\bibinfo{series}{Lecture Notes in
  Physics}}, pp. \bibinfo{pages}{105--181}.

\bibitem[{\citenamefont{Tsumura and Kunihiro}(2008)}]{Tsumura:2007wu}
\bibinfo{author}{\bibfnamefont{K.}~\bibnamefont{Tsumura}} \bibnamefont{and}
  \bibinfo{author}{\bibfnamefont{T.}~\bibnamefont{Kunihiro}},
  \bibinfo{journal}{Phys. Lett.} \textbf{\bibinfo{volume}{B668}},
  \bibinfo{pages}{425} (\bibinfo{year}{2008}), \eprint{0709.3645}.

\bibitem[{\citenamefont{Van and Biro}(2008)}]{Van:2007pw}
\bibinfo{author}{\bibfnamefont{P.}~\bibnamefont{Van}} \bibnamefont{and}
  \bibinfo{author}{\bibfnamefont{T.~S.} \bibnamefont{Biro}},
  \bibinfo{journal}{Eur. Phys. J. ST} \textbf{\bibinfo{volume}{155}},
  \bibinfo{pages}{201} (\bibinfo{year}{2008}), \eprint{0704.2039}.

\bibitem[{\citenamefont{Van and Biro}(2012)}]{Van:2011yn}
\bibinfo{author}{\bibfnamefont{P.}~\bibnamefont{Van}} \bibnamefont{and}
  \bibinfo{author}{\bibfnamefont{T.~S.} \bibnamefont{Biro}},
  \bibinfo{journal}{Phys. Lett.} \textbf{\bibinfo{volume}{B709}},
  \bibinfo{pages}{106} (\bibinfo{year}{2012}), \eprint{1109.0985}.

\bibitem[{\citenamefont{Tsumura and Kunihiro}(2013)}]{Tsumura:2012ss}
\bibinfo{author}{\bibfnamefont{K.}~\bibnamefont{Tsumura}} \bibnamefont{and}
  \bibinfo{author}{\bibfnamefont{T.}~\bibnamefont{Kunihiro}},
  \bibinfo{journal}{Phys. Rev.} \textbf{\bibinfo{volume}{E87}},
  \bibinfo{pages}{053008} (\bibinfo{year}{2013}), \eprint{1206.3913}.

\bibitem[{\citenamefont{Bjorken}(1983)}]{Bjorken:1982qr}
\bibinfo{author}{\bibfnamefont{J.~D.} \bibnamefont{Bjorken}},
  \bibinfo{journal}{Phys. Rev.} \textbf{\bibinfo{volume}{D27}},
  \bibinfo{pages}{140} (\bibinfo{year}{1983}).

\bibitem[{\citenamefont{Danielewicz and Gyulassy}(1985)}]{Danielewicz:1984ww}
\bibinfo{author}{\bibfnamefont{P.}~\bibnamefont{Danielewicz}} \bibnamefont{and}
  \bibinfo{author}{\bibfnamefont{M.}~\bibnamefont{Gyulassy}},
  \bibinfo{journal}{Phys. Rev.} \textbf{\bibinfo{volume}{D31}},
  \bibinfo{pages}{53} (\bibinfo{year}{1985}).

\bibitem[{\citenamefont{Heller et~al.}(2012)\citenamefont{Heller, Janik, and
  Witaszczyk}}]{Heller:2011ju}
\bibinfo{author}{\bibfnamefont{M.~P.} \bibnamefont{Heller}},
  \bibinfo{author}{\bibfnamefont{R.~A.} \bibnamefont{Janik}}, \bibnamefont{and}
  \bibinfo{author}{\bibfnamefont{P.}~\bibnamefont{Witaszczyk}},
  \bibinfo{journal}{Phys. Rev. Lett.} \textbf{\bibinfo{volume}{108}},
  \bibinfo{pages}{201602} (\bibinfo{year}{2012}), \eprint{1103.3452}.

\bibitem[{\citenamefont{Heller and Spalinski}(2015)}]{Heller:2015dha}
\bibinfo{author}{\bibfnamefont{M.~P.} \bibnamefont{Heller}} \bibnamefont{and}
  \bibinfo{author}{\bibfnamefont{M.}~\bibnamefont{Spalinski}},
  \bibinfo{journal}{Phys. Rev. Lett.} \textbf{\bibinfo{volume}{115}},
  \bibinfo{pages}{072501} (\bibinfo{year}{2015}), \eprint{1503.07514}.

\bibitem[{\citenamefont{Heller et~al.}(2013)\citenamefont{Heller, Janik, and
  Witaszczyk}}]{Heller:2013fn}
\bibinfo{author}{\bibfnamefont{M.~P.} \bibnamefont{Heller}},
  \bibinfo{author}{\bibfnamefont{R.~A.} \bibnamefont{Janik}}, \bibnamefont{and}
  \bibinfo{author}{\bibfnamefont{P.}~\bibnamefont{Witaszczyk}},
  \bibinfo{journal}{Phys. Rev. Lett.} \textbf{\bibinfo{volume}{110}},
  \bibinfo{pages}{211602} (\bibinfo{year}{2013}), \eprint{1302.0697}.

\bibitem[{\citenamefont{Buchel et~al.}(2016)\citenamefont{Buchel, Heller, and
  Noronha}}]{Buchel:2016cbj}
\bibinfo{author}{\bibfnamefont{A.}~\bibnamefont{Buchel}},
  \bibinfo{author}{\bibfnamefont{M.~P.} \bibnamefont{Heller}},
  \bibnamefont{and} \bibinfo{author}{\bibfnamefont{J.}~\bibnamefont{Noronha}},
  \bibinfo{journal}{Phys. Rev.} \textbf{\bibinfo{volume}{D94}},
  \bibinfo{pages}{106011} (\bibinfo{year}{2016}), \eprint{1603.05344}.

\bibitem[{\citenamefont{Denicol and Noronha}(2016)}]{Denicol:2016bjh}
\bibinfo{author}{\bibfnamefont{G.~S.} \bibnamefont{Denicol}} \bibnamefont{and}
  \bibinfo{author}{\bibfnamefont{J.}~\bibnamefont{Noronha}}
  (\bibinfo{year}{2016}), \eprint{1608.07869}.

\bibitem[{\citenamefont{Heller et~al.}(2016)\citenamefont{Heller, Kurkela, and
  Spalinski}}]{Heller:2016rtz}
\bibinfo{author}{\bibfnamefont{M.~P.} \bibnamefont{Heller}},
  \bibinfo{author}{\bibfnamefont{A.}~\bibnamefont{Kurkela}}, \bibnamefont{and}
  \bibinfo{author}{\bibfnamefont{M.}~\bibnamefont{Spalinski}}
  (\bibinfo{year}{2016}), \eprint{1609.04803}.

\bibitem[{\citenamefont{Basar and Dunne}(2015)}]{Basar:2015ava}
\bibinfo{author}{\bibfnamefont{G.}~\bibnamefont{Basar}} \bibnamefont{and}
  \bibinfo{author}{\bibfnamefont{G.~V.} \bibnamefont{Dunne}},
  \bibinfo{journal}{Phys. Rev.} \textbf{\bibinfo{volume}{D92}},
  \bibinfo{pages}{125011} (\bibinfo{year}{2015}), \eprint{1509.05046}.

\bibitem[{\citenamefont{Aniceto and Spali{\'n}ski}(2016)}]{Aniceto:2015mto}
\bibinfo{author}{\bibfnamefont{I.}~\bibnamefont{Aniceto}} \bibnamefont{and}
  \bibinfo{author}{\bibfnamefont{M.}~\bibnamefont{Spali{\'n}ski}},
  \bibinfo{journal}{Phys. Rev.} \textbf{\bibinfo{volume}{D93}},
  \bibinfo{pages}{085008} (\bibinfo{year}{2016}), \eprint{1511.06358}.

\bibitem[{\citenamefont{Florkowski et~al.}(2016)\citenamefont{Florkowski,
  Ryblewski, and Spali{\'n}ski}}]{Florkowski:2016zsi}
\bibinfo{author}{\bibfnamefont{W.}~\bibnamefont{Florkowski}},
  \bibinfo{author}{\bibfnamefont{R.}~\bibnamefont{Ryblewski}},
  \bibnamefont{and}
  \bibinfo{author}{\bibfnamefont{M.}~\bibnamefont{Spali{\'n}ski}},
  \bibinfo{journal}{Phys. Rev.} \textbf{\bibinfo{volume}{D94}},
  \bibinfo{pages}{114025} (\bibinfo{year}{2016}), \eprint{1608.07558}.

\bibitem[{\citenamefont{Florkowski et~al.}(2017)\citenamefont{Florkowski,
  Heller, and Spalinski}}]{Florkowski:2017olj}
\bibinfo{author}{\bibfnamefont{W.}~\bibnamefont{Florkowski}},
  \bibinfo{author}{\bibfnamefont{M.~P.} \bibnamefont{Heller}},
  \bibnamefont{and} \bibinfo{author}{\bibfnamefont{M.}~\bibnamefont{Spalinski}}
  (\bibinfo{year}{2017}), \eprint{1707.02282}.

\bibitem[{\citenamefont{Liddle et~al.}(1994)\citenamefont{Liddle, Parsons, and
  Barrow}}]{Liddle:1994dx}
\bibinfo{author}{\bibfnamefont{A.~R.} \bibnamefont{Liddle}},
  \bibinfo{author}{\bibfnamefont{P.}~\bibnamefont{Parsons}}, \bibnamefont{and}
  \bibinfo{author}{\bibfnamefont{J.~D.} \bibnamefont{Barrow}},
  \bibinfo{journal}{Phys. Rev.} \textbf{\bibinfo{volume}{D50}},
  \bibinfo{pages}{7222} (\bibinfo{year}{1994}), \eprint{astro-ph/9408015}.

\bibitem[{\citenamefont{Gubser}(2010)}]{Gubser:2010ze}
\bibinfo{author}{\bibfnamefont{S.~S.} \bibnamefont{Gubser}},
  \bibinfo{journal}{Phys. Rev.} \textbf{\bibinfo{volume}{D82}},
  \bibinfo{pages}{085027} (\bibinfo{year}{2010}), \eprint{1006.0006}.

\bibitem[{\citenamefont{Gubser and Yarom}(2011)}]{Gubser:2010ui}
\bibinfo{author}{\bibfnamefont{S.~S.} \bibnamefont{Gubser}} \bibnamefont{and}
  \bibinfo{author}{\bibfnamefont{A.}~\bibnamefont{Yarom}},
  \bibinfo{journal}{Nucl. Phys.} \textbf{\bibinfo{volume}{B846}},
  \bibinfo{pages}{469} (\bibinfo{year}{2011}), \eprint{1012.1314}.

\bibitem[{\citenamefont{Marrochio et~al.}(2015)\citenamefont{Marrochio,
  Noronha, Denicol, Luzum, Jeon, and Gale}}]{Marrochio:2013wla}
\bibinfo{author}{\bibfnamefont{H.}~\bibnamefont{Marrochio}},
  \bibinfo{author}{\bibfnamefont{J.}~\bibnamefont{Noronha}},
  \bibinfo{author}{\bibfnamefont{G.~S.} \bibnamefont{Denicol}},
  \bibinfo{author}{\bibfnamefont{M.}~\bibnamefont{Luzum}},
  \bibinfo{author}{\bibfnamefont{S.}~\bibnamefont{Jeon}}, \bibnamefont{and}
  \bibinfo{author}{\bibfnamefont{C.}~\bibnamefont{Gale}},
  \bibinfo{journal}{Phys. Rev.} \textbf{\bibinfo{volume}{C91}},
  \bibinfo{pages}{014903} (\bibinfo{year}{2015}), \eprint{1307.6130}.

\bibitem[{\citenamefont{Nopoush et~al.}(2015)\citenamefont{Nopoush, Ryblewski,
  and Strickland}}]{Nopoush:2014qba}
\bibinfo{author}{\bibfnamefont{M.}~\bibnamefont{Nopoush}},
  \bibinfo{author}{\bibfnamefont{R.}~\bibnamefont{Ryblewski}},
  \bibnamefont{and}
  \bibinfo{author}{\bibfnamefont{M.}~\bibnamefont{Strickland}},
  \bibinfo{journal}{Phys. Rev.} \textbf{\bibinfo{volume}{D91}},
  \bibinfo{pages}{045007} (\bibinfo{year}{2015}), \eprint{1410.6790}.

\bibitem[{\citenamefont{Noronha and Denicol}(2015)}]{Noronha:2015jia}
\bibinfo{author}{\bibfnamefont{J.}~\bibnamefont{Noronha}} \bibnamefont{and}
  \bibinfo{author}{\bibfnamefont{G.~S.} \bibnamefont{Denicol}},
  \bibinfo{journal}{Phys. Rev.} \textbf{\bibinfo{volume}{D92}},
  \bibinfo{pages}{114032} (\bibinfo{year}{2015}), \eprint{1502.05892}.

\bibitem[{\citenamefont{Hatta et~al.}(2014{\natexlab{a}})\citenamefont{Hatta,
  Noronha, and Xiao}}]{Hatta:2014gqa}
\bibinfo{author}{\bibfnamefont{Y.}~\bibnamefont{Hatta}},
  \bibinfo{author}{\bibfnamefont{J.}~\bibnamefont{Noronha}}, \bibnamefont{and}
  \bibinfo{author}{\bibfnamefont{B.-W.} \bibnamefont{Xiao}},
  \bibinfo{journal}{Phys. Rev.} \textbf{\bibinfo{volume}{D89}},
  \bibinfo{pages}{051702} (\bibinfo{year}{2014}{\natexlab{a}}),
  \eprint{1401.6248}.

\bibitem[{\citenamefont{Hatta et~al.}(2014{\natexlab{b}})\citenamefont{Hatta,
  Noronha, and Xiao}}]{Hatta:2014gga}
\bibinfo{author}{\bibfnamefont{Y.}~\bibnamefont{Hatta}},
  \bibinfo{author}{\bibfnamefont{J.}~\bibnamefont{Noronha}}, \bibnamefont{and}
  \bibinfo{author}{\bibfnamefont{B.-W.} \bibnamefont{Xiao}},
  \bibinfo{journal}{Phys. Rev.} \textbf{\bibinfo{volume}{D89}},
  \bibinfo{pages}{114011} (\bibinfo{year}{2014}{\natexlab{b}}),
  \eprint{1403.7693}.

\bibitem[{\citenamefont{Behtash et~al.}(2018)\citenamefont{Behtash,
  Cruz-Camacho, and Martinez}}]{Behtash:2017wqg}
\bibinfo{author}{\bibfnamefont{A.}~\bibnamefont{Behtash}},
  \bibinfo{author}{\bibfnamefont{C.~N.} \bibnamefont{Cruz-Camacho}},
  \bibnamefont{and} \bibinfo{author}{\bibfnamefont{M.}~\bibnamefont{Martinez}},
  \bibinfo{journal}{Phys. Rev.} \textbf{\bibinfo{volume}{D97}},
  \bibinfo{pages}{044041} (\bibinfo{year}{2018}), \eprint{1711.01745}.

\bibitem[{\citenamefont{Denicol and Noronha}(2018)}]{Denicol:2018pak}
\bibinfo{author}{\bibfnamefont{G.~S.} \bibnamefont{Denicol}} \bibnamefont{and}
  \bibinfo{author}{\bibfnamefont{J.}~\bibnamefont{Noronha}}
  (\bibinfo{year}{2018}), \eprint{1804.04771}.

\bibitem[{\citenamefont{Schenke et~al.}(2012)\citenamefont{Schenke, Tribedy,
  and Venugopalan}}]{Schenke:2012wb}
\bibinfo{author}{\bibfnamefont{B.}~\bibnamefont{Schenke}},
  \bibinfo{author}{\bibfnamefont{P.}~\bibnamefont{Tribedy}}, \bibnamefont{and}
  \bibinfo{author}{\bibfnamefont{R.}~\bibnamefont{Venugopalan}},
  \bibinfo{journal}{Phys. Rev. Lett.} \textbf{\bibinfo{volume}{108}},
  \bibinfo{pages}{252301} (\bibinfo{year}{2012}), \eprint{1202.6646}.

\bibitem[{\citenamefont{Torrieri and Mishustin}(2008)}]{Torrieri:2008ip}
\bibinfo{author}{\bibfnamefont{G.}~\bibnamefont{Torrieri}} \bibnamefont{and}
  \bibinfo{author}{\bibfnamefont{I.}~\bibnamefont{Mishustin}},
  \bibinfo{journal}{Phys. Rev.} \textbf{\bibinfo{volume}{C78}},
  \bibinfo{pages}{021901} (\bibinfo{year}{2008}), \eprint{0805.0442}.

\bibitem[{\citenamefont{Rajagopal and Tripuraneni}(2010)}]{Rajagopal:2009yw}
\bibinfo{author}{\bibfnamefont{K.}~\bibnamefont{Rajagopal}} \bibnamefont{and}
  \bibinfo{author}{\bibfnamefont{N.}~\bibnamefont{Tripuraneni}},
  \bibinfo{journal}{JHEP} \textbf{\bibinfo{volume}{03}}, \bibinfo{pages}{018}
  (\bibinfo{year}{2010}), \eprint{0908.1785}.

\bibitem[{\citenamefont{Denicol et~al.}(2015)\citenamefont{Denicol, Gale, and
  Jeon}}]{Denicol:2015bpa}
\bibinfo{author}{\bibfnamefont{G.~S.} \bibnamefont{Denicol}},
  \bibinfo{author}{\bibfnamefont{C.}~\bibnamefont{Gale}}, \bibnamefont{and}
  \bibinfo{author}{\bibfnamefont{S.}~\bibnamefont{Jeon}},
  \bibinfo{journal}{PoS} \textbf{\bibinfo{volume}{CPOD2014}},
  \bibinfo{pages}{033} (\bibinfo{year}{2015}), \eprint{1503.00531}.

\bibitem[{\citenamefont{Bae et~al.}(2012)\citenamefont{Bae, Biswas, and
  Tadmor}}]{TadmorBesovGevrey}
\bibinfo{author}{\bibfnamefont{H.}~\bibnamefont{Bae}},
  \bibinfo{author}{\bibfnamefont{A.}~\bibnamefont{Biswas}}, \bibnamefont{and}
  \bibinfo{author}{\bibfnamefont{E.}~\bibnamefont{Tadmor}},
  \bibinfo{journal}{Archive for Rational Mechanics and Analysis}
  \textbf{\bibinfo{volume}{205}}, \bibinfo{pages}{963} (\bibinfo{year}{2012}).

\bibitem[{\citenamefont{Cao et~al.}(1999)\citenamefont{Cao, Rammaha, and
  Titi}}]{TitiGevreyNavier}
\bibinfo{author}{\bibfnamefont{C.}~\bibnamefont{Cao}},
  \bibinfo{author}{\bibfnamefont{M.~A.} \bibnamefont{Rammaha}},
  \bibnamefont{and} \bibinfo{author}{\bibfnamefont{E.~S.} \bibnamefont{Titi}},
  \bibinfo{journal}{Zeitschrift f\"ur Angewandte Mathematik und Physik (ZAMP)}
  \textbf{\bibinfo{volume}{50}}, \bibinfo{pages}{341} (\bibinfo{year}{1999}).

\bibitem[{\citenamefont{Ferrari and Titi}(1998)}]{TitiGevreyParabolic}
\bibinfo{author}{\bibfnamefont{A.~B.} \bibnamefont{Ferrari}} \bibnamefont{and}
  \bibinfo{author}{\bibfnamefont{E.~S.} \bibnamefont{Titi}},
  \bibinfo{journal}{Communications in Partial Differential Equations}
  \textbf{\bibinfo{volume}{23}}, \bibinfo{pages}{424} (\bibinfo{year}{1998}).

\bibitem[{\citenamefont{Foias and Temam}(1989)}]{TemamGevrey}
\bibinfo{author}{\bibfnamefont{C.}~\bibnamefont{Foias}} \bibnamefont{and}
  \bibinfo{author}{\bibfnamefont{R.}~\bibnamefont{Temam}}, \bibinfo{journal}{J.
  Funct. Anal.} \textbf{\bibinfo{volume}{87}}, \bibinfo{pages}{359}
  (\bibinfo{year}{1989}).

\bibitem[{\citenamefont{Lichnerowicz}(1965)}]{Lich_MHD_paper}
\bibinfo{author}{\bibfnamefont{A.}~\bibnamefont{Lichnerowicz}},
  \bibinfo{journal}{C. R. Acad. Sci. Paris} \textbf{\bibinfo{volume}{260}},
  \bibinfo{pages}{4449} (\bibinfo{year}{1965}).

\bibitem[{\citenamefont{Lichnerowicz}(1967)}]{Lichnerowicz_MHD_book}
\bibinfo{author}{\bibfnamefont{A.}~\bibnamefont{Lichnerowicz}},
  \emph{\bibinfo{title}{Relativistic Hydrodynamics and Magnetohydrodynamics:
  Lectures on the Existence of Solutions}} (\bibinfo{publisher}{W. A.
  Benjamin}, \bibinfo{address}{New York}, \bibinfo{year}{1967}).

\bibitem[{\citenamefont{Friedrich and Rendall}(2000)}]{FriRenCauchy}
\bibinfo{author}{\bibfnamefont{H.}~\bibnamefont{Friedrich}} \bibnamefont{and}
  \bibinfo{author}{\bibfnamefont{A.~D.} \bibnamefont{Rendall}},
  \bibinfo{journal}{Lect. Notes Phys.} \textbf{\bibinfo{volume}{540}},
  \bibinfo{pages}{127} (\bibinfo{year}{2000}).

\bibitem[{\citenamefont{Anile and Pennisi}(1987)}]{AnilePennisiMHD}
\bibinfo{author}{\bibfnamefont{A.~M.} \bibnamefont{Anile}} \bibnamefont{and}
  \bibinfo{author}{\bibfnamefont{S.}~\bibnamefont{Pennisi}},
  \bibinfo{journal}{Ann. Inst. H. Poincar\'e Phys. Th\'eor.}
  \textbf{\bibinfo{volume}{46}}, \bibinfo{pages}{27} (\bibinfo{year}{1987}),
  ISSN \bibinfo{issn}{0246-0211},
  \urlprefix\url{http://www.numdam.org/item?id=AIHPB_1987__46_1_27_0}.

\bibitem[{\citenamefont{Christodoulou}(2009)}]{ChristodoulouBlackHoles}
\bibinfo{author}{\bibfnamefont{D.}~\bibnamefont{Christodoulou}},
  \emph{\bibinfo{title}{The formation of black holes in general relativity}},
  EMS Monographs in Mathematics (\bibinfo{publisher}{European Mathematical
  Society (EMS), Z\"urich}, \bibinfo{year}{2009}), ISBN
  \bibinfo{isbn}{978-3-03719-068-5}.

\bibitem[{\citenamefont{Majda}(1984)}]{MajdaCompressibleFlow}
\bibinfo{author}{\bibfnamefont{A.}~\bibnamefont{Majda}},
  \emph{\bibinfo{title}{Compressible fluid flow and systems of conservation
  laws in several space variables}}, vol.~\bibinfo{volume}{53} of
  \emph{\bibinfo{series}{Applied Mathematical Sciences}}
  (\bibinfo{publisher}{Springer-Verlag, New York}, \bibinfo{year}{1984}), ISBN
  \bibinfo{isbn}{0-387-96037-6},
  \urlprefix\url{http://dx.doi.org/10.1007/978-1-4612-1116-7}.

\bibitem[{\citenamefont{Coutand and
  Shkoller}(2007)}]{CoutandShkollerFreeBoundary}
\bibinfo{author}{\bibfnamefont{D.}~\bibnamefont{Coutand}} \bibnamefont{and}
  \bibinfo{author}{\bibfnamefont{S.}~\bibnamefont{Shkoller}},
  \bibinfo{journal}{J. Amer. Math. Soc.} \textbf{\bibinfo{volume}{20}},
  \bibinfo{pages}{829} (\bibinfo{year}{2007}), ISSN \bibinfo{issn}{0894-0347}.

\bibitem[{\citenamefont{Coutand et~al.}(2013)\citenamefont{Coutand, Hole, and
  Shkoller}}]{CoutandHoleShkollerLimit}
\bibinfo{author}{\bibfnamefont{D.}~\bibnamefont{Coutand}},
  \bibinfo{author}{\bibfnamefont{J.}~\bibnamefont{Hole}}, \bibnamefont{and}
  \bibinfo{author}{\bibfnamefont{S.}~\bibnamefont{Shkoller}},
  \bibinfo{journal}{SIAM J. Math. Anal.} \textbf{\bibinfo{volume}{45}},
  \bibinfo{pages}{3690} (\bibinfo{year}{2013}), ISSN \bibinfo{issn}{0036-1410}.

\bibitem[{\citenamefont{Disconzi and Ebin}(2014)}]{DisconziEbinFreeBoundary2d}
\bibinfo{author}{\bibfnamefont{M.~M.} \bibnamefont{Disconzi}} \bibnamefont{and}
  \bibinfo{author}{\bibfnamefont{D.~G.} \bibnamefont{Ebin}},
  \bibinfo{journal}{Comm. Partial Differential Equations}
  \textbf{\bibinfo{volume}{39}}, \bibinfo{pages}{740} (\bibinfo{year}{2014}),
  ISSN \bibinfo{issn}{0360-5302}.

\bibitem[{\citenamefont{Disconzi and Ebin}(2016)}]{DisconziEbinFreeBoundary3d}
\bibinfo{author}{\bibfnamefont{M.~M.} \bibnamefont{Disconzi}} \bibnamefont{and}
  \bibinfo{author}{\bibfnamefont{D.~G.} \bibnamefont{Ebin}},
  \bibinfo{journal}{Journal of Differential Equations}
  \textbf{\bibinfo{volume}{261}}, \bibinfo{pages}{821} (\bibinfo{year}{2016}).

\bibitem[{\citenamefont{Lindblad}(2005{\natexlab{a}})}]{LindbladFreeBoundary}
\bibinfo{author}{\bibfnamefont{H.}~\bibnamefont{Lindblad}},
  \bibinfo{journal}{Ann. of Math. (2)} \textbf{\bibinfo{volume}{162}},
  \bibinfo{pages}{109} (\bibinfo{year}{2005}{\natexlab{a}}), ISSN
  \bibinfo{issn}{0003-486X}.

\bibitem[{\citenamefont{Lindblad}(2005{\natexlab{b}})}]{Lindblad-FreeBoundaryCompressbile}
\bibinfo{author}{\bibfnamefont{H.}~\bibnamefont{Lindblad}},
  \bibinfo{journal}{Comm. Math. Phys.} \textbf{\bibinfo{volume}{260}},
  \bibinfo{pages}{319} (\bibinfo{year}{2005}{\natexlab{b}}), ISSN
  \bibinfo{issn}{0010-3616}.

\bibitem[{\citenamefont{Jang et~al.}(2016)\citenamefont{Jang, LeFloch, and
  Masmoudi}}]{JangLeFlochMasmoudi}
\bibinfo{author}{\bibfnamefont{J.}~\bibnamefont{Jang}},
  \bibinfo{author}{\bibfnamefont{P.~G.} \bibnamefont{LeFloch}},
  \bibnamefont{and} \bibinfo{author}{\bibfnamefont{N.}~\bibnamefont{Masmoudi}},
  \bibinfo{journal}{Journal of Differential Equations}
  \textbf{\bibinfo{volume}{260}}, \bibinfo{pages}{5481} (\bibinfo{year}{2016}).

\bibitem[{\citenamefont{Had{\v{z}}i{\'c}
  et~al.}(2015)\citenamefont{Had{\v{z}}i{\'c}, Shkoller, and
  Speck}}]{HadzicShkollerSpeck}
\bibinfo{author}{\bibfnamefont{M.}~\bibnamefont{Had{\v{z}}i{\'c}}},
  \bibinfo{author}{\bibfnamefont{S.}~\bibnamefont{Shkoller}}, \bibnamefont{and}
  \bibinfo{author}{\bibfnamefont{J.}~\bibnamefont{Speck}},
  \bibinfo{journal}{arXiv:1511.07467 [math.AP]}  (\bibinfo{year}{2015}).

\bibitem[{\citenamefont{Rendall}(1992)}]{RendallFluidBodies}
\bibinfo{author}{\bibfnamefont{A.~D.} \bibnamefont{Rendall}},
  \bibinfo{journal}{J. Math. Phys.} \textbf{\bibinfo{volume}{33}},
  \bibinfo{pages}{1047} (\bibinfo{year}{1992}), ISSN \bibinfo{issn}{0022-2488},
  \urlprefix\url{http://dx.doi.org/10.1063/1.529766}.

\end{thebibliography}

\end{document}